\documentclass[11pt]{report}
\usepackage[
  dissertation
 ,final
 ,raggedbottom
]{USCthesis}

\usepackage[export]{adjustbox} % for frame option in \includegraphics
\usepackage{amsmath}
\usepackage{amssymb}
\usepackage{array}
\usepackage[utf8]{inputenc} % load inputenc before csquotes
\usepackage[english]{babel}
\usepackage{booktabs}
\usepackage{color, colortbl}
\usepackage{csquotes}
\usepackage{efbox}
\usepackage{enumitem}
\usepackage[shortcuts]{extdash} % use `\-/' to hyphenate words/phrases that have a dash in them
\usepackage[tt=false]{libertine} % libertine's \ttfamily isn't that great
\usepackage[T1]{fontenc} % load fonts before fontenc
\usepackage[symbol]{footmisc}
\usepackage[
  showframe = false,% draw a border around textwidth
  pass      = true, % force 8.5"x11" pagesize
]{geometry}
\usepackage{graphicx}
\usepackage{hyphenat}
\usepackage{ifthen}
\usepackage{lipsum}
\usepackage{multirow}
\usepackage{parnotes}
\usepackage{pdflscape} % rotate some pages in an {landscape} environment
\usepackage{pifont}
\usepackage{ragged2e}
\usepackage{seqsplit}
\usepackage{siunitx}
\usepackage{subcaption}
\usepackage[graphicx]{realboxes}
\usepackage{tabularx}
\usepackage{xcolor}
\usepackage{makecell}
\usepackage{todonotes}
\usepackage{xspace}
\usepackage[all]{nowidow}
\usepackage{url}
\newcommand{\cmark}{\ding{51}}%
\newcommand{\xmark}{\ding{55}}%
\usepackage{xurl}
\usepackage[
  breaklinks    = true,
  colorlinks    = true,
  hypertexnames = false,
  pdfpagelabels = false,
  citecolor     = {blue!80!black},
  linkcolor     = {blue!80!black},
  urlcolor      = {blue!80!black},
]{hyperref} % load hyperref as the last package

\usepackage{natbib}
\renewcommand\cite{\citep}	% to get "(Author Year)" with natbib    
\addto\extrasenglish{%

}

\usepackage{etoolbox}
\newtoggle{draft}
\settoggle{draft}{true} % change toggle for draft or final versions

\iftoggle{draft}{
  \overfullrule=10pt                       % highlight overfull hboxes
}{
  \PassOptionsToPackage{final}{showlabels} % hide labels on figures, etc
}

\newtoggle{thesis}
\settoggle{thesis}{true}

\newcolumntype{L}[1]{>{\raggedright\let\newline\\\arraybackslash\hspace{0pt}}m{#1}}
\newcolumntype{C}[1]{>{\centering\let\newline\\\arraybackslash\hspace{0pt}}m{#1}}
\newcolumntype{R}[1]{>{\raggedleft\let\newline\\\arraybackslash\hspace{0pt}}m{#1}}

\begin{document}

% title should be all caps
\title{SOCIALLY-INFORMED CONTENT ANALYSIS OF ONLINE HUMAN BEHAVIOR}

% use your full name!
% https://cs.stanford.edu/~knuth/news19.html
% "Let's celebrate everybody's full names"
\author{Julie Jiang}

% major should be all caps
\majorfield{COMPUTER SCIENCE}

% date should be May, August, or December (when degrees are conferred)
\submitdate{May 2024}

%%% preface %%%%%%%%%%%%%%%%%%%%%%%%%%%%%%%%%%%%%%%%%%%%%%%%%%%%%%%%%%%%
\begin{preface}
  % \prefacesection{Dedication}
  \begin{center}
    \thispagestyle{empty}
    \vspace*{\fill}
      \textit{To my little sister.}
    \vspace*{\fill}
\end{center}

  \prefacesection{Acknowledgments}
  This dissertation represents not just the culmination of my PhD research but also the symbolic distillation of five fantastic years of my 20s. For this, I have many people to thank. This journey of professional and personal growth has been made possible only by the constellation of people who have generously shared their mentorship, wisdom, camaraderie, friendship, and love with me. At the core of this dissertation on human behavior is my belief that the very essence of our humanity lies within our interpersonal interactions. It is with great privilege and honor that I extend my heartfelt thanks within these pages to those who have enriched my life.

First and foremost, I thank my amazing advisor, Emilio Ferrara. His remarkable research acumen, sharp wit, and occasional dry humor have been a constant source of inspiration. Without his guidance, I would not have carved the path to becoming a computational social scientist, forging my own research journey. Emilio’s insightful, constructive feedback, coupled with his generous encouragement, has been invaluable. I thank Kristina Lerman, whose pivotal role in my emerging research career cannot be overstated. Her endless stream of new and exciting research ideas continually impresses me. My appreciation also extends to my dissertation committee members, Pablo Barberá and Marlon Twyman, whose thoughtful and engaging comments have made this dissertation extra awesome, as well as my previous committee members and collaborators, Xiang Ren and Jesse Thomason, who helped me co-author some of the proudest works I have published. Special thanks go to Joe Walter for his ingenious research ideas, exceedingly helpful and kind feedback, and witty emails that never fail to make me laugh. I have learned immensely from each and every one of you, and for that, I am extremely grateful.

I extend my deepest thanks to my mentors from beyond USC. From Tufts, my gratitude goes to Li-Ping Liu, Soha Hassoun, and Nate Bragg for setting the foundations of my research career. From Snap Research, I thank Francesco Barbieri, Leo Neves, Maarten Bos, Ron Dotsch, Neil Shah, Nils Murrugarra-llerena, Vitor Sousa, and Shubham Vij. From Spotify Research, I thank Sam Way, Ang Li, Aditya Ponnada, and Ben Lacker. It has been my privilege to have such incredible mentors guiding me.

Throughout my PhD journey, I have had the privilege of meeting a handful of researchers who have also become dear friends. My heartfelt thanks go to Luca Luceri, one of my closest colleagues and friends, from whom I have learned the virtues of patience and dedication in being a supportive mentor, as well as the importance of maintaining a healthy work-life balance. I am grateful to have Bijean Ghafouri, who serves as a wise sounding board when I find myself too absorbed in my thoughts and who is, just as importantly, an extremely fun friend outside of work. My thanks also go to Ashok Deb, whose steadfast optimism and kindness never cease to humble me. I am incredibly fortunate to have each of you to share in every small triumph and to commiserate over every paper rejection. I also thank my lab mates and other colleagues I met through work—Herbert Chang, Emily Chen, Alex Spangher, Alex Bisberg, Goran Murić, Jinyi Ye, Shen Yan, Yilei Zeng, David Chu, Patrick Gerard, Francesco Pierri, Gabriela Pinto, Priyanka Dey, Charles Beckham, Eun Cheol Choi, Adam Badawy, Basel Shbita, Yulin Yu, Wen Xie, Xiao Fu, Indira Sen, Giuseppe Russo, Manoel Horta Ribeiro, and Dominik Bär. Our productive research collaborations and our random rants about the PhD program have made my time here infinitely more enjoyable and fulfilling.

I am also profoundly grateful for the lifelong friends I have made. This journey would have been impossible without the support of Min, Phil, and Rox, my three closest friends from college. I cherish our in-depth, late-night discussions, where we chatted about everything in life and more. Your unwavering support has seen me through the best and worst of times. Though we have all changed since our college days--a bit older, probably wiser, and perhaps more jaded--I treasure our closeness despite the physical distances. Min, your genuineness, trustworthiness, and openness make you an irreplaceable friend in my life. Phil, your high-achieving yet easy-going personality continually inspires me, and I know we will always have each other's back. Rox, your confidence is infectious, and your ability to listen has always brought comfort to me. I am also fortunate to have made incredible friends in LA. Special thanks to Mercedes, whose kindness and vibrancy light up every room, and Maja, who is often the most considerate and funniest person in the room. I want to give a shout-out to my friends from college, LA, and beyond, who have shared countless memorable holidays and excursions with me, including Jenn, Lydia, Steph, Ian, Rachel, Florence, Noah, Olivia W., David Y., Brittany, Iszzy, Eeuny, Gab, the other Gab, Audrey, Daniel, Kyara, Ixchel, Gianluca, Jordan, Tijana, Charles, Kim, Melody, Jaz, Felix, and Sunny. Every day feels extra special when I am with you.

Above all, I thank my family. My academic pursuits were made possible only by their constant support.  I am especially grateful for my mom, Grace, who has enthusiastically cheered me on since day one, and my dad, Michael, who is always the pillar we depend on. You always believed in me, even when I doubted myself. My sister Jamie, no longer the ``baby'' of the family but forever so in my heart, deserves special thanks. Knowing that a younger sibling might be looking up to me provides a perpetual source of motivation, especially considering she is not only intelligent and bright but also, more importantly, one of the most empathetic, kind, and driven people I know. Jamie, while I may be a role model to you, it's your qualities that inspire me to be the best I can be. And finally, I cannot mention my family without talking about my grandparents. My \textit{wai gong wai po} raised me with nothing but love, patience, dedication, and a lot of really great homemade food. I love you all.

Los Angeles, California

April 6, 2024

Julie

% I thank Los Angeles. Even though LA is nothing like it’s depicted in La-La-Land—the lack of a Ryan Goslings to meet at random parties was a bit of a disappointment for me when I first came here—LA has gradually become home for me. I love the beach, the year-round temperate weather, the overpriced but high-quality restaurant food, and the cute bars around the city. I like LA so much that I don’t even mind driving on the 405. In my mind, LA is the city of “you can be whatever you want to be,” and I think it did it for me.

% I thank music. There were some very hard times during my PhD--some of them were because of mean Reviewer 2’s, and some had nothing to do with my PhD but nonetheless happened during my PhD. Music has always been a constant pillar of my life, carrying me through every obstacle. I am grateful for having healthy ears, for classical composers like Beethoven, for house music DJs like Parra for Cuva, for Spotify, for my piano, for my Airpods, and for my beloved Bose speakers.

  {
  \hypersetup{hidelinks} % color all links black in the preface
  \tableofcontents
  \listoftables
  \listoffigures
  }

  \prefacesection{Abstract}
  The explosive growth of social media has not only revolutionized communication but also brought challenges such as political polarization, misinformation, hate speech, and echo chambers. This dissertation employs computational social science techniques to investigate these issues, understand the social dynamics driving negative online behaviors, and propose data-driven solutions for healthier digital interactions. I begin by introducing a scalable social network representation learning method that integrates user-generated content with social connections to create unified user embeddings, enabling accurate prediction and visualization of user attributes, communities, and behavioral propensities. Using this tool, I explore three interrelated problems: 1) COVID-19 discourse on Twitter, revealing polarization and asymmetric political echo chambers; 2) online hate speech, suggesting the pursuit of social approval motivates toxic behavior; and 3) moral underpinnings of COVID-19 discussions, uncovering patterns of moral homophily and echo chambers, while also indicating moral diversity and plurality can improve message reach and acceptance across ideological divides. These findings contribute to the advancement of computational social science and provide a foundation for understanding human behavior through the lens of social interactions and network homophily.

\end{preface}

%%% introduction %%%%%%%%%%%%%%%%%%%%%%%%%%%%%%%%%%%%%%%%%%%%%%%%%%%%%%%
% \chapter{Introduction}
% \label{ch:introduction}

% \graphicspath{}

\chapter{Introduction}

The explosive growth of social media platforms in recent years has facilitated an unparalleled level of online communication and interaction between people. Staggering statistics highlight this growth: nearly half the US adults use Instagram, and even more use Facebook and YouTube \cite{pew2024americans}. Globally, as of 2024, 4.8 billion people are using social media, accounting for 60\% of the global population and 93\% of all internet users \cite{sej2023stats}. This growth shows no plateauing, with TikTok being among the fastest-growing social media apps from 2021 to 2023, accruing a 12\% growth in user base \cite{pew2024americans}. Investors remain optimistic about the future of social media tech companies, as evidenced by Reddit's recent valuation of \$6.4 billion for its initial public offering in March 2024 \cite{nyt2024reddit}. Moreover, new social media apps continue to emerge, catering to ever-evolving consumer demands, such as Bluesky (2021), BeReal (2020), and Threads (2023).

Nowadays, social media use is ubiquitous and intertwined with many aspects of our lives. It addresses a fundamental part of human beings, which is the desire for interpersonal communication \cite{baumeister2017need}. With the advances in technology, social media apps offer an easy way to stay connected with friends and family, as well as to follow the lives of influencers and public figures. It also serves as a source of entertainment, news, and education. For many individuals, particularly those who grew up in the digital era, social media has become an indispensable tool for maintaining relationships and engaging with the world around them. Platforms such as Facebook, Instagram, WeChat, Snapchat, TikTok, YouTube, Reddit, Twitter (now X), and LinkedIn play a crucial role in connecting people with friends and family, providing access to captivating content, and enabling participation in academic and professional communities. As social media continues to evolve and adapt to user needs, its importance in our daily lives is likely to grow further.

% amplifying information through systems like retweet chains \cite{asur2011trends}.

However, social media has also faced significant criticism, so much so that many people have voluntarily undertaken ``digital detoxes'' to resist access to social media apps and focus on their non-digital lives \cite{syvertsen2020digital}. Social media can negatively impact users' mental health. For example, internal research at Meta revealed Instagram's negative impact on the mental health of young girls, propagating disillusioned beliefs about body image \cite{wsj2021fb}. Equally troubling is the issue of addiction, as heavy social media use can negatively impact one's mental health, relationships, and academic or professional achievement \cite{kuss2011online}. 

Additionally, the lack of adequate content moderation to filter harmful, abusive, or illegal material poses serious risks for users. Online hate speech is estimated to be very pervasiveness \cite{siegel2020online}; one cross-national survey estimates that a staggering half of young adults experience some form of hateful messages \cite{keipi2016online}. Another example is the ``Elsagate'' controversy on YouTube, where malicious actors exploited the recommendation algorithm to spread inappropriate videos targeting children \cite{nyt2017elsagate}. 

Moreover, social media can serve as a conduit for the spread of misinformation \cite{delvicario2016spreading}--the unintentional spread of false information--and disinformation \cite{tucker2018social}--the intentional spread of false information. This has severe consequences for matters of national security and election integrity \cite{marwick2017media}. For instance, the high-profile Cambridge Analytica and Facebook scandal that broke the news in 2018 in which millions of user profiles were leaked for target political advertising \cite{nyt2018cambridge}, or the state-backed propaganda by Russia's Internet Research Agency (IRA) that used internet ``trolls'', or fake social media accounts, to interfere in the 2016 US elections. \cite{bastos2019donald}. Social media can also act as a filter bubble and echo chamber, supplying users with the type of information they already agree with, worsening ideological divides, and amplifying beliefs \cite{barbera2020social}. The greater consequences of this have already been seen in matters of political polarization \cite{kubin2021role} and extremism radicalization \cite{thompson2011radicalization}. Relatedly, user privacy is also a concern as social media companies accumulate extensive user data \cite{zhang2014creepy}, as was most recently seen in the recent 2024 US House initiative to ban TikTok due to privacy concerns \cite{nyt2024tiktok}.

Despite these significant drawbacks, we are unable to truly part ways with social media. It has evolved into a constant source of connection—an arguably necessary component interwoven into both our personal and professional lives. Even if an individual were to abstain from social media, the rest of the world remains deeply connected and engaged with these platforms. This inescapable reality is not without its merits, however. Social media \textit{can} be an invaluable platform for fast, convenient communication. It can serve as a useful tool for learning \cite{duffy2008using} and education \cite{greenhow2019social}. And anecdotally, the entertaining content shared on these platforms, such as amusing dog videos, may even provide a mood boost and improve mental well-being.

If we cannot escape the digital world, can we make it better? This question drives my research, aiming to identify problems on social media and offer data-driven suggestions for improvement. Throughout this dissertation, I hope to use my knowledge to explore specific issues and potential solutions related to social media usage and its impacts. While a comprehensive solution is beyond the scope of a single work, this research aims to contribute one piece towards the broader goal of enhancing the social media experience and mitigating its negative consequences.

% \section{Motivation}
% I will focus on the \textit{network} part: how we can use one user to understand the behavior of another. As it follows, a core theme propagating my entire research is the concept of network homophily \cite{mcpherson2001birds}. It refers to the tendency of individuals to associate and bond with others who are similar to themselves, resulting in social network bonds of users who share similar characteristics, opinions, emotions, or other attributes. Homophily can, in part, explain why we are socially attached to those we are attached to and why we have some ideologies. Homophilic ties can forge echo chambers reinforcing stances, facilitate the spread of unchecked belief, and harden ideological divisions. I use the theory of homophily throughout this work to devise research hypotheses, develop machine learning modeling methods, and qualitatively interpret our findings. 

\textit{This dissertation is motivated by the desire to make online social network platforms better by mitigating the undesirable, toxic, and ill-intentioned consequences that often arise.} To this end, I consider a few cases of negative online human behavior to examine the severity of the problem, provide potential explanations for their occurrence, and predict human behavior. By leveraging the large-scale nature of social media data and employing empirical, quantitative analyses, I aim to understand user motivations, social dynamics, community effects, and influence pathways.

\section{Computational Social Science}

The availability of large-scale online behavioral traces has fueled research in computational social science (CSS), a fusion of computer science and social science \cite{lazer2009computational,cioffi2014introduction,wallach2018computational,zhang2020data}. Operating at the intersection of empirical, social, and computational sciences, CSS provides both a theoretical framework for understanding social phenomena at a macro level and a methodological tool for computationally validating and testing new theories \cite{cioffi2021scope}. Its scope is broad, encompassing computational foundations on humans and social systems, information extraction, social network analysis, social complexity, and social simulations, as outlined by \citet{cioffi2021scope}. Due to its highly interdisciplinary nature, CSS has applications in various domains such as psychology, political science, communications, economics, and public policy \cite{edelmann2020computational,zhang2020data}.
% The scope of CSS is diverse \cite{cioffi2021scope}, but 
% This field is uniquely characterized by its capacity to analyze large-scale data and address `macro' level social phenomenon empirically \cite{lazer2009computational}. 

In this work, my focus lies on the empirical, data-driven aspect of CSS, leveraging the novelty of ``big data'' computing and advancements in computer science \cite{lazer2009computational,lazer2020computational}. The methodology employed here varies widely, including data analysis techniques, statistical methods, network science, traditional machine learning, and deep learning \cite{zhang2020data}. As such, rapid technological advances in deep learning have also been instrumental in this emerging field's growth. CSS researchers are able to use data-driven methods to analyze individual and collective human dynamics. 

It is not lost on me the irony of utilizing CSS approaches, a field greatly bolstered by large-scale social media data, to address issues that have arisen because of social media. Nonetheless, CSS methods provide a valuable opportunity to glean empirical, data-driven insights that can inform strategies for mitigating the negative consequences of modern online communication. The vast troves of social media data harbor the potential to illuminate crucial mechanisms underlying human behavior on these platforms. By harnessing computational methods to analyze this rich data, I aim to uncover insights that can guide solutions to tackle the problems emerging in online systems.

\section{Social Network Homophily}

A core theme underlying my research is the concept of network homophily--the tendency of individuals to associate and bond with others who are similar to themselves, resulting in social network connections between users who share comparable characteristics, opinions, emotions, or other attributes \cite{mcpherson2001birds}. In network theory, this is also called assortative mixing, where nodes preferentially attach themselves to other nodes with similar characteristics \cite{newman2003mixing}. 

While homophily implies that ``similarity breeds connection'' \cite{mcpherson2001birds}, a closely related theory is social contagion, which posits the opposite: one user's behavior influences their network connections to do the same \cite{christakis2013social}, similar to how one might contract COVID from a close contact. Social contagion can occur as a simple contagion, with adoption probability increasing monotonically with the number of exposures, or as a complex contagion involving more intricate dynamics \cite{hodas2014simple}. Homophily and social contagion represent contrasting causal relationships. However, the purpose of this research is not to argue for the presence of one mechanism over the other. As \citet{shalizi2011homophily} aptly stated in their paper title, ``homophily and contagion are generically confounded in observational social network studies.'' Regardless of whether network connections are formed through homophily or contagion, the observable outcome remains the same: individuals with similar characteristics tend to be situated closer to each other in the network. Therefore, the primary objective of this work is to leverage the theory of homophily to model and interpret the data at hand.

Throughout my dissertation, I will exercise caution to avoid making unsubstantiated causal claims. The data utilized in this research is observational in nature, and drawing causal inferences from such data is, at best, pseudo-causal and, at worst, entirely unjustified. Establishing robust causal relationships would necessitate interventional studies, which are beyond the scope and capabilities of an independent academic researcher. Even barring practical limitations, conducting such experiments is ethically questionable, as it would involve subjecting users to potentially harmful content such as radicalism or hate speech.

Despite the inability to conclusively determine the underlying mechanism of homophily, the theory itself can still be effectively utilized. Specifically, I leverage the concept of homophily to formulate research hypotheses and interpret the findings of this study. The literature on user homophily in social networks provides numerous examples of its application. For instance, on Twitter, users tend to gravitate towards others who share similar political ideologies \cite{conover2011political,barbera2015tweeting}. Similarly, on Facebook, users who spread conspiracy theories and those who disseminate scientific news are situated in polarized and homogenous communities \cite{delvicario2016spreading}. Additionally, one study suggests that hateful messaging is also a homophilous trait \cite{nagar2022homophily}. Moreover, homophily manifests through the lens of moral psychology, as demonstrated by \citet{dehghani2016purity}, who found that users sharing certain moral foundations are more likely to be closely connected in their social network. These examples of homophily in social networks inspire thought-provoking interpretations of human behaviors and opinions within the context of social network structures and can thus illuminate online social dynamics.

In addition to its utility in interpreting social phenomena, I will also draw from the concept of homophily in developing a novel graph representation learning method for social networks. Graph representation learning, also known as network representation learning, has emerged as a promising machine learning approach for extracting insights from structured graph data. This technique aims to learn vector representations, or embeddings, for each node in a graph, capturing relevant information for downstream tasks like node classification or link prediction \cite{hamilton2020graph}. Here, I treat users as nodes and social network interactions as edges, applying a network representation learning algorithm to derive latent user embeddings.

Many graph representation learning methods rely on and leverage the homophily property \cite{grover2016node2vec,hamilton2017inductive,kipf2017semi,veličković2018graph}. These methods work to embed connected users, assuming that they are more similar, according to the homophily theory, closer together in the latent space. Consequently, the resulting embedding vectors will exhibit greater similarity for connected users. It is worth noting that while homophily is not strictly necessary for some graph representation learning approaches like graph convolutional networks, which can derive meaningful representations from heterophilous connections, graphs lacking homophily still present challenges for graph representation learning \cite{ma2022is}. However, in the context of social networks, homophily predominates \cite{mcpherson2001birds}.

\section{Technical Challenges}

A key aspect of this work is utilizing the social network to capture the rich relational signals from user-user interaction data. I aim to go beyond simply considering the number of people a user interacts with and tap into the identity of connected users to better understand individual users. This motivates using graph representation learning techniques such as graph neural networks \cite{goyal2018graph,hamilton2020graph}. By representing users as nodes and user interactions or connections as edges, network representation learning methods encode community properties, structural roles, and other properties in user embeddings that are easily accessible for machine learning models.

Regrettably, these techniques often encounter difficulties in scaling computationally to handle large datasets \cite{zhang2018network}, a challenge particularly pronounced by the scale of empirical social network data \cite{wu2020comprehensive,ma2022graph}. The sheer volume of big data frequently surpasses the capabilities of contemporary technology in terms of both hardware resources and processing power \cite{fan2014challenges}. Additionally, their methodological constraints limit their compatibility with popular distributed training methods such as parallelization \cite{serafini2021scalable}. Furthermore, many of these methods struggle with inductive inferencing on new users without necessitating retraining \cite{goyal2018graph}.

To address this challenge, I propose a novel social network representation learning method, \textit{Social-LLM}, which integrates user content with social connections to create unified user representations. This approach enables accurate prediction and insightful visualizations of user attributes, communities, and behavioral propensities. Social-LLM leverages the simplest form of connections, where one edge connects two users, to semi-supervise user embedding training via a Siamese network structure. The model aims to embed two connected users closer in the latent dimension space. To facilitate the training process, I also utilize large language models \cite{zhao2023llmsurvey}. As social network data is mostly textual, the latest language model technology can be used to obtain initial user embeddings based on texts, providing the model with a substantial head-start in performance without relying on network information.

Social-LLM offers several practical advantages. First, by using pairs of nodes connected by an edge as inputs, the data can be easily processed with stochastic gradient descent via batching, resulting in a time complexity that is linear with respect to the number of edges. Second, since Social-LLM is semi-supervised using edges, node labels are not required. This is particularly beneficial as node labels can be costly to annotate for large datasets and may be inherently noisy and ambiguous due to the nature of many social science problems. Moreover, annotations require the problem to be carefully thought out ahead of time, tailoring to a single problem and limiting the generalizability of a dataset. Additionally, the learned Social-LLM user embeddings can be highly flexible and used as features in downstream prediction tasks, such as predicting user partisanship or clustering users for community analysis. The versatility and reusability of these embeddings make them extremely useful. Finally, Social-LLM does not require any social network data at the inference stage, substantially expanding its usefulness in producing inductive user representations when applying the model to out-of-sample users.

\section{Research Overview }
In this dissertation, I explore the dynamics of online communication within the realm of pressing social and political matters by focusing on three research topics. A central theme that binds them together is the investigation into social dynamics and polarization across digital platforms, with a special emphasis on the capacity of social media to enhance or mitigate specific behaviors. Although each study focuses on a distinct facet of these dynamics, together, they illuminate the complex manner in which online social networks shape user behavior, the spread of information, and ideological division.

\paragraph{Polarization in COVID-19 Discourse.} Focusing on the politicization of COVID-19 discourse on Twitter, this research explores the dynamics of political polarization. I find significant polarization and the existence of two distinct and asymmetric echo chambers. The left-leaning users are in a larger echo chamber, while the right-leaning users form more densely connected and isolated echo chambers. This emphasizes the role of social media in reinforcing pre-existing beliefs and alludes to challenges in penetrating these echo chambers with alternative information regarding a public health crisis.

\paragraph{Online Hate Speech.} This research focuses on the role of social approval in driving online toxicity, suggesting that the pursuit of social approval, rather than a direct intent to harm, motivates users to engage in online hate speech. It emphasizes the impact of social networks and engagements (likes, retweets, etc.) on escalating or reducing toxicity, providing insights into how social dynamics on platforms contribute to the spread of hate speech. Coupled with the observation that hateful behavior is homophilous, this research has important implications for understanding how hateful behavior is ``networked'' and socially motivated.

\paragraph{Moralization and COVID-19.} This research delves into the moral underpinnings of online discussions about COVID-19, showing how moral psychology and political ideologies together shape users' communication. It reveals patterns of moral homophily and the existence of a moral echo chamber. Promisingly, it also suggests that moral diversity and plurality can improve the reach and acceptance of messages across ideological divides.

\bigskip
Collectively, these studies shed light on how social media platforms serve as fertile grounds for the reinforcement and amplification of existing beliefs, the formation of homophilous networks, and the escalation of behaviors like toxicity and polarization. They underscore the complex interplay between individual psychology, social approval mechanisms, political ideologies, and network structures in shaping online discourse. These insights are crucial for developing strategies to mitigate polarization, combat online hate, and promote the dissemination of information. By providing a comprehensive understanding of the social dynamics at play within online environments, this dissertation contributes to the growing body of research aimed at fostering healthier online interactions and more informed public discourse.

\section{Contributions}

This dissertation analyzes the dynamics of online communication against the backdrop of significant social and political issues, emphasizing the role of social media in magnifying behaviors and ideologies. Its contributions encompass a technical innovation as well as three explorations of computational social science problems that shed light on harmful and fruitful communication practices online.

The technical cornerstone of this work, presented in Part \ref{part1}, is a scalable social network representation learning algorithm. This algorithm excels at generating insightful user representations, thereby facilitating subsequent analyses such as user classification and community detection. I first introduce the preliminary model, Retweet-BERT, in Chapter \ref{chp:retweetbert} as a proof-of-concept. It leverages user profile descriptions and retweet interactions for modeling political partisanship. I then implement enhancements in a Social-LLM, discussed in Chapter \ref{chp:socialllm}, to incorporate multimodal user content and a broader spectrum of network interactions. I also comprehensively evaluate Social-LLM in user detection to showcase its utility in user modeling tasks.

Part \ref{part2} conducts online behavior analysis to tackle distinct yet interrelated problems within online communication. In Chapter \ref{chp:poli_covid}, I investigate the political echo chambers that arise from COVID-19 pandemic discussions, revealing how digital spaces can become echo chambers that could reinforce pre-existing beliefs. In Chapter \ref{chp:hate}, I examine networked hate speech behavior, uncovering the driving force of social approval behind online toxicity. In Chapter \ref{chp:moral_covid}, I analyze COVID-19 discussions through the lens of moral psychology to offer insights into communication dynamics based on moral inclinations, painting a nuanced picture of online interaction extending beyond political ideology.

This dissertation underscores the need to dissect online dynamics to devise effective strategies for accurate information dissemination and public health messaging, especially within polarized and echo chamber-prone environments. It highlights the significance of leveraging social network data in computational social science, demonstrating how users' social circles can reveal communication cues critical for understanding online behavior.

These research studies primarily rely on social networks built on interactions, such as retweets. This choice is driven by my focus on interactive and highly contextualized networks surrounding specific events and topics. While my work may not delve into more stable networks, such as friendship or following networks, it does yield clear behavioral insights into how people engage and allocate their attention within social media.

My findings offer actionable insights for various stakeholders. Researchers can develop targeted interventions to help mitigate the negative consequences of polarization, hate speech, and moral misalignment, ultimately promoting more constructive and inclusive online interactions. Social media platforms can leverage these findings to update their policies, intervention strategies, and platform architectures in their efforts to foster healthier online environments. Lawmakers can use this research to inform regulations that address the challenges posed by the modern digital landscape. Educators can incorporate the insights into digital literacy curricula, empowering individuals to navigate online spaces safely and effectively.

More broadly, the significance of this work extends beyond the realm of social media alone. It strives to harness the potential of computational social science to address the multifaceted challenges presented by the digital age. While social media serves as a primary focus, the methodologies and insights developed here have broader applicability across various domains where human interaction intersects with digital platforms. By shedding light on the intricate dynamics of online behavior, this research contributes to the development of strategies aimed at promoting healthier and more productive online communities through computational social science.

% The research presented in this dissertation has the potential to make significant contributions to our understanding of online communication dynamics and their impact on society. By analyzing the content of user discussions on social media, the networks of their conversations, and the socially reinforcing signals they receive, I aim to unravel the mechanisms that contribute to unhealthy and harmful communication. I provide empirical evidence and offer insights into why modern online social media, despite its increasing accessibility and widespread adoption, can lead to miscommunication and misunderstanding. 

% Ultimately, this work yields data-driven recommendations for cultivating healthier online information ecosystems, inclusive discourse, reduced polarization risks, and socially-positive platform affordances. It demonstrates how interdisciplinary computational approaches can reveal the root causes driving harmful digital phenomena. This knowledge can inform designs promoting open ideological diversity. My work could inform educating online users about how to consume information, as well as to guide platform designers and policymakers to design online systems that promote good communication habits.

Some chapters in this dissertation are adapted from papers that have been published \cite{jiang2021social,retweetbert} or are available online as preprints \cite{jiang2023socialllm,jiang2023social,jiang2024moral}.

% he rich user-user interaction data (i.e., the social network) could be extremely useful when modeling human behavior \cite{de2010birds}. 

% As we shift toward an increasingly digital society, research focuses on the human aspect of our digital world,

%%% chapter %%%%%%%%%%%%%%%%%%%%%%%%%%%%%%%%%%%%%%%%%%%%%%%%%%%%%%%%%%%%

\part{Social Network Representation Learning} 
\label{part1}

Social media data has provided researchers with an exciting avenue for empirical data-driven mining of human behavior \cite{freeman2004development,lazer2009computational}, enabling the tracking of mass sentiment, health trends, political polarization, mis/disinformation spread, and information diffusion. Social network data comprises two crucial elements: content—\textit{what} people share —and network--who, when, and how frequently users interact with each other. The text-based content aspect of social network data is easier to process computationally due to recent advancements in large language models (LLMs).

Graph representation learning has emerged as a tool to learn information from networks effectively \cite{hamilton2020graph}. Graph representation learning would convert each node in the network to a vector representation that could capture useful latent properties, such as global community properties and local node structural properties. Common approaches to accomplish this include random walk and graph neural network algorithms \cite{hamilton2017inductive,goyal2018graph}. Random walk methods preserve the probability of one node visiting another on a random walk over the graph \cite{deepwalk,grover2016node2vec}. Graph neural networks use a message-passing framework to exchange information between connected nodes, aggregating $k$-hop neighborhood embeddings around a node during embedding optimization \cite{hamilton2017inductive,kipf2017semi}.

% wu2020comprehensive
However, the accessibility of big data often surpasses the capabilities of modern technology in terms of hardware resources and processing capabilities \cite{fan2014challenges}. While advancements in graph neural networks have yielded significant progress in network representation, these methods often struggle to mitigate training overhead. They could suffer from the ``neighbor explosion'' problem as the time and memory required to model neighborhoods grows exponentially with the depth \cite{hamilton2017inductive,duan2022a}. They also cannot use conventional approaches to speed up training, such as distributed training, due to the cost of coordination required to update node embeddings across devices \cite{serafini2021scalable}. To circumvent this practical constraint, many algorithms rely on sampling or decoupling-based approaches \cite{serafini2021scalable,duan2022a}. Additionally, most graph learning algorithms are, inherently by design, unable to produce embeddings for unseen and unconnected nodes, limiting their capacity to generalize to out-of-sample users without retraining \cite{grover2016node2vec,kipf2017semi}. This challenge is particularly pronounced when modeling the network ties of large-scale social networks \cite{ma2022graph}, which are much larger and sparser compared to other commonly modeled networks (such as protein-protein interaction or citation networks). 

In this part, we will introduce a pragmatic social network representation learning method designed specifically to address these problems using social network connections and user content features. In our social network setting, we treat users as nodes and social network connections between two users as edges. Under the assumption of social network homophily \cite{mcpherson2001birds}, we optimize user representations by making users who are connected by edge more similar to one another. We also leverage the recent advancements in large language models \cite{zhao2023llmsurvey} by incorporating the user's textual content. 

This model contains several advantages. First, it is scalable to train since the training time complexity is linearly proportional to the number of edges, and it also enables us to use stochastic gradient descent for batch-wise processing. Second, the model is self-supervised and requires no labels, which also, in turn, makes the learned representations reusable for any task. This is particularly advantageous as annotating human social network data can be costly, noisy, and inherently ambiguous. Finally, the method is inductive since the social network is no longer needed during the inference stage. We are able to produce representations for any user, provided the suitable user content features, during inference.

We begin with a preliminary version of the model called \textit{Retweet-BERT}. This model serves as proof of concept, using only user profile description features and retweet connections. We evaluate this model on the task of user partisanship prediction. The success of this model motivates the \textit{Social-LLM}, a more general-purpose solution that can accommodate multimodel user content features and heterogeneous edge types. We conduct a full-scale evaluation of Social-LLM on seven different real-world datasets.

\chapter{Retweet-BERT: Learning from Language Features and Retweet Diffusions}
\label{chp:retweetbert}
\section{Introduction}
The rise of social media has brought forth an era of unprecedented connectivity, enabling individuals to engage in online discussions on a wide range of topics. However, this increased connectivity has also amplified the issue of political polarization, which has become a major driver of many online conversations. This chapter introduces a preliminary model for social network representation learning, focusing specifically on the problem of user partisanship prediction. By leveraging the vast amounts of data generated through online interactions, we aim to develop a tool that can accurately predict users' political affiliations based on their social network structure and behavior.

Political polarization has been shown to play a significant role in various online discussions, ranging from election-related topics, such as the 2010 US congressional midterm elections \cite{conover2011political} or the 2012 US presidential election \cite{barbera2015tweeting}, to seemingly unrelated issues like the 2020 global health crisis.  \cite{calvillo2020political,jiang2020political}. Literature suggests that political affiliations may have an impact on people's favorability of public health preventive measures (e.g., social distancing, wearing masks) \cite{jiang2020political}, vaccine hesitancy \cite{hornsey2020donald,peretti2020future}, and conspiracy theories \cite{uscinski2020people}. 

Divisive politicized discourse can be fueled by the presence of \textit{echo chambers}, where users are mostly exposed to information that aligns with ideas they already agree with, further reinforcing one's positions due to confirmation bias \cite{garrett2009echo,barbera2015tweeting}. Political polarization can contribute to the emergence of echo chambers \cite{conover2011political,cinelli2020echo}, which may accelerate the spread of misinformation and conspiracies \cite{delvicario2016spreading,shu2017fake,motta2020right,muric2021covid,rao2021political}. 
\begin{figure}
    \centering
    \includegraphics[width=0.5\linewidth]{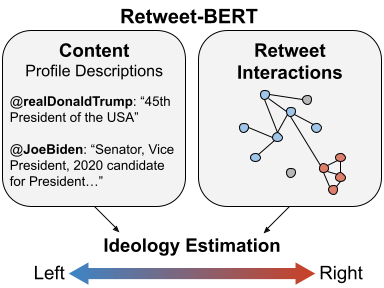}
    \caption{The two key motivating components of Retweet-BERT.}
    \label{fig:rbert_motivation}
\end{figure}
To facilitate research in online polarization, such as the COVID-19 infodemic, we present \textit{Retweet-BERT}, a lightweight tool to accurately detect user ideology in large Twitter datasets (illustrated in Figure \ref{fig:rbert_motivation}). Our method simultaneously captures \textit{(i)} semantic features about the user's textual content in their profile descriptions (e.g., affiliations, ideologies, sentiment, and linguistics) and \textit{(ii)} the patterns of diffusion of information -- i.e., the spread of a given message on the social network -- and how they can contribute to the formation of particular network structures (e.g., echo chambers). Prior works on polarization primarily focus on only one of these aspects \cite{conover2011political,conover2011predicting,barbera2015tweeting,wong2016quantifying,preotiuc2017beyond}. 

There are two important assumptions behind Retweet-BERT. One is that the act of retweets implies endorsement \cite{boyd2010tweet}, which further implies support for another's ideology \cite{wong2016quantifying}. The other is that people who share similar ideologies also share similar textual content in their profile descriptions, including not only similar keywords (e.g., ``\textit{Vote Blue!}'') and sentiment but also linguistics. The idea of \textit{linguistic homophily} among similar groups of people has been documented and explored in the past \cite{yang2017overcoming}. People who adopt similar language styles have a higher likelihood of friendship formation \cite{kovacs2020language}. 

Retweet-BERT leverages both network structure and language cues to predict user ideology. Our method is simple, intuitive, and scalable. The two steps to Retweet-BERT are: 
\begin{enumerate}
    \item \textbf{Training} in an unsupervised manner on the full dataset by learning representations based on users' profile descriptions and retweet interactions,
    \item \textbf{Fine-tuning} the model for polarity estimation on a smaller labeled subset.
\end{enumerate}

\begin{figure}
    \centering
    \includegraphics[width=0.7\linewidth]{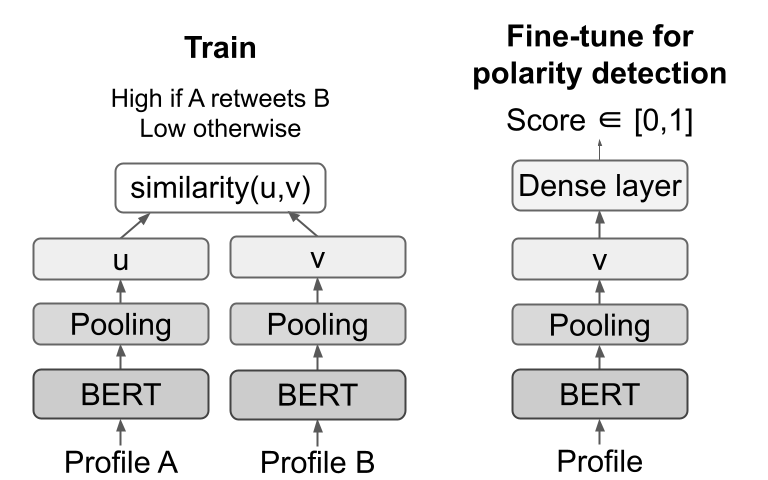}
    \caption{Illustration of the proposed Retweet-BERT. We first train it in an unsupervised manner on the retweet network (left) using a Siamese network structure, where the two BERT networks share weights. We then train a new dense layer on top to predict polarity on a labeled dataset (right).}
    \label{fig:retweetbert}
\end{figure}

An illustration of Retweet-BERT is shown in Figure \ref{fig:retweetbert}. Crucially, our method does not require human annotations. Instead, we label a small set of users heuristically based on hashtags and mentions of biased new media outlets, as was done in prior works \cite{conover2011predicting,badawy2018analyzing,addawood2019linguistic}. In addition, since we only use profile descriptions instead of all of the users' tweets, Retweet-BERT can be easily deployed.

The datasets we use are two large-scale Twitter datasets collected in recent years. The COVID-19 Twitter dataset was collected from January to July 2020 for 232,000 active users. We demonstrate that Retweet-BERT attains 96\% cross-validated macro-F1 on this dataset and outperforms other state-of-the-art methods based on large language models, graph embedding, etc. We also perform extensive evaluations of our model on a second Twitter dataset on the 2020 presidential elections to showcase the reliability of Retweet-BERT (97\% macro-F1).

In sum, the contributions of this chapter are:
\begin{itemize}
    \item We present Retweet-BERT, a simple and elegant approach for estimating user ideology based on linguistic homophily and social network interactions.
    \item We conduct experiments and manual validations to highlight the effectiveness of Retweet-BERT on two recent public Twitter datasets compared to baselines: COVID-19 and the 2020 US presidential elections.
\end{itemize}

\section{Related Work}
\subsection{Ideology Detection}
There is growing interest in estimating expressed ideologies. Many works focused on opinion mining and stance detection \cite{somasundaran2009recognizing,walker2012stance,abujabara2013identifying,hasan2014taking,sridha2015joint,darwish2020unsupervised}. Of particular interest are political ideology detection of textual data  \cite{sim2013measuring,iyyer2014political,bamman2015open} as well as of Twitter users \cite{conover2011predicting,conover2011political,barbera2015tweeting,wong2016quantifying,yang2016social,preotiuc2017beyond,badawy2018analyzing,badawy2019falls,xiao2020timme}. There are two general strategies for identifying Twitter user ideologies: content-based and network-based. Content-based strategies are concerned with the user's tweets and other textual data. An earlier study used hashtags in tweets to classify users' political ideologies \cite{conover2011predicting}. \citet{preotiuc2017beyond} applied word embedding on tweets to detect tweets of similar topics. Network-based strategies leverage cues from information diffusion to inform ideological differences. These models observe that users interact more with people they share similar ideologies with \cite{yang2016social}. Interactions can be retweets \cite{wong2016quantifying} or followings \cite{barbera2015birds}. \citet{xiao2020timme} formulated a multi-relational network using retweets, mentions, likes, and follows to detect binary ideological labels. Other works used a blend of both content- and network-based approaches \cite{badawy2019falls}. Hashtag-based methods were combined with label propagation to infer the leanings of users from the retweet network
\cite{conover2011predicting,conover2011political,badawy2018analyzing}. Closely related to our methodology, \citet{darwish2020unsupervised} clustered users by projecting them on a space jointly characterized by their tweets, hashtags, and retweeted accounts; however, this algorithm comes at a high computational cost.

\subsection{Socially-Infused Text Mining} 
More related to our method is a recent line of work that learns from socially infused text data. 
\citet{li2019encoding} combined user interactions and user sharing of news media to predict the bias of new articles. \citet{pan2016tri} used node structure, node content, and node labels to learn node representations to classify categories of scientific publications. \citet{yang2017overcoming} used social interactions to improve sentiment detection by leveraging the idea of linguistics homophily. \citet{johnson2017leveraging} used lexical, behavioral, and social information to categorize tweets from politicians into various topics of political issues. These works provide promising results for combining social network data and textual data. 

\paragraph{\normalfont \textbf{This model:}} Retweet-BERT is unique from the approaches described above in two substantial ways: (i) it combines both language features, particularly pretrained large language models for natural language processing, and social network features for a more comprehensive estimation of user ideology, and (ii) it is scalable to large datasets without supervision.

\section{Data}
\label{ref:rtbert_data}
We use two recent large-scale Twitter datasets. The primary dataset is on COVID-19 (\textsc{Covid-Political}) from January 21 to July 31, 2020 (v2.7) \cite{chen2020covid}. All tweets collected contain COVID-related keywords. We also use a secondary dataset on the 2020 presidential election (\textsc{Election-2020}) collected from March 1 to May 31, 2020 \cite{chen2021election2020}. Both datasets are publicly available. Each tweet contains user metadata, including their profile description, the number of followers, the user-provided location, etc. Users can be verified, which means they are authenticated by Twitter in the interest of the public.

Although a number of Twitter accounts have since been banned by Twitter---notably, @realDonaldTrump was suspended in January 2021 \cite{twitterbantrump}---our data collection was done in real-time, and therefore all tweets by banned accounts are still in our dataset.

\subsection{Content Cues: Profiles} For the purposes of this chapter, we do not use tweet contents but rather user profile descriptions. In addition to different users posting various numbers of tweets, our main assumption is that profile descriptions are more descriptive of a user's ideology than tweets. The profile description is a short biography that is displayed prominently when clicking on a user. It usually includes personal descriptors (e.g., ``\textit{Father}'', ``\textit{Governor}'', ``\textit{Best-selling author}'') and, when appropriate, the political ideology or activism they support (e.g., ``\textit{Democratic}'', ``\textit{\#BLM}''). Capped at 160 characters, these descriptions have to be short, which motivates users to convey essential information about themselves clearly, succinctly, and attractively. Previous work established a positive link between the number of followers and the character length of the user \cite{mention2018twitter}, which would suggest that more influential users will have a more meaningful profile.

\subsection{Interaction Cues: Retweet Network} In this chapter, we use \textit{retweets} to build the interaction network. Retweets refer only to tweets that were shared verbatim. Retweets are distinct from \textit{quoted tweets}, which are essentially retweets with additional comments. We do not use the \textit{following} network as it is rarely used due to the time-consuming nature of its data collection \cite{martha2013study}. The retweet network $G_R$ is a weighted, directed graph where vertices $V$ are users and edges $E$ are retweet connections. An edge $(u,v)\in E$ indicates that user $u$ retweeted from user $v$, and the weight $w(u,v)$ represents the number of retweets.

\subsection{Data Preprocessing}
We removed inactive users and users who are likely not in the U.S. (see Appendix \ref{appendix:2} for details). Users in our dataset must have posted more than one tweet. To remove biases from potential bots infiltrating the dataset \cite{ferrara2020types}, we calculate bot scores using \citet{botometer}, which assigns a score from 0 (likely human) to 1 (likely bots), and remove the top 10\% of users by bot scores as suggested by \citet{ferrara2020types}. 
The \textsc{Covid-Political} dataset contains 232,000 users with 1.4 million retweet interactions. The average degree of the retweet network is 6.15. Around 18k users ($\approx 8\%$) are verified. The \textsc{Election-2020} dataset contains 115,000 users and 3.6 million retweet interactions.

\section{Method} \label{sec:polarity_estimation}
This section describes our proposed method to estimate the polarity of users as a binary classification problem. We first use heuristics-based methods to generate ``pseudo"-labels for two polarized groups of users, which are used as seed users for training and evaluating polarity estimation models. We then introduce several baseline models followed by Retweet-BERT.

\subsection{Pseudo-Label Generation} \label{sec:poli_seed_users}
We consider two reliable measures to estimate political leanings for some users, which can be used for model training and automatic, large-scale evaluation. These measures will be used to generate ``pseudo" political leaning labels for a subset of users (i.e., \textit{seed users}). These seed users will be used as the training set of users. 

\subsubsection{Hashtag-Based Method} The first method involves annotating the 50 most popular hashtags used in user profiles as left- or right-leaning, depending on what political party or candidate they support (or oppose). 17 of these hashtags are classified as left-leaning (e.g., \textit{\#Resist}) and 12 as right-leaning (e.g., \textit{\#MAGA}). The list of hashtags can be found in Appendix \ref{appendix:2}. Users are labeled left-leaning if their profiles contain more left-leaning than right-leaning hashtags and vice versa. We do not consider hashtags appearing in tweets because hashtags in tweets can be used to reply to opposing ideology content  \cite{conover2011political}. Instead, following prior work  \cite{badawy2018analyzing,addawood2019linguistic}, we assume that hashtags appearing in users' self-reported profile descriptions are better indicators of their true ideological affiliations.

\subsubsection{News Media-Based Method} 
\begin{sloppypar} The second method utilizes media outlets mentioned in users' tweets through mentions or retweets \cite{badawy2019falls,bovet2019influence,ferrara2020characterizing}. Following \citet{ferrara2020characterizing}, we determined 29 prominent media outlets on Twitter. Each media outlet's political bias is evaluated by the non-partisan media watchdog \textsl{AllSides.com} on a scale of 1 to 5 (\textit{left}, \textit{center-left}, \textit{neutral}, \textit{center-right},  \textit{right}). If a user mentions any of these media outlets, either by retweeting the media outlet's Twitter account or by link sharing, the user is considered to have endorsed that media outlet. Given a user who has given at least two endorsements to any of these media (to avoid those who are not extremely active in news sharing), we calculate their media bias score from the average of the scores of their media outlets. A user is considered left-leaning if their media bias score is equal to or below 2 and right-leaning if their score is equal to or above 4.
\end{sloppypar}
\subsubsection{Pseudo-Labeling Seed Users} Using a combination of the profile hashtag method and the media outlet method, we categorized 79,370 ($\approx34\%$ of all) users as either left- or right-leaning. The first, hashtag-based, method alone was only able to label around 16,000 users, while the second, media-based, method labeled around 49,000 users. The two methods overlapped in labeling around 10,000 users. In case of any disagreements between the two methods, which were exceedingly rare at only 200 instances, we defer to the first hashtag-based method. These users are considered \textit{seed users} for political leaning estimation. 75\% of these seed users are left-leaning, a finding consistent with previous research which revealed that there are more liberal users on Twitter \cite{pew2019sizing}. In our secondary \textsc{Election-2020} dataset, we tagged 75,301 seed users.

\subsubsection{Pseudo-Labeling Validation} This pseudo-labeling method is limited in its capacity for labeling \textit{all} users (i.e., low coverage ratio, covering only 34\% of all users), but it serves as a good starting point for its simplicity. We validated this labeling strategy by annotating 100 randomly sampled users from the main \textsc{Covid-Political} dataset. Two authors independently annotated the data by considering both the tweets and the profile descriptions to determine the users' political leaning, keeping political neutrality to the extent possible. We then discussed and resolved any annotation differences until reaching a consensus. We attained a substantial inter-annotator agreement (Cohen’s Kappa) of 0.85. 96 users' annotated labels agree with the pseudo-labels, and 4 users' labels cannot be conclusively determined manually. The high agreement with the pseudo-labels makes us highly confident in the precision of our pseudo-label approach. 

\subsection{Methods for Polarity Estimation}\label{sec:lm}
While pseudo-labels can assign confident political leaning labels to a third of all users, they cannot determine the political leaning of the rest. To predict political leanings for all users, we explore several representation learning methods based on users' profile description and/or their retweet interactions. In all of our methods in this section and the one that follows (our proposed method), We do not consider users' tweets. This is because the datasets contain sampled tweets based on keywords and do not encompass any user's full tweet histories. Considering tweets in isolation can bias an algorithm for political leaning detection.

\subsubsection{Word Embeddings}
As baselines, we use pre-trained Word2Vec \cite{mikolov2013distributed} and GloVe \cite{pennington2014glove} word embeddings from Gensim \cite{gensim}. The profile embeddings are formed by averaging the embeddings of the profile tokens.

\subsubsection{Large Language Models}
Large language models (LLMs), particularly the BERT family of models \cite{devlin2019bert,liu2019roberta,sanh2019distilbert}, are pre-trained language models that have led to significant performance gains across many NLP tasks. We experiment with two different ways to apply LLMs for our task: (1) \textit{averaging} the output embeddings of all words in the profile to form profile embeddings, and (2) \textit{fine-tuning} an LLM through the initial token embedding of the sentence (e.g., \texttt{[CLS]} for BERT, \texttt{<s>} for RoBERTa) with a sequence classification head. We use the sequence classification head by \citet{huggingface}, which adds a dense layer on top of the pooled output of the LLM's initial token embedding.

\citet{reimers2019sbert} proposed Sentence-BERT (SBERT), which is a Siamese network optimized for sentence-level embeddings. Sentence-BERT outperforms naive transformer-based methods for sentence-based tasks while massively reducing the time complexity. We directly retrieve profile embeddings for each user using Sentence-BERT's pre-trained model for semantic textual similarity.

\subsubsection{Network-Based Models} We explore network-based models such as node2vec \cite{grover2016node2vec}, which learns node embeddings based on structural similarity and homophily, and label propagation, which deterministically propagates labels using the network. Neither of these models can classify isolated nodes in the network. We also experiment with GraphSAGE \cite{hamilton2017inductive}, an inductive graph neural network method that utilizes node attributes to enable predictions for isolated nodes. We use the aforementioned profile embeddings as node attributes.

All profile or network embeddings are subsequently fit with a logistic regression model for the classification task. Hyperparameter-tuning details can be found in Appendix \ref{appendix:2}. The profiles are pre-processed and tokenized according to the instructions for each language model.

% \paragraph{Limitations}
\bigskip
With the exception of GraphSAGE, all of these aforementioned methods use either the textual features of the profile description or the network content, but not both. Purely network-based models will do poorly for nodes with only a few connections and may only be suitable for non-isolated nodes. Purely text-based models will do poorly when there are insufficient textual features to inform the models. 

\subsection{Retweet-BERT Framework}\label{sec:rbert}

% Inspired by SBERT \cite{reimers2019sbert},
\subsubsection{Combining Textual and Social Content}
% \begin{sloppypar}
To overcome the aforementioned issues, we propose Retweet-BERT (Figure~\ref{fig:retweetbert}), a sentence embedding model that incorporates the retweet network. We base our model on the assumption that users who retweet each other are more likely to share similar ideologies. As such, the intuition of our model is to encourage the profile embeddings to be more similar for users who retweet each other. Retweet-BERT is trained in two steps. The first step involves training in an unsupervised manner on the retweet network, and the second step involves supervised fine-tuning on the labeled dataset for classification. Similar to the training of Sentence-BERT \cite{reimers2019sbert}, the unsupervised training step of Retweet-BERT uses a Siamese network structure. Specifically, using any of the aforementioned models that can produce sentence-level embeddings, we apply it to a profile description to obtain the profile embedding $s_i$ for user $i$. For every positive retweet interaction from user $i$ to $j$ (i.e., $(i,j)\in E$), we optimize the objective:
\begin{equation}
    \sum_{k\in V, (i, k)\not\in E}\max(||s_i - s_j|| - ||s_i-s_k|| + \epsilon, 0),
\end{equation}
where $||\cdot||$ is a distance metric and $\epsilon$ is a margin hyperparameter. We follow the default configuration as in Sentence-BERT \cite{reimers2019sbert}, which uses the Euclidean distance and $\epsilon=1$. We then freeze the learned weights and add a new layer on top to fine-tune on a labeled dataset for classification. 
% \end{sloppypar}

\subsubsection{Negative Sampling} To optimize the training procedure during the unsupervised training step, we employ negative sampling. We explore two types of negative sampling strategies. The first is a simple negative sampling (\texttt{one-neg}), in which we randomly sample one other node $k$ for every anchor node in each iteration \cite{mikolov2013distributed}. For simplicity, we assume all nodes are uniformly distributed. The second is multiple negative sampling (\texttt{mult-neg}), in which the negative examples are drawn from all other examples in the same batch \cite{henderson2017efficient}. For instance, if the batch of positive examples are $[(s_{i1}, s_{j1}), (s_{i2}, s_{j2}), ..., (s_{in}, s_{jn})]$, then the negative examples for $(s_{ik}, s_{jk})$, the pair at index $k$, are $\{s_{jk'}\}$ for $k'\in[1,n]$ and $k'\neq k$.

It is worth noting that Retweet-BERT disregards the directionality of the network and only considers the immediate neighbors of all nodes. In practice, we find that doing so balances the trade-off between training complexity and testing performance. Building on the convenience of Sentence-BERT for sentence embeddings, we use the aforementioned Sentence-BERT models pre-trained for semantic textual similarity as the base model for fine-tuning.

\section{Results}\label{sec:characterize_partisan}

We conduct two sets of evaluations to compare the methods: 1) cross-validation over the pseudo-labeled seed users as an automatic, large-scale evaluation and 2) in-house human evaluation on a set of held-out users as a complementary evaluation to the first one. We use the macro-averaged F1 score as the primary metric due to data imbalance. We note that due to our setup, many of the aforementioned related works are not directly comparable. We do not use the following network \cite{barbera2015birds,xiao2020timme}. We also do not use manual labeling \cite{wong2016quantifying} or additional external sources to determine user ideology \cite{wong2016quantifying,preotiuc2017beyond}. We do include a comparison with the label propagation method used in \citet{conover2011predicting,conover2011political,badawy2018analyzing} on the held-out users.

Finally, the best model (ours) is selected to classify all the remaining users (non-seed users) to obtain their polarity leaning labels in the \textsc{Covid-Political} dataset. These labels are used to conduct a case study of polarization COVID-19 on Twitter.

\subsection{Automatic Evaluation on Seed Users}

\begin{table*}
    \centering
    \small
    % \footnotesize
    %  \fontsize{9.0pt}{10.0pt} \selectfont
    % \scalebox{0.8}
    \caption{Retweet-BERT results for political leaning classification on seed users on the main \textsc{Covid-Political}  dataset ($N=79,000$) and the secondary \textsc{Election-2020} dataset ($N=75,000$). The best F1 (macro) scores for each model type are shown in bold, and the best overall scores are underlined. Retweet-BERT outperforms all other models on both datasets.
    %(\texttt{COVID}: $N=53,829$).
    }
    {
    \begin{tabular}{p{0.5cm}lcccccccc}
       
    \toprule 
    & & && \multicolumn{3}{c}{\textsc{Covid-Political}} & \multicolumn{3}{c}{\textsc{Election-2020}}\\
    \cmidrule(lr){5-7} \cmidrule(lr){8-10}
    & \textbf{Model}  & \textbf{Profile}& \textbf{Network} & \textbf{Acc.} & \textbf{AUC} & \textbf{F1} & \textbf{Acc.} & \textbf{AUC} & \textbf{F1}\\
    \midrule
    \multicolumn{2}{l}{\textit{Random and Majority}} \\
        & Random   & \xmark & \xmark & 0.585 & 0.501 & 0.706 & 0.499 & 0.499 & 0.506\\
        & Majority & \xmark & \xmark &  0.706 & 0.500 & 0.828 & 0.508 & 0.500 & 0.674 \\
    \midrule     
    \multicolumn{2}{l}{\textit{Average word embeddings}} \\
        & Word2Vec-google-news-300 & \cmark& \xmark  & 0.852 & 0.877 & 0.907 & 0.831 & 0.906 & 0.839\\
        & GloVe-wiki-gigaword-300  & \cmark& \xmark  & 0.856 & 0.875 &\textbf{0.909} & 0.835 & 0.908 & \textbf{0.844} \\
        
    \midrule 
    \multicolumn{2}{l}{\textit{Average LLM embeddings}} \\
        &BERT-base-uncased  & \cmark & \xmark & 0.859 & 0.882 & 0.910 &  0.837 & 0.912 & 0.844\\
        &BERT-large-uncased & \cmark & \xmark & 0.862 & 0.887 & 0.911 & 0.842 & 0.913 & 0.848 \\
        &DistilBERT-uncased & \cmark & \xmark & 0.863 & 0.888 & 0.912 & 0.845 & 0.919 & 0.851\\
        &RoBERTa-base       & \cmark & \xmark &0.870& 0.898 & 0.917 & 0.853 & 0.925 & \textbf{0.859} \\
        &RoBERTa-large      & \cmark & \xmark &0.882 & 0.914 & \textbf{0.924} & - & - & -\\
    \midrule 
    \multicolumn{2}{l}{\textit{Fine-tuned LLMs}} \\
        &BERT-base-uncased  & \cmark &\xmark  &0.900 & 0.932 & \textbf{0.934} &0.902 & 0.963 &\textbf{0.906}\\ 
        &DistilBERT-uncased & \cmark& \xmark  &0.899 & 0.931 & \textbf{0.934} & 0.899 & 0.962 & 0.904 \\
        &RoBERTa-base       & \cmark& \xmark &0.893&0.916 & 0.930 & 0.888 & 0.953 & 0.895\\
    \midrule 
    \multicolumn{2}{l}{\textit{Sentence-BERT}} \\
        &SBERT-large-uncased & \cmark& \xmark &  0.869 &    0.890&0.916 & 0.849 &0.924 & 0.855 \\
        &S-DistilBERT-uncased  & \cmark& \xmark &0.864 &    0.885 & 0.913 & 0.843 & 0.917 & 0.849\\
        &S-RoBERTa-large &  \cmark& \xmark & 0.879 & 0.903 &\textbf{0.922}  &0.874 & 0.944 & \textbf{0.878}\\

    \midrule 
    \multicolumn{2}{l}{\textit{Graph embedding}}\\
        & Node2vec* &\xmark& \cmark  & 0.928 & 0.955  &\textbf{0.949} & 0.882 & 0.944 & \textbf{0.883}\\
        & GraphSAGE + RoBERTa-base & \cmark& \cmark  & 0.789 & 0.725 & 0.873 & - & - & -\\
    \midrule
    \multicolumn{2}{l}{\textit{Retweet-BERT} (our model)} \\
        &Retweet-DistilBERT-one-neg  & \cmark & \cmark & 0.900  &  0.933 & 0.935 & - & -  & -\\
        &Retweet-DistilBERT-mult-neg & \cmark &\cmark  &0.935 & 0.965 & \underline{\textbf{0.957}} &  0.973 & 0.984 & \underline{\textbf{0.973}}\\ 
        &Retweet-BERT-base-mult-neg &\cmark & \cmark  & 0.934 & 0.966  & \underline{\textbf{0.957}} & 0.971 & 0.984 & 0.971\\
    \bottomrule
    \multicolumn{10}{l}{* Node2vec, a transductive-only model, can only be applied to non-isolated users in the retweet network}
    \end{tabular}
    }

    \label{tab:class_results}
\end{table*}
\subsubsection{Baselines} 
We conduct a 5-fold cross-validation on the seed users (i.e., full set of training users) comparing Retweet-BERT with baselines. In addition, we also use a random label predictor (based on the distribution of the labels) and a majority label predictor model as additional baselines. Table \ref{tab:class_results} shows the cross-validated results for political leaning classification on the seed users, Overall, the models perform comparatively similarly between the two datasets. Of all models that do not consider the retweet network, fine-tuned LLMs are demonstrably better. Averaging LLM outputs and fine-tuning Sentence-BERTs lead to similar results. For LLMs that have a \textit{base} and \textit{large} variant, where the \textit{large} version has roughly twice the number of tunable parameters as the \textit{base}, we see very little added improvement with the \textit{large} version, which may be attributed to having to vastly reduce the batch size due to memory issues, which could hurt performance.\footnote{\url{https://github.com/google-research/bert\#out-of-memory-issues}} DistilBERT, a smaller and faster version of BERT, produces comparable or even better results than BERT or RoBERTa. Though the network-based model, node2vec, achieves good performance, it can only be applied on nodes that are not disconnected in the retweet network. While GraphSAGE can be applied to all nodes, it vastly underperforms compared to other models due to its training complexity and time efficiency \cite{wu2020comprehensive}.

Our proposed model, Retweet-BERT, delivers the best results using the DistilBERT base model and the multiple negatives training strategy on both datasets. Other Retweet-BERT variants also achieve good results, which shows our methodology can work robustly with any base language model.

\begin{table}
    \centering
    \caption{Retweet-BERT results on 85 users with human-annotated political-leaning labels from a random sample of 100 users without seed labels. Retweet-BERT outperforms all models. 
    }
    \label{tab:results_annotated_unlabeled}
    \begin{tabular}{lccl}
        \toprule 
        \textbf{Model}  & \textbf{Profile}& \textbf{Network} & \textbf{F1}\\ %  & AUC  \\ 
        % accuracy
        \midrule
        RoBERTa-large (\textit{average}) & \cmark& \xmark    & 0.892\\ % & 0.883 
        BERT-base-uncased (\textit{fine-tuned}) & \cmark& \xmark  & 0.908 \\ %  & 0.869
        S-RoBERTA-large (\textit{SBERT}) & \cmark& \xmark   & 0.909 \\ % & 0.861
        Label Propagation* & \xmark& \cmark   & 0.910\\ % & 0.881
        node2vec* & \xmark& \cmark& 0.922 \\ % & 0.890
        Retweet-BERT-base-mult-neg & \cmark & \cmark   & 0.932 \\ %\textbf{0.917}
        \bottomrule
        
    \end{tabular}

\end{table}
\subsection{Human Evaluation on Held-out Users} 
For further validation, we manually annotated the political leanings of 100 randomly sampled users \textit{without} seed labels. We annotated these users as either left- or right-leaning based on their tweets and their profile descriptions. We were unable to determine the political leanings of 15 people.  We take the best model from each category in Table \ref{tab:class_results} and evaluate them on this labeled set. In this experiment, we also include label-propagation, a simple but efficient method to propagate pseudo-labels through the network commonly used in past work \cite{conover2011predicting,conover2011political,badawy2018analyzing}. However, label propagation and also node2vec only predict labels for nodes connected to the training network (i.e., they are transductive), but 10 nodes were not connected and thus were excluded from their evaluation. The results are reported in Table \ref{tab:results_annotated_unlabeled} for the 85 labeled users. With a macro-F1 of 0.932, Retweet-BERT outperforms all baselines, further strengthening our confidence in our model.

\section{Conclusion}

In this chapter, we proposed Retweet-BERT, a simple and elegant method to estimate user political leanings based on social network interactions (the social) and linguistic homophily (the textual). We evaluated our model on two recent Twitter datasets and compared it with other state-of-the-art baselines to show that Retweet-BERT achieves highly competitive performance (96\%-97\% macro-F1 scores). Our experiments demonstrated the importance of including both the textual and the social components. Additionally, we proposed a modeling pipeline that does not require manual annotation but only a training set of users labeled heuristically through hashtags and news media mentions. In Chapter \ref{chp:poli_covid}, we demonstrate how we applied Retweet-BERT to understand the polarization and the partisan divide of COVID-19 discourse on Twitter.

The effectiveness of Retweet-BERT is mainly attributed to the use of both social and textual data. Using both modalities led to significant improvement gains over using only one. This finding has also been validated in other contexts \cite{pan2016tri,johnson2017leveraging,yang2017overcoming,li2019encoding}, but ours is the first to apply this line of thought to detecting user ideology on social media. Our model can be utilized by researchers to understand the political and ideological landscapes of social media users. 

Though we apply Retweet-BERT specifically to the retweet network on Twitter, we note that it can be extended to \textit{any} data with a social network structure and textual content, which is essentially any social media. Though we use hashtags as the method to initiate weak labels in place of manual supervision, other methods can be used depending on the social network platform, such as user-declared interests in community groups (e.g., Facebook groups, Reddit Subreddits, YouTube channels).

This method is not without limitations. Since our method relied on mining both user profile descriptions and the retweet network, it was necessary to remove users who did not have profile descriptions or sufficient retweet interactions (see Appendix  \ref{appendix:2}). As such, our dataset only contains some of the most active and vocal users. The practical use of our model, consequently, should only be limited to active and vocal users of Twitter. Additionally, we acknowledge that Retweet-BERT is most accurate on datasets of polarizing topics where users can be distinguished almost explicitly through verbal cues. This is driven by two reasons. First, polarizing datasets makes it clearer to evaluate detection performance. Second, and more importantly, the applications of Retweet-BERT are realistically more useful when applied to controversial or polarizing topics. Since our detection method relies on users revealing explicit cues for their political preference in their profile descriptions or their retweet activities, we focus on the top 20\% (most likely right-leaning) and the bottom 20\% (most likely left-leaning) when conducting the case study on the polarization of COVID-19 discussions. The decision to leave most users out is \textit{intentional}: we only want to compare users for which Retweet-BERT is most confident in predicting political bias. Detecting user ideology is a difficult and largely ambiguous problem, even for humans \cite{elfardy2016addressing}. \citet{cohen2013classifying} raised concerns that it is harder to predict the political leanings of the general Twitter public, who are much more ``modest'' in vocalizing their political opinions. Thus, we focus our efforts on detecting the more extreme cases of political bias in an effort to reduce false positives (predicting users as politically biased when, in fact, they are neutral) over false negatives (predicting users as politically neutral when, in fact, they are biased). 

Another major limitation is that Retweet-BERT is restricted to using only user profile descriptions and retweet interactions, leaving out many more detailed and fine-grained aspects of social network data. In the next chapter, we discuss how to extend Retweet-BERT, a model that relies on retweets and LLM embeddings of the profile description, to a more comprehensive model that considers all social network interactions and content cues.

% \section{Ethical Statement} We believe our work has the potential to be used in combating misinformation and conspiracy spread, as well as identifying communication patterns between and within polarized communities. However, we are aware of the ways our work can be misused. For instance, malicious actors can use our work to politically target certain groups of users and propagate the spread of misinformation. As such, we encourage researchers to use these tools in a way that is beneficial for society. 

\chapter{Social-LLM: Learning from Social Network Data}
\label{chp:socialllm}

\section{Introduction}

In the previous chapter, we introduced Retweet-BERT, a simple and scalable solution for learning user representation from their profile descriptions and retweet connections. It builds on the assumption of social network homophily, which suggests that users connected with retweets are more likely to be characteristically and linguistically similar \cite{mcpherson2001birds,kovacs2020language}, to develop user embeddings that encode socially connected users in closer latent spaces. We evaluated its feasibility in detecting user political partisanship. 

\begin{figure*}[t]
    \centering
    \includegraphics[width=\linewidth]{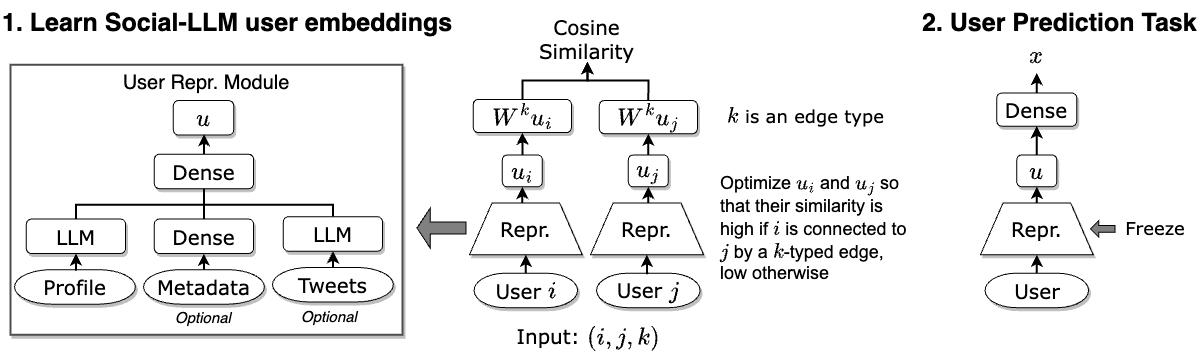}
    \caption{Overview of the Social-LLM method.}
    \label{fig:sllm_overview}
\end{figure*}

However, the concept of the method can be generalized to more complex data and is not limited to any specific tasks. In this chapter, we introduce Social-LLM, a natural extension of Retweet-BERT. Social-LLM can use multimodal user features, meaning different types of user content features, and can also draw information from heterogeneous types of network interactions. Similar to Retweet-BERT, Social-LLM remains a self-supervised representation learning method that does not need labels. It is also inductive, generalizing to any users with content features during the inference stage. We conduct a thorough evaluation of Social-LLMs on 7 real-world, large-scale social media datasets showcasing its applicability for a diverse range of user detection tasks. We also showcase the utility of using Social-LLM embeddings for visualization. We provide an overview of the model and its application in Figure \ref{fig:sllm_overview}

% Social-LLM is a pragmatic approach for modeling large-scale social network data by leveraging localized social network interactions. W We also harness the power of LLMs under the assumption of linguistic homophily in social networks \cite{kovacs2020language}.We generalizes to heterogeneous social network interactions and user content features. 
% , including political polarization, online hate speech, account suspension, and morality. % In recent years, there has been no shortage of novel graph embedding or graph neural network methods. However, they all face a critical scalability issue when it comes to large-scale datasets. 
% In this chapter, 
% Our summary of contributions is as follows:
% \begin{itemize}
% \item We propose Social-LLM, a scalable social network representation model that combines user content cues with social network cues for inductive user detection tasks.
% \item We conduct a thorough evaluation of Social-LLMs on 7 real-world, large-scale social media datasets of various topics and detection tasks.
% \item We showcase the utility of using Social-LLM embeddings for visualization.
% \end{itemize}
\section{Related Work}

User detection is a crucial element in computational social science research, encompassing a spectrum of areas such as identifying ideologies \cite{barbera2015tweeting,jiang2020political}, spotting inauthentic accounts \cite{botometer,masood2019spammer}, flagging toxic behavior \cite{ribeiro2018characterizing,jiang2023social}, recognizing influencers  \cite{rios2019semantically}, and assessing vulnerability to misinformation \cite{aral2012identifying,ye2023susceptibility}.
% ....\todo[inline]{cite} % Existing methods mainly face one of two shortcomings: (1) they don't utilize state-of-the-art results enabled by LLMs, (2) they don't consider the social network data beyond simple network statistical features. We address the first shortcoming by simply including LLMs in our modeling procedure, but that is hardly a novelty
While considerable information lies in the social network structure itself, most user characterization methods that utilize network features only consider high-level statistics like node centrality measures \cite{saravia2016midas,botometer,masood2019spammer}, failing to fully capture the complex relational patterns among users. Other studies tackle the task of political ideology prediction using Bayesian ideal point estimation \cite{barbera2015birds} or label propagation \cite{conover2011predicting,badawy2018analyzing}, but such methods require a subset of labeled users, which both come at a data acquisition cost and results in solutions tailored at very specific problems.
% In a survey paper on spam and fake user detection methods, \citet{masood2019spammer} summarized that most methods use simple content features (\textit{e.g.,} number of followers) and network features up to the point of node centrality measures but do not accommodate more complex content features or network ties. Similarly, \citet{davis2016botornot} proposed one of the most highly utilized bot detection models for Twitter, which considers more than 1,000 features but does not harness the full powers of the social network relationship data. Other works similarly compiled user and network features for mental health status detection \cite{saravia2016midas,mcmanus2015mining}.

In recent years, graph representation learning techniques have been proposed as the state-of-the-art method to model complex network information in an unsupervised manner \cite{grover2016node2vec,liu2018heterogeneous,hamilton2020graph}.
These methods capture higher-order proximity and community structure from network topology but often demand substantial computational resources and heightened hardware requirements during training that can limit scalability \cite{zhang2018network}. Some approaches alleviate the complexity with sampling strategies \cite{serafini2021scalable,ma2022graph}, though this inherently trades off representational power \cite{wu2020comprehensive}. In this chapter, we preserve the full social network structure but utilize it in the simplest manner by considering only first-order proximity (i.e., the edges between users). We demonstrate that such a simple method is sufficient for accurate user characterization on large-scale social media datasets without sacrificing computational traceability.

Our proposed Social-LLM method utilizes multi-relational user interaction data along with content-based user features. The most similar prior works are TIMME \cite{xiao2020timme}, a scalable graph neural network (GNN) for user classification that leverages multi-relational social networks, and GEM \cite{liu2018heterogeneous}, another heterogeneous GNN designed for malicious account detection. While these GNN methods can incorporate user content as node features, because they inherently rely on modeling the network structure, they cannot be applied inductively to out-of-sample users without retraining. In contrast, Social-LLM can be applied inductively on any unseen user during inference. 

Social-LLM builds upon our previous Retweet-BERT model \cite{retweetbert}, which learns user embeddings encoding political orientation by optimizing them to be similar for users who retweet each other's content. Retweet-BERT only requires user profile text and retweet interactions as input yet demonstrates effective political leaning prediction compared to other approaches. Social-LLM generalizes Retweet-BERT by modeling more extensive heterogeneous social relations beyond just retweets and incorporating other user content information.

% graph is too big to fit on device, even distributed/parallelism doesn't work because you can't partition the graph. recommend sampling 
% \cite{serafini2021scalable}
% \cite{wu2021survey} scalability trade off
\section{Method}

We propose Social-LLM, a model that leverages network homophily and user features to learn user representations scalably. This representation model can then be applied inductively for a range of downstream user detection tasks. Social-LLM draws from two types of social network features: content cues from each user and network cues from social interactions. 

\subsection{Content Cues}
The content cues are derived mainly from the textual content on their social media but can also be from other contextual metadata. We primarily utilize users' profile descriptions, which are self-provided mini-biographies. For most user detection task purposes, users' biography encodes a substantial amount of personal information with personal descriptors (e.g., ``\textit{Senator}'') and, in some cases, self-identities and beliefs (e.g., ``\textit{Democratic}''). Capped at 160 characters, these descriptions have to be short, incentivizing users to convey essential information they wish to share about themselves succinctly and attractively. The use of Twitter profile descriptions, not the tweet texts, has proved useful in a large number of computational social science research \cite{rogers2021using,jiang2022pronouns,retweetbert}. 
% piao2017inferring, thelwall2021male
% A study also found that the number of followers a user has is correlated with the length of their profiles, suggesting that most influential and popular users have meaningful profile descriptions \cite{mention2018}. 
From a practical standpoint, using user profiles instead of all of the tweets by a user also vastly reduces the complexity of the computation problem as well as alleviates data collection challenges. In addition to profile descriptions, we also leverage, when applicable, user metadata features (e.g., follower counts, account creation date, etc.) and user tweets. 

\subsection{Network Cues}
Online social media platforms offer a variety of ways to engage with one another, such as by following, liking, or re-sharing. These acts of social interaction can be gathered to form social networks. The Twitter API enables us to obtain three types of social interactions: retweeting, mentioning, and following. Though the following network is perhaps the most useful indication of user interaction, it is rarely in empirical research used due to the API rate limits set by Twitter \cite{martha2013study}. As such, following other Twitter research (e.g., \citeauthor{conover2011political}, \citeyear{conover2011political}; \citeauthor{ferrara2016rise}, \citeyear{ferrara2016rise}), we use the retweet and mention networks in this research. \textit{Retweet} refers to the act of re-sharing another user's tweet directly, without any additional comments. \textit{Mention} includes all other acts of mentioning (using `@') another user, including retweeting with comments (i.e., quoting), replying, or otherwise referencing another user in a tweet. We draw a distinction between retweets and mentions because they may represent two distinct motivations of social interaction: retweeting is usually understood as an endorsement \cite{boyd2010tweet,metaxas2015retweets} while mentioning could be used to criticize publicly \cite{hemsley2018tweeting}.

\subsection{Social-LLM Framework}

We train Social-LLM in an unsupervised manner to learn user representations in a $d-$dimensional embedding space (Figure \ref{fig:sllm_overview}, step 1). Once we train the user representation module, we can apply the user representation module to obtain user embeddings. Additional layers can be trained on top of any downstream user detection task (Figure \ref{fig:sllm_overview}, step 2). 

\subsubsection{User Representation Module} 
The user representation module takes in any user content features and produces a user embedding. Most importantly, we use pre-trained LLM models for any textual content, such as their profile description. This LLM model would be trainable in order to allow for fine-tuning in our training process. Other inputs, such as the user metadata features, will go through a deep neural network. We can also model tweet text through text embeddings, but to ensure equal input length, we precompute the text embeddings and average them. We concatenate these outputs into one single embedding and apply another dense layer to produce a single $d$-dimensional embedding $u_i$ for user $i$.

\subsubsection{Unsupervised Training Via Siamese Architecture} 
\begin{sloppypar}
The user representation module is wrapped in a Siamese model architecture in a manner similar to Sentence-BERT \cite{reimers2019sbert}, which employs identical representation modules on two sentences and optimizes the similarity of their embeddings if the sentences are deemed semantically similar. In our case, we apply an identical representation module on the user content cue and optimize the resulting embeddings based on the network cues. A training instance of Social-LLM is a tuple $(i, j, k)$ where $i$ and $j$ are two users who are connected by a social network interaction (i.e., an edge) of type $k$. We want to train the user embeddings $u_i$ and $u_j$ so that they are as similar as possible. Sentence-BERT \cite{reimers2019sbert} and Retweet-BERT (Chapter \ref{chp:retweetbert}) achieve this by optimizing the cosine similarity of embeddings. However, we also want to consider (1) multiple edge types--modeling retweets distinct from mentions--and (2) directionality--user $A$ retweeting from user $B$ is not the same as user $B$ retweeting from user $A$. To account for multiple edge types, we initialize a learnable weight matrix $W^{k}$ for every edge type $k$. To account for directionality, we can use separate weight matrices $W^{k_{\text{in}}}$ and $W^{k_{\text{out}}}$ for the in- and out-edges. We then calculate the cosine similarity scores between $W^ku_{i}$ and $W^ku_{j}$, or $W^{k_{\text{in}}}u_i$ and $W^{k_{\text{in}}}u_j$ in the directional case, as the final output. We can also account for edge weights by weighting each training instance proportionally to their weight. 
\end{sloppypar}

\subsubsection{Multiple Negatives Ranking Loss} 
We optimize with a ranking loss function, pitching positive examples against negative examples. All edges in the graph serve as positive examples, and all other pairs of users who are \textit{not} connected by an edge can serve as negative examples. To speed up the training procedure, we use the multiple negatives loss function \cite{henderson2017efficient} used in Retweet-BERT (Chapter \ref{chp:retweetbert}). Essentially, all other pairs of users in the same batch serve as negative examples. For instance, if the input batch contains positive examples $[(i_1, j_1), (i_2, j_2), ...]$, then $\{(i_x, j_y)\}$ for all $x\neq y$ are negative examples. This would encourage users who are truly connected in the graph to have more similar representations than users who are not. To minimize training complexity, we alternate the training of different types of edges in a round-robin fashion. For example, if we want to accommodate for both $k=\text{retweet}$ and $k=\text{mention}$ edges, we will train one batch of retweet edges, followed by one batch of mention edges in a round-robin manner.

\subsubsection{Downstream Task Application}
The Social-LLM model produces reusable user representation that can be used on any downstream user prediction tasks (Figure \ref{fig:sllm_overview}, step 2). We can fine-tune the representation module further or freeze the layers and add task-specific fine-tuning layers on top. 

\subsection{Advantages and Disadvantages}
Social-LLM builds on traditional user detection methods by adding social network components. There are two main advantages of Social-LLM over similar GNN approaches.
\begin{itemize}
    \item \textbf{Ease of training}: The time complexity of step 1 is $\mathcal O(|E|)$, and that of step 2 is even quicker at $\mathcal O(|V|)$. Crucially, since we forgo training the complete graph and only focus on edges as if they are independent, we can fit very large datasets via batching.
    \item \textbf{Inductive capabilities}: Since step 2 of the framework no longer relies on the network, we can extend our model to produce embeddings for any users, provided we have their content information, without needing their network information and refitting the whole model. This is called inductive learning, and most graph-embedding approaches either cannot natively support this (e.g., \citeauthor{grover2016node2vec}, \citeyear{grover2016node2vec}; \citeauthor{zhang2019prone}, \citeyear{zhang2019prone}), or they do so at a significantly higher training complexity \cite{hamilton2017inductive}.
    \item \textbf{Reusability}: The Social-LLM embedding training process is separate from the downstream applications so that we can reuse the embeddings for various applications, including user detection tasks, user clustering, and user visualization.
\end{itemize} 

The advantages of Social-LLM come with costs. Notably, we sacrifice precision and thoroughness for speed and efficiency. Our model focuses only on first-order proximity, or users who are connected immediately by an edge. This undoubtedly loses valuable information from the global network structures or higher-order proximities. However, as we will demonstrate in this chapter, in the cases of many user detection problems on social networks, it is \textit{sufficient} to model the localized connections for a \textit{cheap boost} to performance compared to a framework that does not use the social network at all. For these large but sparse real-world social network datasets, the more powerful graph embedding methods may require a lot more training time, memory footprint, or hardware resources for a marginal gain in performance.

\section{Data}
\label{sec:all_data}
We describe the dataset we use to validate our approach. The first two datasets, \textsc{Covid-Political} and \textsc{Election2020}, were described in Chapter \ref{chp:retweetbert}. These datasets focusing only on using user profile descriptions and retweet interactions to predict political partisanship. We include them in this chapter again for completeness.
% For the purpose of being self-contained, we describe the data processing steps done in \citet{retweetbert}. 
To demonstrate the additional capabilities of Social-LLM, we introduce several new datasets below that encompass more heterogeneous user metadata and network features. Our new datasets add diversity to the types of labels, prediction methods, time spans, and data sizes to demonstrate the robustness of our approach. Table \ref{tab:data_stats} displays the summary statistics of our dataset.

% \begin{table*}
%     \centering
%     \footnotesize
%     \caption{Summary statistics of our Twitter datasets.}
%     \begin{tabular}{lrrrcccrrr}
%     \toprule
%     & &  \textbf{\# Retweet}&  \textbf{\# Mention} & \textbf{Profile} & \textbf{Metadata} &\textbf{Tweet} & \textbf{Time}&& \textbf{Pred.} \\
%     \textbf{Dataset} &\textbf{ \# Users} & \textbf{Edges}  & \textbf{Edges} & \textbf{Desc.} & \textbf{Features} &\textbf{Texts} & \textbf{Span} & \textbf{Label(s)} & \textbf{Type} \\
%     \midrule
%     \textsc{Covid-Political} & 78,672 & 180,928 & - & \cmark& \cmark& \xmark & 6 Months & Partisanship (1) & Cls.\\
%     \textsc{Election2020} &  78,932 & 2,818,603 & - & \cmark& \xmark& \xmark &3 Months &Partisanship (1) & Cls. \\
%     \textsc{COVID-Morality} & 119,770 & 609,845 & 639,994 & \cmark& \cmark& \xmark &2 Years &Morality (5) & Reg.\\
%     \textsc{Ukr-Rus-Suspended} & 56,440 & 135,053& 255,476& \cmark& \cmark& \cmark &1 Month &Suspension (1) & Cls. \\
%     \textsc{Ukr-Rus-Hate} & 82,041 & 166,741 & 414,258 & \cmark& \cmark & \xmark &1 Month &Toxicity (5) &  Reg. \\
%     \textsc{Immigration-Hate-08} & 5,759 & 63,097 & 83,870 & \cmark& \cmark & \xmark& All times &Toxicity (5) & Reg.\\
%     \textsc{Immigration-Hate-05} &2,188 & 4,827 & 7,993 & \cmark& \cmark & \xmark& All times & Toxicity (5)  &Reg.  \\
%     \bottomrule
%     \end{tabular}
% \label{tab:data_stats}
% \end{table*}

\begin{table*}
    \footnotesize
    \caption{Summary statistics of our Twitter datasets.}
    \begin{tabular}{lrrrcccrrr}
    \toprule
    & &  \textbf{\# Rt}&  \textbf{\# Mn} & \textbf{Profile} & \textbf{Meta} &\textbf{Tweet} & \textbf{Time}&& \textbf{Pred.} \\
    \textbf{Dataset} &\textbf{ \# Users} & \textbf{Edges}  & \textbf{Edges} & \textbf{Desc.} & \textbf{Feat.} &\textbf{Texts} & \textbf{Span} & \textbf{Label(s)} & \textbf{Type} \\
    \midrule
    \textsc{Covid-Political} & 79,370 & 180,928 & - & \cmark& \cmark& \xmark & 6 Mo & Partisanship (1) & Cls.\\
    \textsc{Election2020} &  75,301 & 2,818,603 & - & \cmark& \xmark& \xmark &3 Mo &Partisanship (1) & Cls. \\
    \textsc{COVID-Morality} & 119,770 & 609,845 & 639,994 & \cmark& \cmark& \xmark &2 Yr &Morality (5) & Reg.\\
    \textsc{Ukr-Rus-Suspended} & 56,440 & 135,053& 255,476& \cmark& \cmark& \cmark &1 Mo &Suspension (1) & Cls. \\
    \textsc{Ukr-Rus-Hate} & 82,041 & 166,741 & 414,258 & \cmark& \cmark & \xmark &1 Mo&Toxicity (6) &  Reg. \\
    \textsc{Immigration-Hate-08} & 5,759 & 63,097 & 83,870 & \cmark& \cmark & \xmark& All &Toxicity (5) & Reg.\\
    \textsc{Immigration-Hate-05} &2,188 & 4,827 & 7,993 & \cmark& \cmark & \xmark& All & Toxicity (5)  &Reg.  \\
    \bottomrule
    \end{tabular}
\label{tab:data_stats}
\end{table*}

\subsection{Politics in COVID Discussion}\label{sec:data_covid_political}
The COVID-19 pandemic left an unprecedented impact on everyone worldwide. Research has shown that COVID-19 was politicized, with partisanship steering online discourse about the pandemic \cite{calvillo2020political,jiang2020political}, motivating our prediction task of detecting user partisanship. Our dataset, \textsc{Covid-Politics}, is based on a real-time collection of COVID-19 tweets \cite{chen2020covid} between January 21 and July 31, 2020, and was further preprocessed in \S\ref{ref:rtbert_data} to remove bots and inactive users \cite{botometer,yang2022botometer}.
% The time period of this dataset is limited to January 21 to July 31, 2020. Following \citet{garimella2018quantifying}, we retain only retweet interactions between users if there are at least two retweet occurrences to filter for stronger indications of retweet endorsement. To eliminate users who were inactive or were not sampled sufficiently to be included in the dataset, we remove users with total degrees (in- and out-degrees) less than 10. Further, to mitigate the influence of social bots, we also removed the top 10\% of users who are most likely to be bots \cite{davis2016botornot,yang2022botometer}. After filtering, 
We include user metadata features of the initial follower count, final follower count, number of tweets, number of original tweets, number of days actively posting, and whether the user is verified. The ground truth partisanship labels from \citet{retweetbert} are derived from two heuristics-labeling approaches, using annotated political hashtags used in user profile descriptions and the partisanship leaning of new media URLs mentioned in users' tweets.
This dataset contains around 79,000 users and 181,000 retweet interactions.
The distribution of users is unbalanced, with 75\% of users labeled as left-leaning.

\subsection{2020 Presidential Election}
\begin{sloppypar}
The 2020 US presidential election took place amidst the backdrop of the COVID-19 pandemic. 
% Former Vice President Joe Biden, the Democratic nominee, defeated the incumbent Republican President Donald Trump.
The \textsc{Election-2020} dataset is based on a real-time collection of tweets regarding the 2020 US presidential election \cite{chen2021election2020} from March 1 to May 31, 2020. This dataset was also preprocessed to remove bots and inactive users  \S\ref{ref:rtbert_data}.
% The dataset was also preprocessed in our prior work \cite{retweetbert}, and we retained only the profile description of the users and retweet interactions, not user metadata features. Similar to the \textsc{Covid-Political} dataset (\S\ref{sec:data_covid_political}), we eliminate users who are likely bots (top 10\% of bot scores), edges that occurred only once, users with degrees less than 10. After filtering, 
This dataset contains around 75,000 users and 2.8 million retweet interactions.
Since this is specifically a dataset on US politics, we are also interested in predicting user partisanship. We use the same partisanship labeling approach as above. The distribution split is even, with around 50\% of the users labeled as left-leaning. 
\end{sloppypar}

\subsection{Morality in COVID Discussion} \label{sec:data_covid_morality}
% \begin{sloppypar}

The \textsc{Covid-Morality} dataset spans tweets on COVID-19 from February 2020 to October 2021 \cite{chen2020covid}. Our task is to predict users' moral foundations based on the Moral Foundation Theory: care, fairness, loyalty, authority, and purity \cite{haidt2004intuitive}. Prior work found social networks exhibit moral homophily, with moral values such as purity predicting network distances \cite{dehghani2016purity}. Studies also link morality to COVID-19 health decisions such as masking and vaccination \cite{chan2021moral,diaz2022reactance,francis2022moral}, suggesting morality's role in online communication patterns \cite{jiang2024moral}.

We use the best off-the-shelf moral value detector \cite{guo2023data}, a data-fusion morality detector technique fine-tuned on this specific dataset, to label tweets with the 10 moral inclinations. Retaining tweets with $>=1$ moral value, we filter for active users with $>=10$ moral tweets per month. From this set, we sample 150,000 users and build their retweet/mention network, keeping edges with weight $>=2$ \cite{retweetbert}. User meta features include account age, follower count, following count, list count, tweet count, favorite count, and whether they are verified. Our multi-output regression task predicts each user's average moral foundation scores. We aggregate the 10 labels into 5 foundations, scoring 1 if both the virtue and the vice are present, 0.5 if only one is present, and 0 otherwise. We combine the virtue and vice scores for each moral foundation because the morality detector separately identifies explicit expressions of both the positive (e.g., \textit{care}) and negative (e.g., \textit{harm}) aspects. However, for our purposes, the presence of either aspect reflects a moral disposition within that particular foundation.
We then calculate the user-level moral foundation score as the average across their tweets. 

% The distributions of the labels are shown in Figure \ref{fig:covid_mf_label_dist}.

\subsection{Ukraine-Russia Suspended Accounts}\label{sec:data_ukr_rus_suspended}
% indiana2022suspicious
After the Ukraine-Russia war erupted in 2022, Russian disinformation campaigns were rampant on social media \cite{pierri2023propaganda}. We use tweets about the conflict from March 2022 \cite{chen2023ukrrus}, where many users were suspended by Twitter,\footnote{\url{https://help.twitter.com/en/managing-your-account/suspended-x-accounts}} often newer accounts exhibiting suspicious behaviors \cite{pierri2023does}. We hypothesize that suspended and non-suspended users have different communication patterns due to divergent motivations.

Our task predicts whether a user was eventually suspended. The raw dataset contains ~10M non-suspended and ~1M suspended users. We filter for users with $>=10$ tweets to remove inactive users. To focus on likely human-operated accounts, we remove users with $>130$ tweets that month (90th percentile) as a rough de-spamming step since we no longer had access to Botometer at this time of research. We then sample an equal proportion of non-suspended users.
% At this time of the research, we no longer had access to the Botometer \cite{davis2016botornot} tool due to it losing Twitter's API access \cite{botometershutdown}. Therefore, we resort to using the maximum tweet amount as a rough elimination criterion to remove spamming users. We choose 130, the 90\% threshold of the number of tweets per user after removing users with less than 10 tweets. Having 130 tweets in one month represents 4.19 tweets per day, which is a reasonable maximum number of tweets an authentic account could post. 
We build the retweet/mention network with edges $>=2$ weight \cite{retweetbert}, removing isolated users. The final \textsc{Ukr-Rus-Suspended} dataset is 58\% suspended users. We use the same user metadata features as in \S\ref{sec:data_covid_morality}. We also include tweet texts since the labels are not derived from the tweets.

\subsection{Ukraine-Russia Hate Speech} \label{sec:data_ukr_rus_hate}
Beyond misinformation, online discourse on the Ukraine-Russia war was riddled with toxicity \cite{thapa2022multi}. As a spin-off from \textsc{Ukr-Rus-Suspended}, we experiment with detecting users' toxicity levels from their Twitter behavior in this \textsc{Ukr-Rus-Hate} dataset. After filtering for users with $>=10$ and $<=130$ tweets, we use the Perspective API\footnote{\url{https://perspectiveapi.com/}} toxicity detector \cite{kim2021distorting,frimer2023incivility} on users' original tweets. We use six toxicity scores (overall, identity attack, insult, profanity, threat) from 0-1. We filter for users with $>=10$ tweets rated for toxicity. Unlike \textsc{Ukr-Rus-Suspended}, we retain all network edges regardless of weight due to smaller network density. User metadata features are the same as \S\ref{sec:data_covid_morality}. The prediction task is to detect each user's overall toxicity level from their tweets, Twitter activity, and network interactions. Modeling toxicity can provide insights into how toxic language and user behaviors propagate through online networks during crisis events.

\begin{figure}
    \centering
    \includegraphics[width=0.5\linewidth]{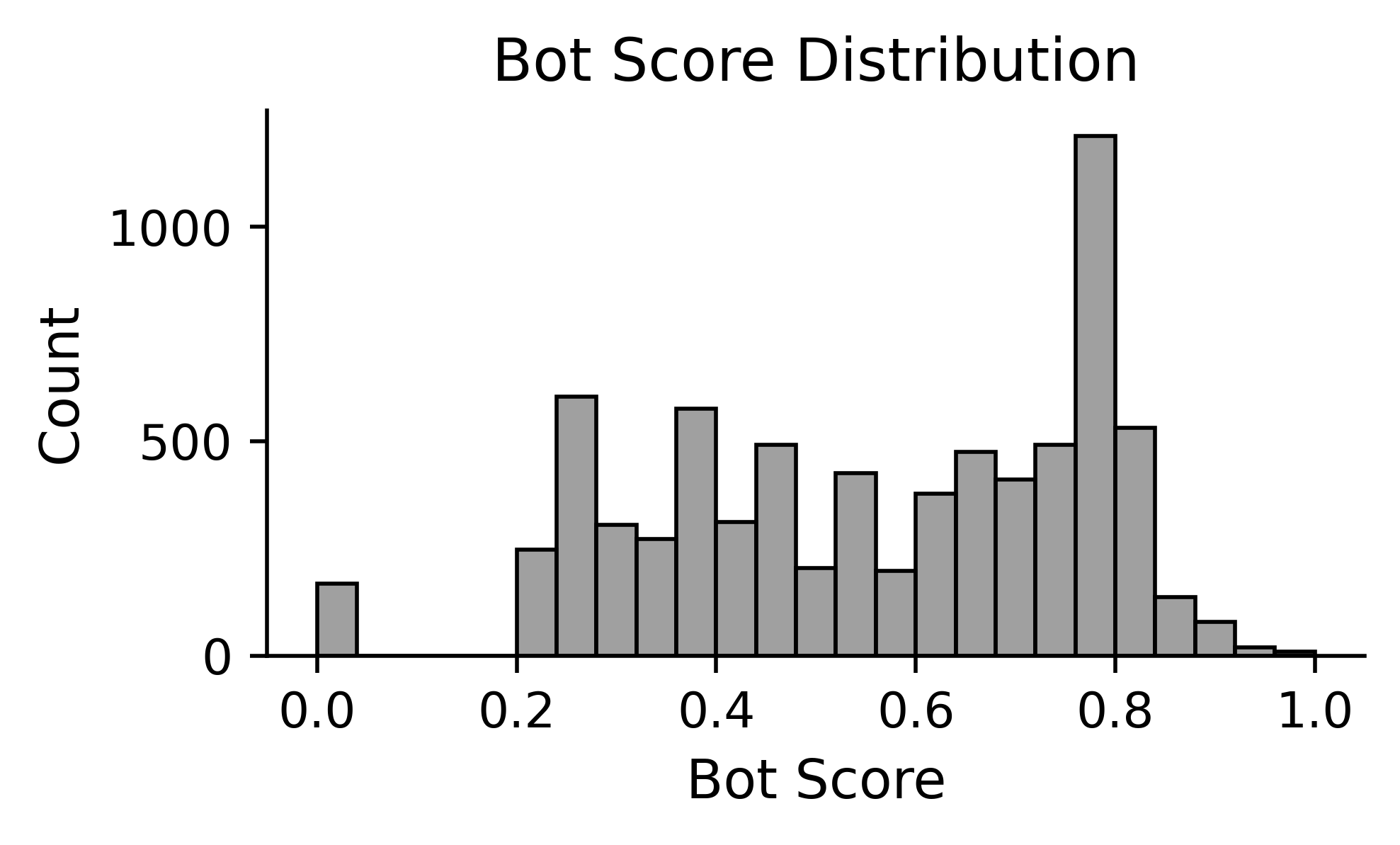}
    \caption{Distribution of users' bot scores prior to user preprocessing for the \textsc{Immigration-Hate} datasets.}
    \label{fig:hate_bot_dist}
\end{figure}

\subsection{Immigration Hate Speech} \label{sec:sllm_data_hate}

We compile an \textsc{Immigration-Hate} dataset by collecting historical tweets from users who previously posted hateful immigration content \cite{bianchi2022njh}. From 18,803 annotated uncivil/intolerant tweets in 2020-2021 \cite{bianchi2022njh}, we successfully rehydrated 8,790 tweets by 7,566 users (the rest of the tweets were no longer available). Considering these hateful users, we used the Twitter API to collect up to their most recent 3,200 tweets, yielding 21 million tweets. Here, we focus only on the 2.9 million original tweets.

We apply the Perspective API toxicity detector (\S\ref{sec:data_ukr_rus_hate}) and aggregate each user's average toxicity scores across their tweets. A Botometer \cite{botometer,yang2022botometer} analysis revealed many bot accounts (Figure \ref{fig:hate_bot_dist}). To mitigate the influence of bots, we remove users using two thresholds of bot score: 0.8, which is a conservative choice given the peak in the distribution of bot scores, and 0.5, which would leave us substantially fewer users but with a higher certainty that they are genuine. 
Since the user set is relatively small, we retain all network edges. User metadata features are the same as \S\ref{sec:data_covid_morality}. The prediction task aims to model each user's propensity for hate speech from their toxicity levels, activity patterns, and network position.

\section{Evaluation}
We evaluate Social-LLM by conducting extensive comparisons with baseline methods for user detection tasks. We also include sensitivity and ablation studies.

\subsection{Baseline Methods}
For a thorough evaluation of our approach, we use a series of state-of-the-art baseline methods divided into three camps: content-based, network-based, and hybrid methods. The content-based and network-based models provide an alternative user embedding that we can utilize in the evaluation procedure (Figure \ref{fig:sllm_overview}, step 2). All input embeddings undergo similar training processes for target task prediction. For the hybrid method, we use TIMME, an end-to-end user detection method that also uses both user features and network features. We conduct a thorough hyperparameter tuning process for all of the baseline models.

\subsubsection{Content-Based Methods}
\begin{sloppypar}

For Content-Based Methods, we primarily investigate using embeddings from pre-trained LLMs. Fine-tuning the LLMs for our specific purpose is also one option; however, doing so on the \textsc{Covid-Politics} and \textsc{Election2020} dataset did not deliver a substantial enough improvement to justify the added training cost. Here, we experiment with the following three LLMs applied to the profile descriptions: 
\begin{enumerate}
    \item RoBERTa-base \cite{liu2019roberta},
    \item BERTweet \cite{bertweet}, a RoBERTa fine-tuned on Twitter data,
    \item SBERT-MPNet, a Sentence-BERT \cite{reimers2019sbert} model based on MPNet \cite{song2020mpnet} and is, as of late 2023, the best-performing Sentence-BERT model.\footnote{\url{https://www.sbert.net/docs/pretrained_models.html} (Accessed November 2023).}
\end{enumerate}
\end{sloppypar}

For datasets with additional metadata features, we also experiment with using only the raw metadata features as the ``user embeddings'' as well as with concatenating the LLM embeddings with the raw metadata features. For \textsc{Ukr-Rus-Suspended}, we additionally experiment with applying the aforementioned three LLMs on users' tweets, averaging one LLM embedding per user.

\subsubsection{Network-Based Methods}
We use two network-based methods as baselines: node2vec \cite{grover2016node2vec} and ProNE \cite{zhang2019prone}. While GraphSAGE \cite{hamilton2017inductive} is another suitable choice for inductive graph representation learning with node attributes, it is, in practice, to train on a large graph within reasonable time limits and can, therefore, underperform \cite{retweetbert}. These network embedding methods support weights and directions but also heterogeneous edge types. Therefore, we run a separate network model on the (1) retweet edges only, (2) mention edges only, and (3) indiscriminately combining retweet and mention edges as one.
% (with edge weights equal to the sum of the retweet edge and mention edge weights). 

\subsubsection{Hybrid Method}
We use TIMME \cite{xiao2020timme} as our hybrid method baseline, providing it with the user content features and network features as our Social-LLM model. The original model was only designed for user classification tasks, but we modified the open-sourced code to enable regression. Since TIMME is designed to be a multi-relational model, we mainly apply it on both retweet and mention edges, but we also experiment with combining these edges indiscriminately. 

\subsection{Experimental Setup}
For every dataset and its corresponding set of user embeddings, we conduct 10 repetitions of the training and evaluation procedure, splitting the dataset randomly using 10 pre-selected seeds into 60\%-20\%-20\% train-val-test splits. The validation sets are used for early stopping and model selection. 
% The model architecture and hyperparameters are fixed for all experiments to maintain cross-comparison. 
The classification tasks are evaluated using Macro-F1, and the regression tasks are evaluated using Pearson's correlation averaged across multiple labels.

\section{Results}
% Below, we discuss our experimental results and interpret our findings. For model selection within the same family of methods, we use the validation sets to select the final model. Most results are presented in Table \ref{tab:results}, which shows the average result over 10 repeated random splits. 
\subsection{Experiments}
Below, we discuss our experimental results and interpret our findings. For model selection within the same family of methods, we use the validation sets to select the final model. The results for the classification tasks are shown in Table \ref{tab:sllm_results_cls}, and the results for the regression tasks are shown in Table \ref{tab:sllm_results_reg}. We display the average value of the metric over 10 repeated random splits.

\begin{table*}[t]
\centering
\small
\caption{Social-LLM results on the classification task datasets evaluated using Macro-F1 scores. The best model for each experiment is in bold, and the best baseline model is underlined.}
\label{tab:sllm_results_cls}
\begin{tabular}{@{}llccwr{1.8cm}wr{1.8cm}wr{1.8cm}}
\toprule
& &\textbf{Content}& \textbf{Network} & \textsc{\textbf{Election}} & \textsc{\textbf{Covid-}} & \textsc{\textbf{Ukr-Rus-}} \\
& & \textbf{Features} & \textbf{Features} & \textsc{\textbf{2020}} & \textsc{\textbf{Political}} & \textsc{\textbf{Suspended}} \\
\midrule
\multicolumn{7}{l}{\textit{Experiment 1: LLMs}}\\
& RoBERTa & \cmark & \xmark & 80.11 & 78.41 & 56.21  \\
& BERTweet & \cmark & \xmark  & 79.31 & 78.42 & 55.69 \\
& SBERT-MPNet & \cmark & \xmark  & \textbf{86.47} & \textbf{82.99} & \textbf{56.79} \\
\midrule
\multicolumn{7}{l}{\textit{Experiment 2 (Main): Baselines vs Social-LLM}}\\
& (a) Profile LLM  & \cmark & \xmark & \underline{86.47} & 82.99 & 56.79  \\
& (a) + (b) Metadata  & \cmark & \xmark & - & 83.26 & 70.75  \\
& (a) + (b) + (c) Tweet LLMs & \cmark & \xmark & - & - & \underline{81.74}  \\
& (d) node2vec  & \xmark & \cmark & - & \underline{88.65} & 72.33  \\
& (e) ProNE  & \xmark & \cmark &76.28 & 64.04 & 77.95 \\
& (f) TIMME & \cmark & \cmark & 84.81 & 81.85 & 72.91 \\
& \textbf{Social-LLM} & \cmark & \cmark& \textbf{97.87} & \textbf{90.82} & \textbf{82.71} \\
% \midrule
& \%$\uparrow$ && & 13\% & 2\% & 1\% \\
\midrule
\multicolumn{7}{l}{\textit{Experiment 3: Ablation on edge types in Social-LLM models}}\\
& RT & \cmark & \cmark & - & - & 70.71 \\
& MN & \cmark & \cmark & - & - &71.32 \\
& RT \& MN (distinct) & \cmark & \cmark & - & - &71.99  \\
& RT + MN (indistinct) & \cmark & \cmark & - & - & \textbf{72.10}  \\
\midrule
\multicolumn{7}{l}{\textit{Experiment 4: Ablation on edge directions and weights in Social-LLM models}}\\
& (best edge combo model) & \cmark & \cmark & 97.78 & 90.68  & 72.10  \\
&  + w & \cmark & \cmark & 97.78& 90.55  & 71.85  \\
&  + d & \cmark & \cmark & \textbf{97.85} & \textbf{90.82}& 71.77 \\
&  + d + w & \cmark & \cmark & 97.82 & 90.42 & \textbf{72.17} \\
\bottomrule
\end{tabular}
\end{table*}

\begin{table*}[t]
\centering
\small
\caption{Social-LLM results on the regression task datasets evaluated using Pearson correlation scores. The best model for each experiment is in bold, and the best baseline model is underlined.}
\label{tab:sllm_results_reg}
\begin{tabular}{@{}llccwr{1.7cm}wr{1.7cm}wr{1.7cm}wr{1.7cm}}
\toprule
& & \textbf{Content} & \textbf{Network} & \textsc{\textbf{Covid-}} & \textsc{\textbf{Ukr-Rus-}} & \multicolumn{2}{r}{\textsc{\textbf{Immigration-Hate-}}}  \\
& & \textbf{Features} & \textbf{Features} &\textsc{\textbf{Morality}} & \textsc{\textbf{Hate}} & \textsc{\textbf{05}} & \textsc{\textbf{08}}\\
\midrule 
\multicolumn{8}{l}{\textit{Experiment 1: LLMs}}\\
& RoBERTa & \cmark & \xmark  & 32.84 & 36.54 & 12.06 & 9.30 \\
& BERTweet & \cmark & \xmark   & 30.72 & 40.38 & 14.33 & 12.03 \\
& SBERT-MPNet & \cmark & \xmark & \textbf{36.77} & \textbf{43.35} & \textbf{17.16} & \textbf{16.76 }\\
\midrule
\multicolumn{8}{l}{\textit{Experiment 2 (Main): Baselines vs Social-LLM}}\\
& (a) Profile LLM  & \cmark & \xmark & 36.77 & 43.35 & 17.16 & 16.76 \\
& (a) + (b) Metadata  & \cmark & \xmark & -  &\underline{ 45.38} & 17.72 & 17.32 \\
& (a) + (b) + (c) Tweet LLMs & \cmark & \xmark  & - &- & - &-  \\
& (d) node2vec  & \xmark & \cmark  & 50.53 & 39.97 & 10.70 & 12.18 \\
& (e) ProNE  & \xmark & \cmark  & \underline{\textbf{51.13}} & \underline{45.38} & 5.47 & 14.30 \\
& (f) TIMME & \cmark & \cmark  & 30.47 & 43.46 & \underline{20.98} & \underline{18.67}\\
& \textbf{Social-LLM} & \cmark & \cmark & 50.15 & \textbf{57.27} & \textbf{21.17} & \textbf{20.11} \\
% \midrule
& \%$\uparrow$ && & -2\% & 26\% & 1\% & 7\% \\
\midrule
\multicolumn{8}{l}{\textit{Experiment 3: Ablation on edge types in Social-LLM models}}\\
& RT & \cmark & \cmark & 46.57 & 48.18 & 18.85  & 18.18 \\
& MN & \cmark & \cmark & 45.33  & 49.55 & 18.73  & 17.92 \\
& RT \& MN (distinct) & \cmark & \cmark & 20.40  & 51.20  & 14.75 & 18.89  \\
& RT + MN (indistinct) & \cmark & \cmark  & \textbf{47.51} & \textbf{50.73}  & \textbf{19.05} & \textbf{18.53}  \\
\midrule
\multicolumn{8}{l}{\textit{Experiment 4: Ablation on edge directions and weights in Social-LLM models}}\\
& (best edge combo model) & \cmark & \cmark  & 47.51 & 50.73 & 18.53  & 19.05  \\
&  + w & \cmark & \cmark  & 46.98 & 51.46 & 18.70  & 18.81  \\
&  + d & \cmark & \cmark & \textbf{50.15}  & \textbf{57.19} & \textbf{18.95}  & 18.77  \\
&  + d + w & \cmark & \cmark  & 46.89 & 49.35 & 17.67 & \textbf{19.21}\\
\bottomrule
\end{tabular}
\end{table*}

\subsubsection{Experiment 1 - Choice of LLMs}
We first experiment with the choice of LLMs by running our prediction tasks using only the profile LLM embeddings as features. This determines both the best baseline method for LLMs and which LLM we should use in our Social-LLM models. We selected the clear winner, SBERT-MPNet, which outperformed RoBERTa and BERTweet on all datasets. We note that, given rapid innovations in NLP and LLMs, SBERT-MPNet may not be the best model or could soon be replaced by a better successor. However, the contribution of Social-LLM is not tied to a single LLM but rather a model training paradigm that can be paired with any LLM. Our choices of LLMs are driven by ease of use, costs, and reproducibility \cite{bosley2023we}.

% \begin{figure*}
%     \centering
%     \includegraphics[width=\linewidth]{figures/exp_training_size.png}
%     \caption{Sensitivity to training sizes (Experiment 7).}
%     \label{fig:exp_training_size}
% \end{figure*}

\subsubsection{Experiment 2 - Main Experiments}
In Tables \ref{tab:sllm_results_cls} and \ref{tab:sllm_results_reg} Experiment 2, we underline the best baseline model and boldface the best model (either a baseline or the Social-LLM model) for each dataset. We also indicate the percentage change in performance using Social-LLM compared to the best-performing baseline. Regarding baseline methods, there is no clear winner among the content-based, network-based, or hybrid models. Network-based models face much higher variability in performance across datasets, pointing at issues when using solely network features. Notably, we observe that Social-LLM is superior in nearly all cases, with improvements ranging from a substantial 26\% to a modest 1\%. Using a one-sided $t$-test, we find that all improvements are statistically significant. Social-LLM performs comparatively worse than the baselines only in \textsc{Covid-Morality}, where the network embedding models are slightly superior, but the Social-LLM model still demonstrates commendable performance. In summary, Social-LLM emerges as the most consistent model, exhibiting robustness across various tasks and data sizes.

% \subsection{Experiment 2: Ensemble Models}

\subsubsection{Experiment 3 - Edge Type Ablation}
We perform an edge-type ablation experiment to evaluate the importance of each edge type on datasets containing both retweets and mention edges. When using only one type of edge, we find they perform comparably. The combination of retweets and mentions as two distinct edge types occasionally results in improved performance but can also lead to deteriorated outcomes. However, combining them indiscriminately as one edge type yields the best performance consistently. This suggests that both retweets and mentions carry important signals, yet the distinctions between the two actions might not be substantial enough to warrant differentiation for our objective tasks.

\subsubsection{Experiment 4 - Edge Weights and Directions}
Using the best edge-type model (RT for \textsc{Election2020} and \textsc{Covid-Political}, and RT + MN for all others), we then experiment with adding edge weights (+ w) and edge directions (+ d). The inclusion of directions always yields better performance, and occasionally, the performance is further enhanced if we stack on weights as well. The importance of directionality emphasizes the value of understanding the flow of information exchange on social networks.

\begin{figure}[t]
    \centering
    \includegraphics[width=0.5\linewidth]{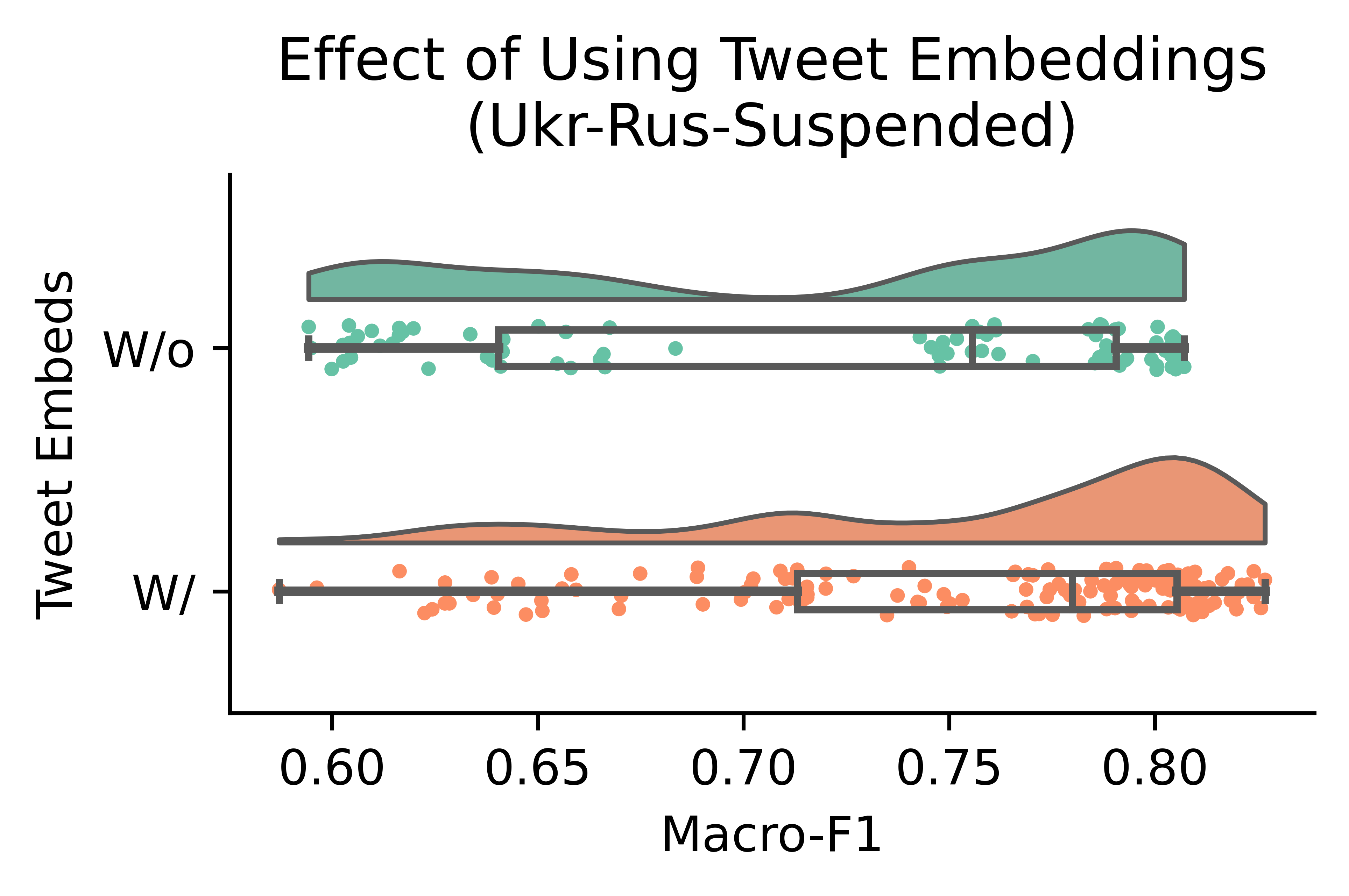}
    \caption{Social-LLM Experient 5: Ablation study of user tweet embeddings on the \textsc{Ukr-Rus-Suspended} dataset.}
    \label{fig:use_tweet_embeds_ukr_rus_suspended}
\end{figure}

\subsubsection{Experiment 5 - User Tweet Embeddings}
On \textsc{Ukr-Rus-Suspended}, we experiment with additionally including user tweet embeddings as features. In Figure \ref{fig:use_tweet_embeds_ukr_rus_suspended}, we see that using user tweet embeddings leads to an average improvement of 4\% Macro-F1 between otherwise identically configured models. This experiment underscores the importance of including user tweets, when applicable and suitable, in user prediction tasks.
\begin{figure}[t]
    \centering
    \includegraphics[width=\linewidth]{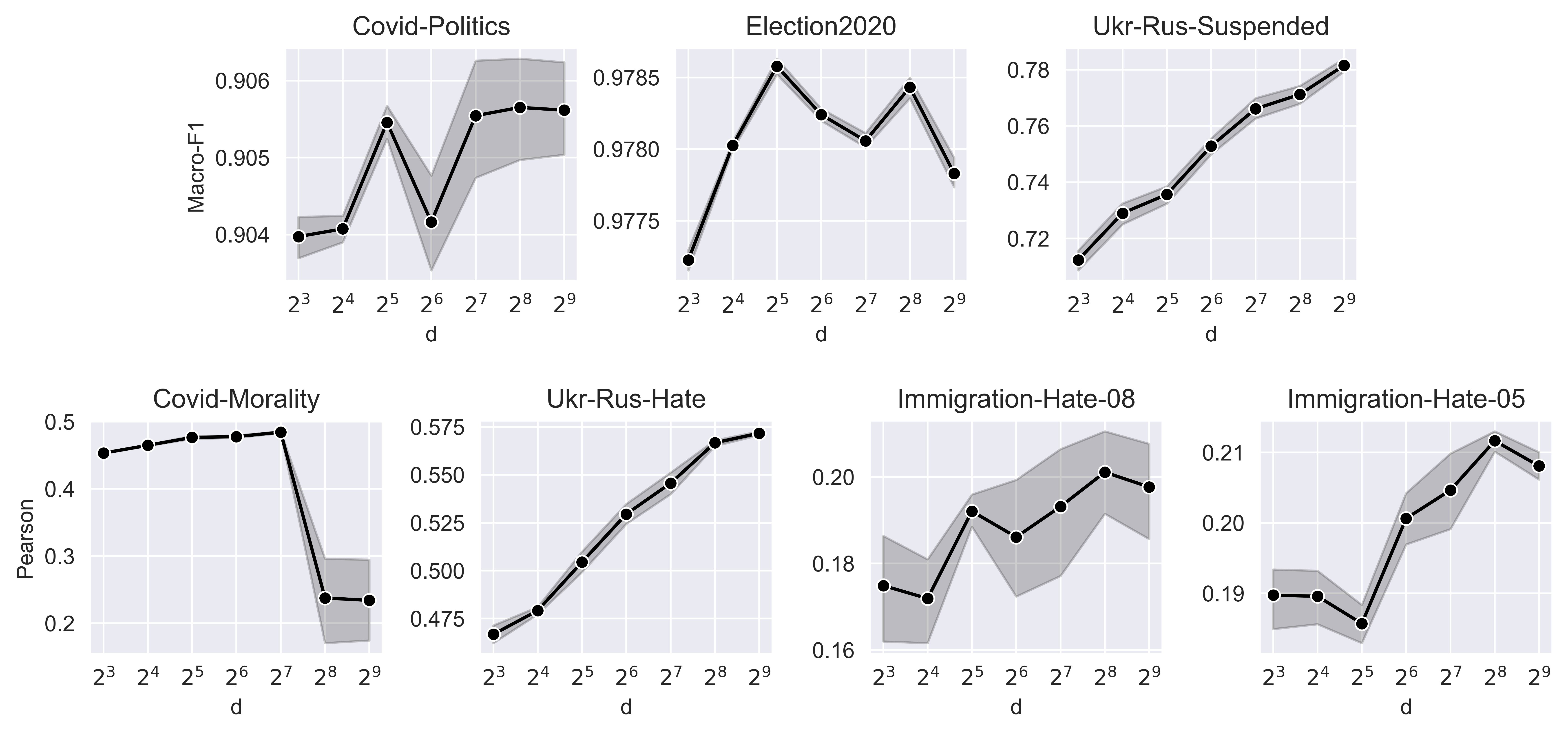}
    \caption{Social-LLM Experiment 6: Sensitivity to embedding dimension $d$.}
    \label{fig:exp_dimension}
\end{figure}
\begin{figure}[t]
    \centering
    \includegraphics[width=\linewidth]{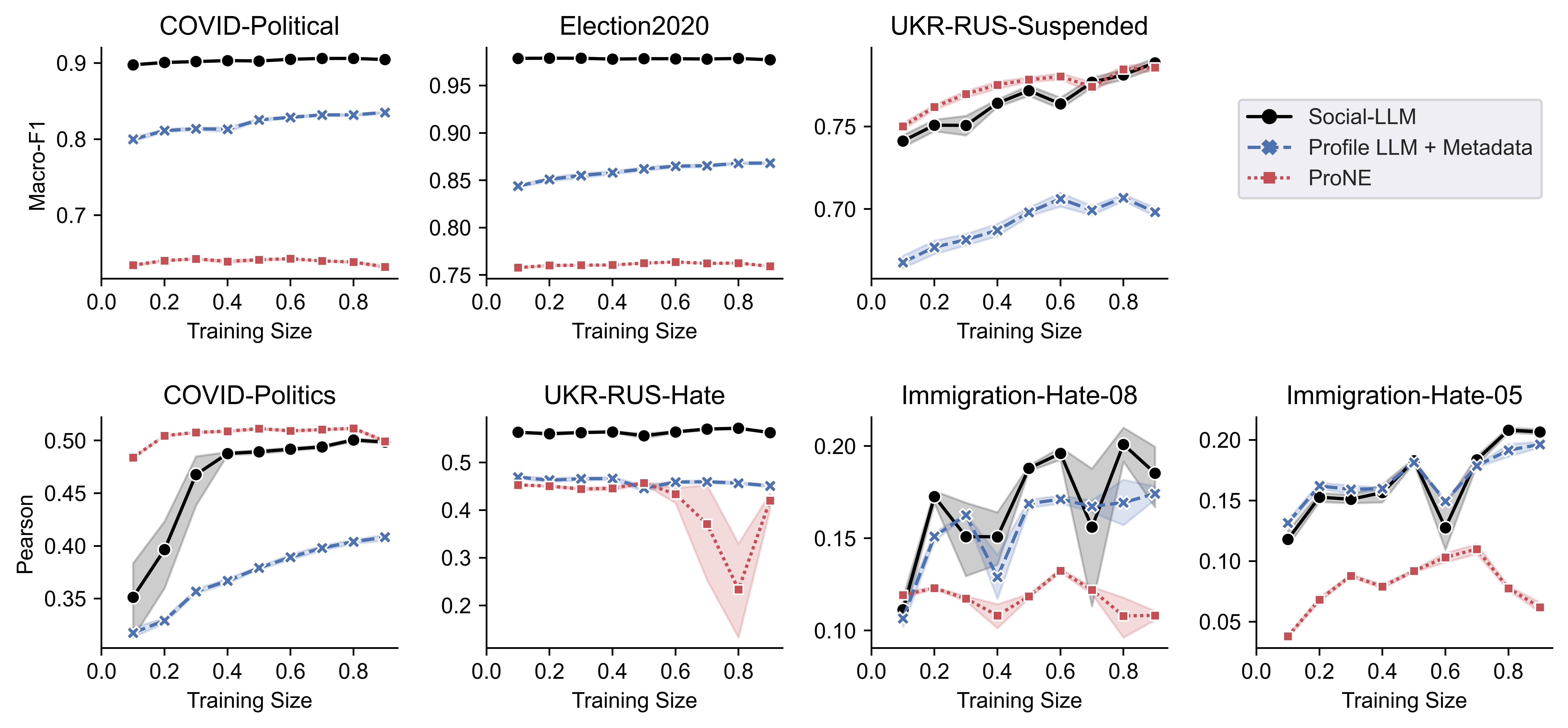}
    \caption{Social-LLM Experiment 7: Sensitivity to training size.}
    \label{fig:exp_training_size}
\end{figure}
\subsubsection{Experiment 6 - Sensitivity to Dimension Size}
For every dataset, we select the Social-LLM model with the best edge type, edge weight, and edge direction configuration to plot the sensitivity to embedding dimension $d$. The results are presented in Figure \ref{fig:exp_dimension}. Performance generally increases with rising dimensions, with $d=258$ being a popular choice; however, we note that Social-LLM usually performs quite well even with very low dimensions.

\subsubsection{Experiment 7: Sensitivity to Training Size}
We compare how Social-LLM fares with baseline methods under varying conditions of training size in Figure \ref{fig:exp_training_size}. On \textsc{Covid-Political}, \textsc{Election2020}, and \textsc{Ukr-Rus-Hate}, Social-LLM consistently yields higher performance even given a very small training size. In other datasets, Social-LLM demonstrates similar or slightly inferior performance compared to the best-performing baseline but is never substantially worse than the best baseline. Given the variability in the baseline models across datasets, the consistency and robustness in Social-LLM performance is noteworthy.

\begin{figure}[t]
    \centering
    \includegraphics[width=0.45\linewidth]{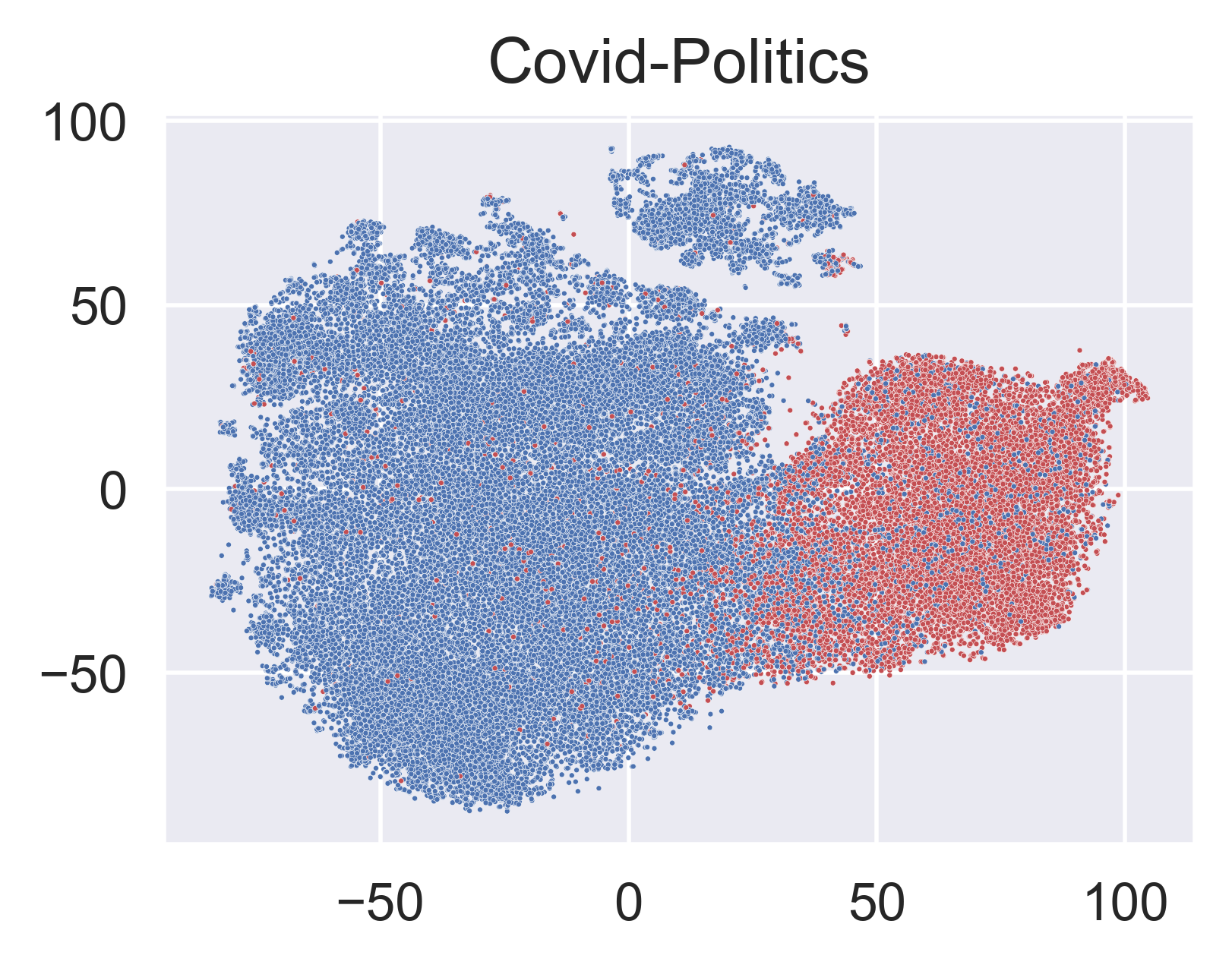}
    \includegraphics[width=0.45\linewidth]{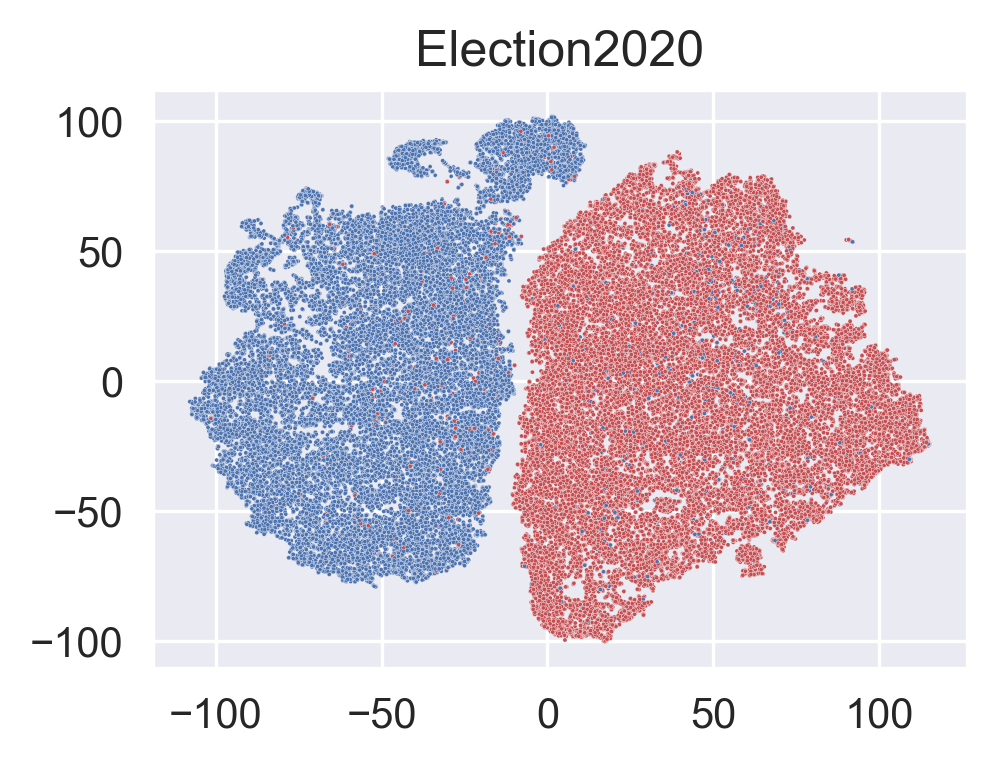}
    \includegraphics[width=0.45\linewidth]{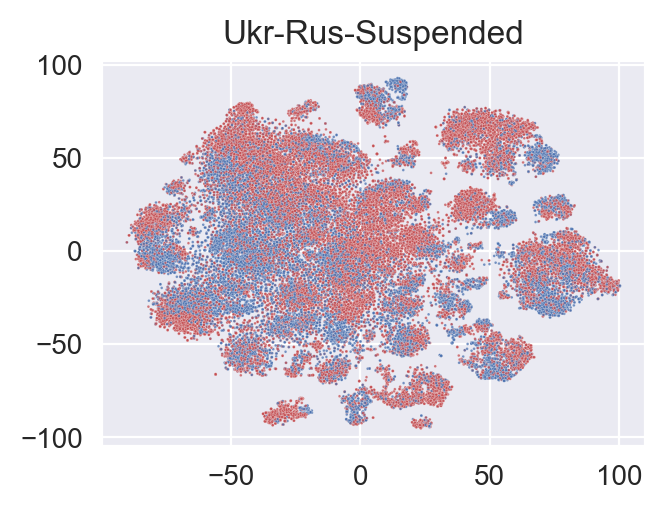}
    \includegraphics[width=0.45\linewidth]{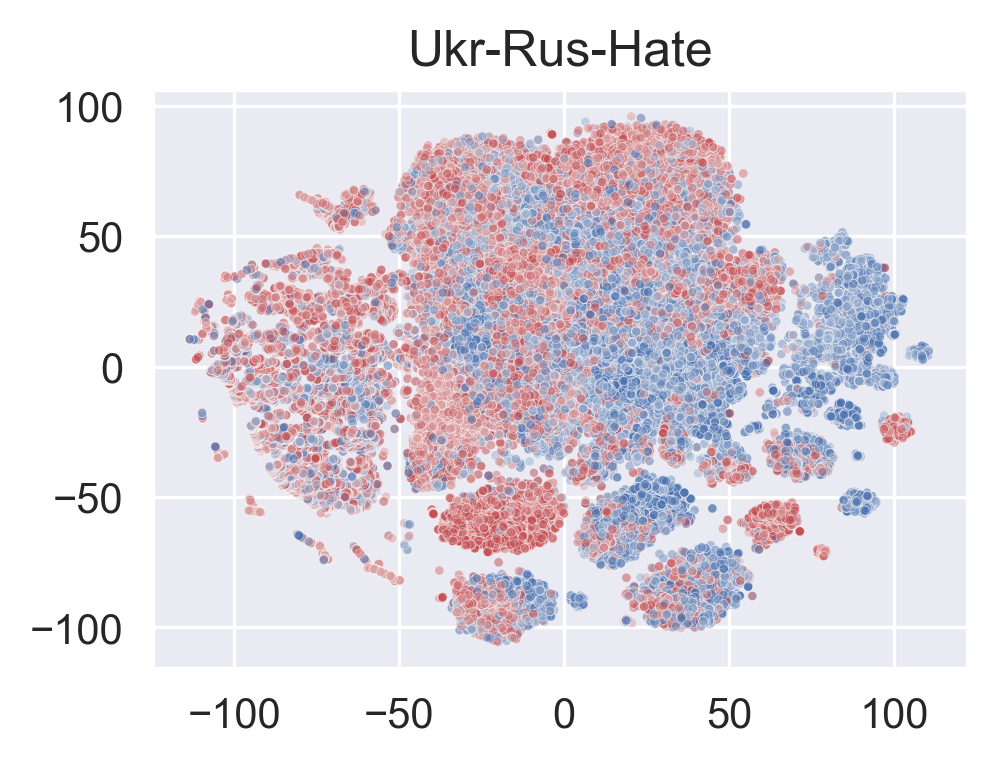}
    \caption{Visualization of Social-LLM embeddings.}
    \label{fig:viz}
\end{figure}
\subsection{Visualization}
We further demonstrate Social-LLM's interpretability by visualizing the user embeddings with t-SNE \cite{tsne}. As shown in Fig  \ref{fig:viz}, the embeddings effectively capture distinctions between political ideologies on \textsc{Covid-Politics}and the \textsc{Election2020} with clear separations between liberal and conservative users. For the \textsc{Ukr-Rus-Suspended dataset}, while there is no global separation between suspended and non-suspended accounts, localized clustering emerges. We also see concentrations of more hateful users against less hateful users in the \textsc{Ukr-Rus-Hate dataset}. These visualizations highlight Social-LLM's ability to encode user attributes and behaviors into an interpretable embedding space, offering insights into the structure and composition of social networks.

\section{Conclusion}
This chapter presents Social-LLM, a scalable method for learning user representations by integrating user content and social network interactions. Our approach combines large language model embeddings of user profiles with a simple yet effective modeling of the social graph structure. Instead of complex high-order proximity, we exploit the inherent sparsity and homophily in social networks by considering only the direct edges between users. Leveraging 7 different large Twitter datasets spanning a diverse range of user detection tasks, we showcase the advantage and robustness of our method. Importantly, once fitted on a social network, Social-LLM can generalize to new users and downstream tasks using only user content features without requiring the original network data. Overall, Social-LLM provides an accurate, scalable, and generalizable framework for characterizing users on social media through their digital footprints.

\bigskip
In the next part of this dissertation, I delve into the qualitative dimensions of this dissertation, focusing on understanding human behavior mining through social network analysis. This part harnesses the capabilities of the Social-LLM method for generating user embeddings, which are then applied in diverse contexts. In Chapter \ref{chp:poli_covid}, I use the embeddings to categorize users by political partisanship to understand the process and impact of political polarization. Chapter \ref{chp:hate} leverages these embeddings as features within a machine learning model that estimates the level of social approval received by hateful users. This approach enables the identification of anomalies where users receive significantly more or less approval than anticipated, offering insights into the mechanisms of social validation in hate speech behavior. In Chapter \ref{chp:moral_covid}, I extend the application of Social-LLM embeddings to datasets enriched with information on users' moral foundations. By employing clustering techniques on these embeddings, the research uncovers distinct communities within the social network characterized by their moral leanings. This exploration contributes to our understanding of the moral underpinnings that influence community formation and interaction online. These varied applications underscore the versatility of Social-LLM embeddings in computational social science and the analysis of online social networks, highlighting their potential to reveal intricate patterns of human behavior and social structuring.
\part{Data-Driven Analysis of Online Human Behavior}
\label{part2}
\chapter{Social Media Polarization and Echo Chambers Surrounding COVID-19}
\label{chp:poli_covid}

\section{Introduction}
As the unprecedented COVID-19 pandemic continues to put millions of people at home in isolation, online communication, especially on social media, is seeing a staggering uptick in engagement \cite{koeze2020virus}. Prior research has shown that COVID-19 has become a highly politicized subject matter, with political preferences linked to beliefs (or disbelief) about the virus \cite{calvillo2020political,uscinski2020people}, support for safe practices \cite{jiang2020political}, and willingness to return to activities \cite{naeim2021effects}. As the United States was simultaneously undergoing one of the largest political events -- the 2020 presidential election -- public health policies may have been undermined by those who disagree politically with health officials and prominent government leaders. As it happens with topics that become politicized, people may fall into \textit{echo chambers} -- the idea that one is only presented with the information they already agree with, thereby reinforcing one's confirmation bias \cite{garrett2009echo,barbera2015tweeting}. 

Social media platforms have been criticized for enhancing political echo chambers and driving political polarization \cite{conover2011political,schmidt2017anatomy, cinelli2020echo}. In part, this is due to a conscious decision made by users when choosing who or what to follow, selectively exposing themselves to contents they already agree with \cite{garrett2009echo}; but this may also be a consequence of the algorithms social media platforms use to attract users \cite{schmidt2017anatomy}. Numerous studies have shown that echo chambers are prevalent on Twitter \cite{conover2011political,an2014sharing,colleoni2014echo,barbera2015tweeting,cossard2020falling}; however, most past works are done on topics that are political in nature.  

In the case of COVID-19, the risks of political polarization and echo chambers can have dire consequences in politicizing this topic that is originally about public health.
The lack of diversity in multi-perspective and evidence-based information can present serious consequences on society by fueling the spread of misinformation \cite{delvicario2016spreading,shu2017fake,motta2020right}. For instance, prior research revealed that conservative users push narratives contradicting public health experts (e.g., anti-mask) and misinformation (e.g., voter fraud) \cite{chen2021covid}. Another research shows that the consumption of conservative media is linked to an increase in conspiracy beliefs \cite{romer2021patterns}. Understanding the degree of polarization and the extent of echo chambers can help policymakers and public health officials effectively relay accurate information and debunk misinformation to the public. 

% \subsection{Research Questions}
In this chapter, we focus on the issue of COVID-19 and present a large-scale empirical analysis of the prevalence of echo chambers and the effect of polarization on social media. Our research is guided by the following research questions surrounding COVID-19 discussions on Twitter:
\begin{itemize}
    \item \textbf{RQ1:} What are the roles of partisan users on social media in spreading COVID-19 information? How polarized are the most influential users? 
    \item \textbf{RQ2:} Do echo chambers exist? And if so, what are the echo chambers, and how do they compare?
\end{itemize}

The technical challenge for addressing these questions is posed by the need to build a scalable and reliable method to estimate user political leanings. To this end, we use the Retweet-BERT model we developed (Chapter \ref{chp:retweetbert}), an end-to-end model that estimates user polarity from their profiles and retweets on a spectrum from left- to right-leaning. Using the estimated polarity scores for all 232,000 Twitter users in our data, we observe and compare the Twitter usage trends of partisan users. Our analyses show that right-leaning users are more vocal in creating original content, more active in broadcasting information (by retweeting), and more impactful through distributing information (by getting retweeted) than their left-leaning counterparts. Moreover, influential users are usually highly partisan, a finding that holds irrespective of the influence measure used.

Finally, we provide evidence that political echo chambers are apparent at both political extremes, though the degrees of cross-ideological interactions are highly asymmetrical: While communication channels remain open between left-leaning and neutral users, right-leaning users are found in a densely-connected political bubble of their own. Information rarely travels in/out of the right-leaning echo chamber.

As our work offers unique insights into the polarization of COVID-19 discussions on Twitter, it carries broader implications for identifying and combating misinformation spread, as well as strengthening the online promotion of public health campaigns. Further, since communication across the two echo chambers functions very differently, we stress that communication effectiveness must be evaluated separately for people in each echo chamber.

\section{Data}
We use a large COVID-19 Twitter dataset collected by \citet{chen2020covid}, containing data from January 21 to July 31, 2020 (v2.7). All tweets collected contain keywords relevant to COVID-19. The tweets can be original tweets, retweets, quoted tweets (retweets with comments), or replies. Each tweet also contains the user's profile description, the number of followers they have, and the user-provided location. Some users are verified, meaning they are authenticated by Twitter in the interest of the public, reducing the chance that they are fake or bot accounts \cite{hentschel2014finding}. All users can optionally fill in their profile descriptions, which can include personal descriptors (e.g., ``\textit{Dog-lover}'', ``\textit{Senator}'', ``\textit{Best-selling author}'') and the political party or activism they support (e.g., ``\textit{Republican}'', ``\textit{\#BLM}''). 

\subsection{Interaction Networks} The retweet network $G_R=(V,E)$ is modeled as a weighted, directed graph. Each user $u\in V$ is a node in the graph, each edge $(u,v)\in E$ indicates that user $u$ has retweeted from user $v$, and the weight of an edge $w(u,v)$ represents the number of retweets. We use the terms retweet interaction and edges of the retweet network interchangeably. Similarly, we construct the mention network $G_M$, where the edges are mentions instead of retweets. A user can be mentioned through retweets, quoted tweets, replies, or otherwise directly mentioned in any tweet.

\subsection{Data Preprocessing}
We restrict our attention to users who are likely in the United States, as determined by their self-provided location \cite{jiang2020political}. Following \citet{garimella2018quantifying}, we only retain edges in the retweet network with weights of at least 2. Since retweets often imply endorsement \cite{boyd2010tweet}, a user retweeting another user more than once would imply stronger endorsement and produce more reliable results. As our analyses depend on user profiles, we remove users with no profile data. We also remove users with degrees less than 10 (in- or out-degrees) in the retweet network, as these are mostly inactive Twitter users. To remove biases from potential bots infiltrating the dataset \cite{ferrara2020types}, we calculate bot scores using \citet{botometer}, which estimates a score from 0 (likely human) to 1 (likely bots), and remove the top 10\% of users by bot scores as suggested by \citet{ferrara2020types}.

Our final dataset contains 232,000 users with 1.4 million retweet interactions among them. The average degree of the retweet network is 6.15. For the same set of users in the mention network, there are 10 million mention interactions, with an average degree of 46.19. Around 18,000, or approximately 8\% of all, users are verified.
%%%

A subset of this dataset is pseudo-labeled and is referred to as the \textsc{Covid-Political} dataset in \S\ref{ref:rtbert_data}. This subset only contains users that we are able to label heuristically as left-leaning or right-leaning from the hashtags and news media URLs they use. This is used to develop a model to estimate user polarity for all users. 

\section{Method}
We use Social-LLM to estimate user political ideology. We first use the Social-LLM model that worked best on the \textsc{Covid-Political} dataset, which is also the Retweet-BERT model in this case, to estimate political ideology. We then train another machine learning model on the user embeddings to estimate user political leaning, using pseudo-labeled seed users as the supervision. Finally, we infer polarity scores for the rest of the users not pseudo-labeled, ranging from 0 (far-left) to 1 (far-right). Since there are more left-leaning seed users in our dataset, the predicted polarity scores are naturally skewed towards 0 (left). Therefore, we bin users by evenly distributed deciles of the polarity scores, with each decile containing exactly 10\% of all users.

\section{Results}
% \label{sec:characterize_partisan}
\subsection{The Roles of Partisan Users}
\label{sec:far_left_right}
We first examine the characteristics of extremely polarized users, defined as the users in the bottom (left-leaning/far-left) or top (right-leaning/far-right) 20\% of the polarity scores. As a point of comparison, we also include neutral users who are in the middle 20\% of the polarity scores. Considering various aspects of user tweeting behaviors, we characterize the Twitter user roles as follows:
\begin{enumerate}
    \item \textit{Information creators}: those who create original content and are usually the source of new information.
    \item \textit{Information braodcasters}: those who foster the distribution of existing content, such as through retweeting other people and promoting the visibility of other's content.
    \item \textit{Information distributors}: those whose contents are likely to be seen by many people, either through passive consumption by their followers or through broadcasting (retweeting) by others.
\end{enumerate}
\begin{figure}[t]
    \centering
    
    \includegraphics[width=\linewidth]{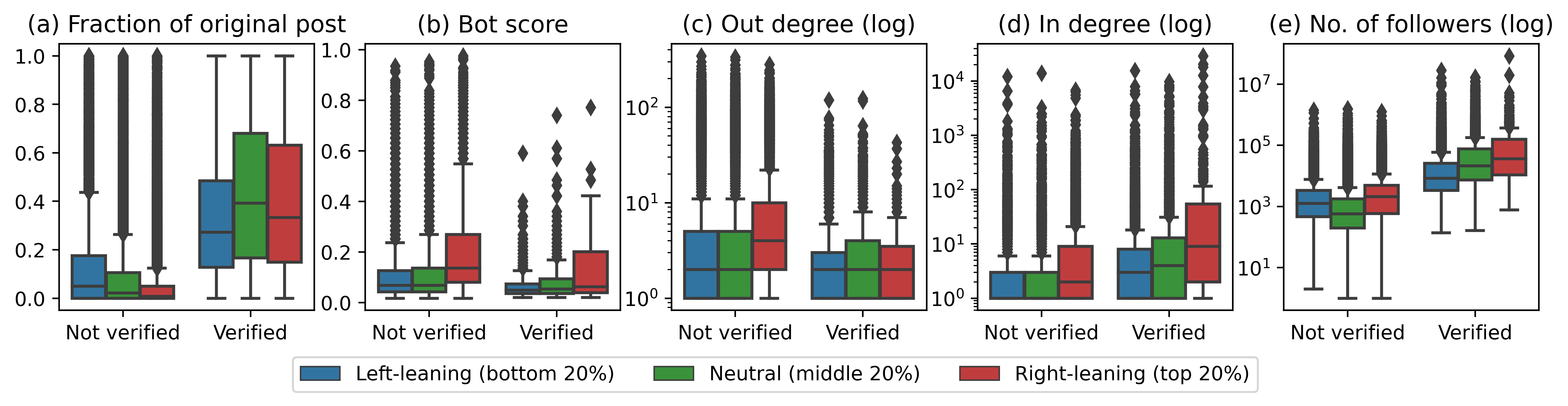}
    \caption{COVID-19 dataset statistics of left-leaning (bottom 20\%), neutral (middle 20\%), and right-leaning (top 20\%) users partitioned by their verification status. The degree distributions are taken from the retweet network. All triplets of distributions (left-leaning, neutral, and right-leaning) are significant using a one-way ANOVA test ($P<0.001$).}
    \label{fig:user_characteristics}
\end{figure}

According to these definitions, a user can be all of these or none of these at the same time. In Figure \ref{fig:user_characteristics}, we plot several Twitter statistics regarding the polarized and neutral users, disaggregated by their verification status. 

Compared to unverified users, verified users are more likely information creators. This is unsurprising, given that verified users can only be verified if they demonstrate they are of public interest and noteworthy. Comparatively, left-leaning verified have the smallest fraction of original posts. However, this is reversed for unverified users, with unverified left-leaning users having the highest fraction of original content and unverified right-leaning users having little to no original content. We note that this may be related to the distribution of bot scores. Figure \ref{fig:user_characteristics}(b) reveals that right-leaning users score significantly higher on the bot scale. Since bots retweet significantly more than normal users \cite{ferrara2016rise}, we cannot rule out the possibility that right-leaning bots are confounding the analysis. However, users scoring the highest on the bot scale have already been removed from the data.

Unverified right-leaning users, in comparison with their left-leaning counterparts, are more likely information broadcasters as they have the highest out-degree distribution (Figure \ref{fig:user_characteristics}(c)). As out-degree measures the number of people a user retweets from, a user with a high out-degree functions critically in information broadcasting. The fact that they also have very little original content (Figure \ref{fig:user_characteristics}(a)) further suggests that unverified right-leaning users primarily retweet from others.

Finally, all right-leaning users function as information distributors regardless of their verification status. Their tweets are much more likely to be shared and consumed by others. Their high in-degree distribution indicates they get retweeted more often (Figure \ref{fig:user_characteristics}(d)), and the higher number of followers they have indicates that their posts are likely seen by more people (Figure \ref{fig:user_characteristics}(e)).

As right-leaning users play larger roles in both the broadcasting and distributing of information, we question if these users form a political echo chamber, wherein right-leaning users retweet frequently from, but only from, users who are also right-leaning. As we will see later in the chapter, we indeed find evidence that right-leaning users form a strong echo chamber.

\begin{figure}[t]
    \centering
    \includegraphics[width=\linewidth]{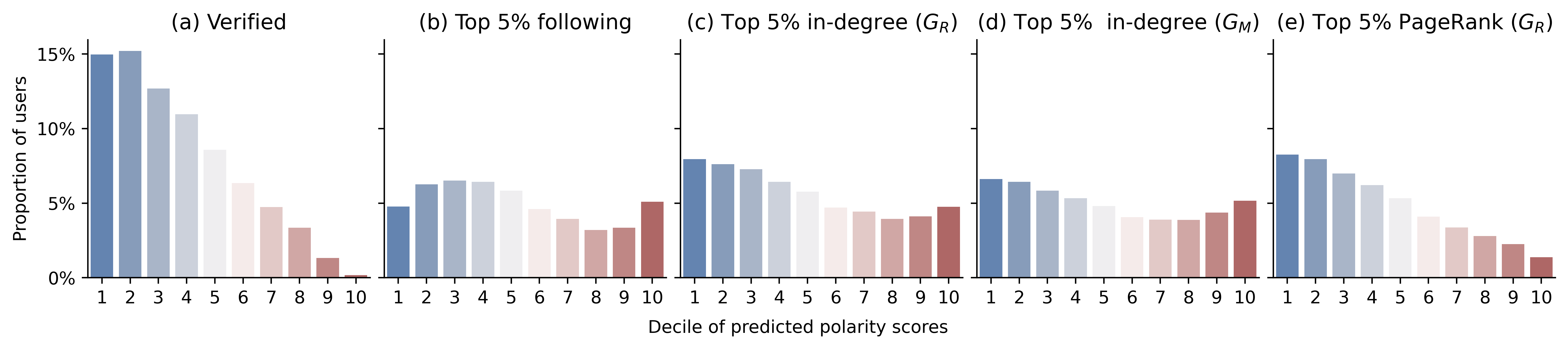}
    \caption{The proportion of users in the COVID-19 dataset in each decile of predicted political bias scores that are (a) verified, (b)  top 5\% in the number of followers, (c) top 5\% of in-degrees in the retweet network (most retweeted by others), (c) top 5\% of in-degrees in the mention network (most mentioned by others), and  (e) top 5\% in PageRank in the retweet network.}.
    \label{fig:proportion_users_poli_decile}
\end{figure}
\subsection{The Polarity of Influencers}\label{sec:polarity_influeners}
The above characterizes the Twitter activities of users who are extremely left or right-biased. However, the majority of the social influence is controlled by a few key individuals \cite{wu2011who,lou2013mining,zhang2015who}. In this section, we consider five measures of social influence: verification status, number of followers, number of retweets, number of mentions, and PageRank in the retweet network \cite{page1999pagerank}. A user is considered influential if they are in the top 5\% of all people according to the measure of influence. Figure \ref{fig:proportion_users_poli_decile} reveals the proportion of users in each decile of polarity score that is influential. We show that, consistent with all of the influence measures above, partisan users are more likely to be found influential.

The verification status is correlated with partisan bias, with the proportion of verified users decreasing linearly as we move from the most left- to the most right-leaning deciles of users (Figure \ref{fig:proportion_users_poli_decile}(a)). 15\% of users in the 1\textsuperscript{st} and 2\textsuperscript{nd} deciles, which are most liberal, are verified, compared to less than 1\% of users in the extremely conservative 10th decile. As verified accounts generally mark the legitimacy and authenticity of the user, the lack of far-right verified accounts opens up the question of whether there is a greater degree of unverified information spreading in the right-leaning community. We stress, however, that our result is cautionary. A closer investigation is needed to establish if there are other politically driven biases, such as a liberal bias from Twitter as a moderating platform, that may contribute to the under-representation of conservative verified users.

While being verified certainly aids visibility and authenticity, users do not need to be verified to be influential. We observe bimodal distributions (U-shaped) in the proportion of users who are influential with respect to their polarity according to three measures of influence: top most followed, retweeted, and mentioned  (Figure \ref{fig:proportion_users_poli_decile}(b)-(d)), indicating that partisan users have more influence in these regards. In particular, far-right users have some of the highest proportion of most-followed users. Far-left users are more likely to be highly retweeted and mentioned, but the far-right also holds considerable influence in those regards.

Lastly, we look at PageRank, a well-known algorithm for measuring node centrality in directed networks \cite{page1999pagerank}. A node with a high PageRank is indicative of high influence and importance. Much like the distribution of verified users, the proportion of users with high PageRank in each polarity decile is correlated with how left-leaning the polarity decile (Figure \ref{fig:proportion_users_poli_decile}(b)), which suggests that left-leaning users hold higher importance and influence. However, this phenomenon may also be an artifact of the much larger left-leaning user base on Twitter.

\begin{figure}
    \centering
    \includegraphics[width=0.75\linewidth]{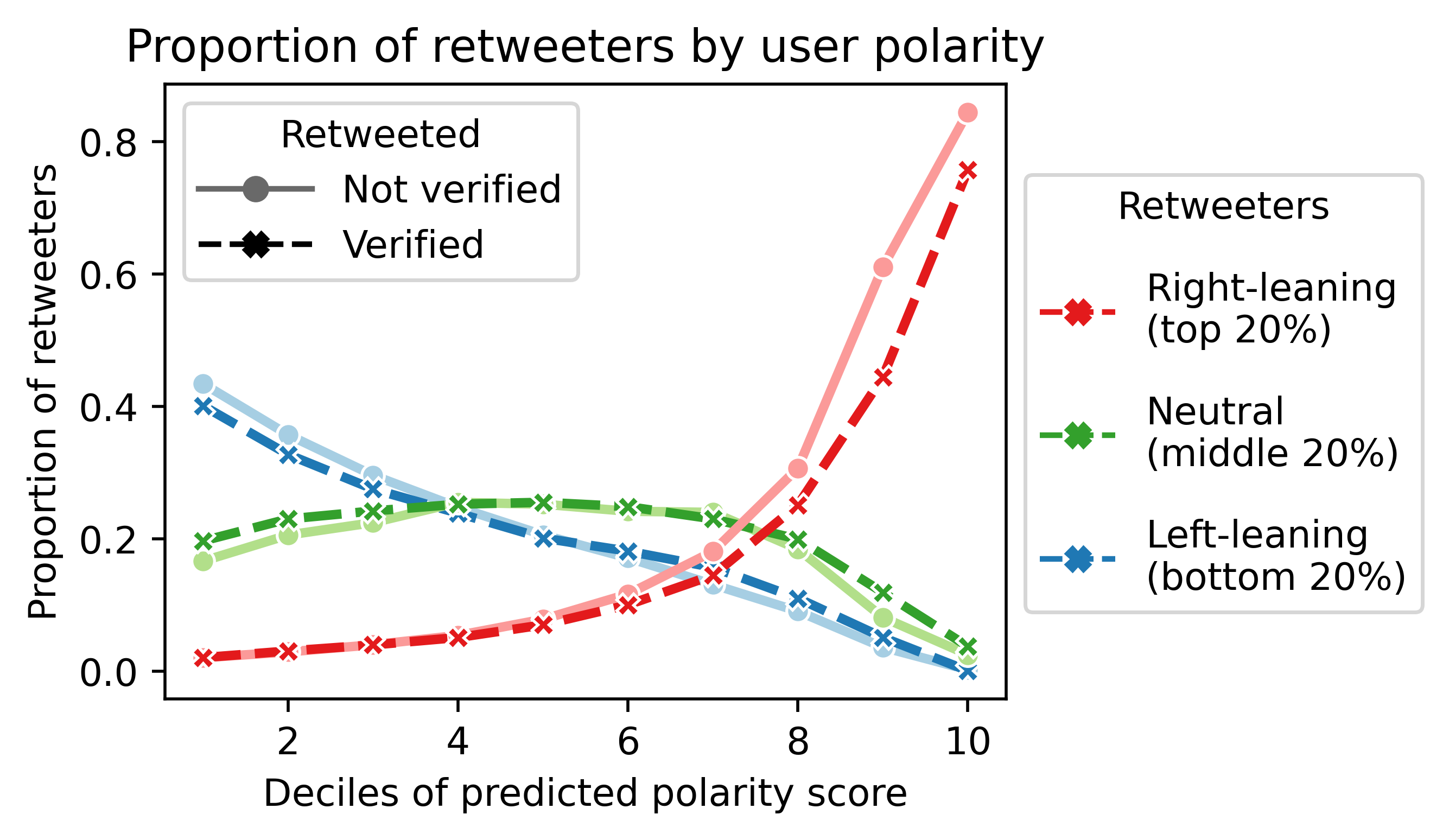}
    \caption{The distribution of left-leaning (bottom 20\% of the polarity scores), center (middle 20\%), and right-leaning (top 20\%) retweeters (y-axis) for users in the COVID-19 dataset across the polarity score deciles (x-axis). The retweeted users are either verified or not verified.}
    \label{fig:decile_retweeters_dist}
\end{figure}

\subsection{Echo Chambers}\label{sec:echo}

% As most influential users are partisan, we question if echo chambers exist how prevalent they are. We first provide evidence of echo chambers (\S\ref{sec:user_audience_polarity}), then examine how cross-ideological information flows between the far-left and far-right (\S\ref{sec:rwc}). Finally, we consider the influence of users who are popular among the left and the right to provide further context on the extent of echo chambers (\S\ref{sec:popular_lr}).
% \subsection{User polarity vs. audience polarity} 

\label{sec:user_audience_polarity} 

As most influential users are partisan, we question if echo chambers exist and how prevalent they are. We begin by exploring the partisan relationship between the retweeted and the retweeter, where the latter is considered as the (immediate) audience of the former. Figure \ref{fig:decile_retweeters_dist} plots the proportion of left-leaning, neutral, or right-leaning retweeters for users in each of the 10 deciles of polarity scores, revealing that users on both ends of the political spectrum reach an audience that primarily agrees with their political stance. In fact, the far-left and far-right users have virtually no retweeters from supporters of the opposite party. However, the echo chamber effect is much more prominent on the far-right. About 80\% of the audience reached by far-right users are also right. In comparison, only 40\% of the audience reached by far-left users are also left. There is little difference in the distribution of retweeters between verified and unverified users.

Since the polarized users are mostly preoccupied in their echo chambers, the politically neutral users (Figure \ref{fig:decile_retweeters_dist}, green) would serve the important function of bridging the echo chambers and allowing for cross-ideological interactions. Most of them (30-40\%) retweet from sources that are also neutral, and around 20\% of them retweet from very liberal sources. When it comes to broadcasting tweets from the far-right, they behave similarly to the far-left retweeters: Almost no neutral users retweet from the far-right. Such observations would imply a much stronger flow of communication between the far-left users and neutral users, whereas the far-right users remain in a political bubble.

\subsection{Random Walk Controversy}
\label{sec:rwc}
Previously, we explored the partisan relationship between users and their immediate audience. To quantify how information is disseminated throughout the Twitter-sphere and its relationship with user polarity, we conduct random walks on the graphs to measure the degree of controversy between any two polarity deciles of users. Our method extends the Random Walk Controversy (RWC) score for two partitions \cite{garimella2018quantifying}, which uses random walks to measure the empirical probability of any node from one polarity decile being exposed to information from another.

A walk begins with a given node and recursively visits a random out-neighbor of the node. It terminates when the maximum walk length is reached or if a node previously seen on the walk is revisited. Following \citet{garimella2018quantifying}, we also halt the walk if we reach an authoritative node, which we define as the top 1000 nodes ($\approx 4\%$) with the highest in-degree in any polarity decile. By stopping at nodes with high in-degrees, we can capture how likely a node from one polarity decile would receive highly endorsed and well-established information from another polarity decile. To quantify the controversy, we measure the RWC from polarity decile $A$ to $B$ by estimating the empirical probability 
\begin{equation}
    \label{eq:rwc}
    \text{RWC}(A, B) = Pr(\text{start in }A |\text{end in }B).
\end{equation}
The probability of walks starting in a partiion is conditional on the walks ending in a given partition to control for varying distribution of high-degree vertices in each polarity decile. RWC yields a probability, with a high RWC$(A,B)$ implying that random walks landing in $B$ started from $A$. Compared to the original work \cite{garimella2018quantifying}, we simplify the definition of RWC as we do not need to consider the varying number of users in each echo chamber.

 \begin{figure}
    \centering
    \includegraphics[width=0.75\linewidth]{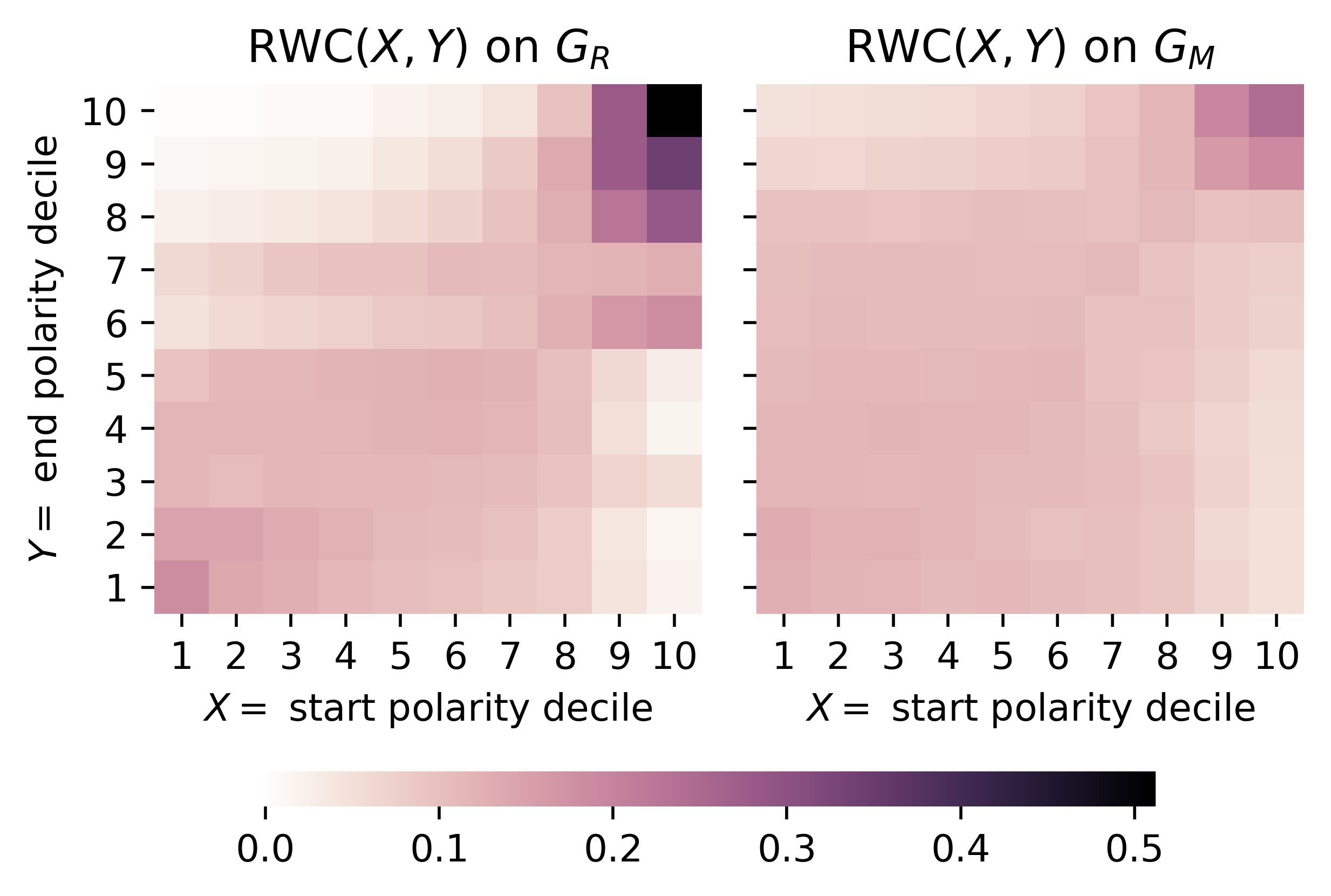}
    \caption{The RWC$(X, Y)$ for every pair of polarity deciles $X$ and $Y$ on the retweet (left) and mention (right) networks using Eq. \ref{eq:rwc}.}
    \label{fig:random_walk}
\end{figure}

We initiate the random walks 10,000 times randomly in each polarity decile for a maximum walk length of 10. The RWC between any two polarity deciles for the retweet and mention networks are visualized in Figure \ref{fig:random_walk}. For both networks, the RWC scores are higher along the diagonal, indicating that random walks most likely terminate close to where they originated. Moreover, the intensities of the heatmap visualizations confirm that there are two separate echo chambers. The right-leaning echo chamber (top right corner) is much denser and smaller than the left-leaning echo chamber (bottom left corner). Any walk in the retweet network that originates in polarity deciles 9 and 10 will terminate in polarity deciles 8 to 10 about 80\% of the time. In contrast, walks that start in deciles 1--7 have a near equal, but overall much smaller, probability of landing in deciles 1--7. In essence, users who are right-leaning form a smaller but stronger echo chamber, while other users form a larger and more distributed echo chamber.

The RWC scores on the mention network confirm the presence of the two echo chambers, but the intensities are reduced. Compared to random walks on the retweet network, those on the mention network are much more likely to end far away. As a result, while there are rarely any cross-ideological retweet interactions, there exists a greater degree of direct communication through mentions, likely done to speak to or criticize the opposing side \cite{conover2011political}. We note that because the RWC scores appear highly symmetrical about the diagonals, there is little difference in the cross-ideological interaction between opposite directions of communication flow.

\begin{figure}
    \centering
    \includegraphics[width=\linewidth]{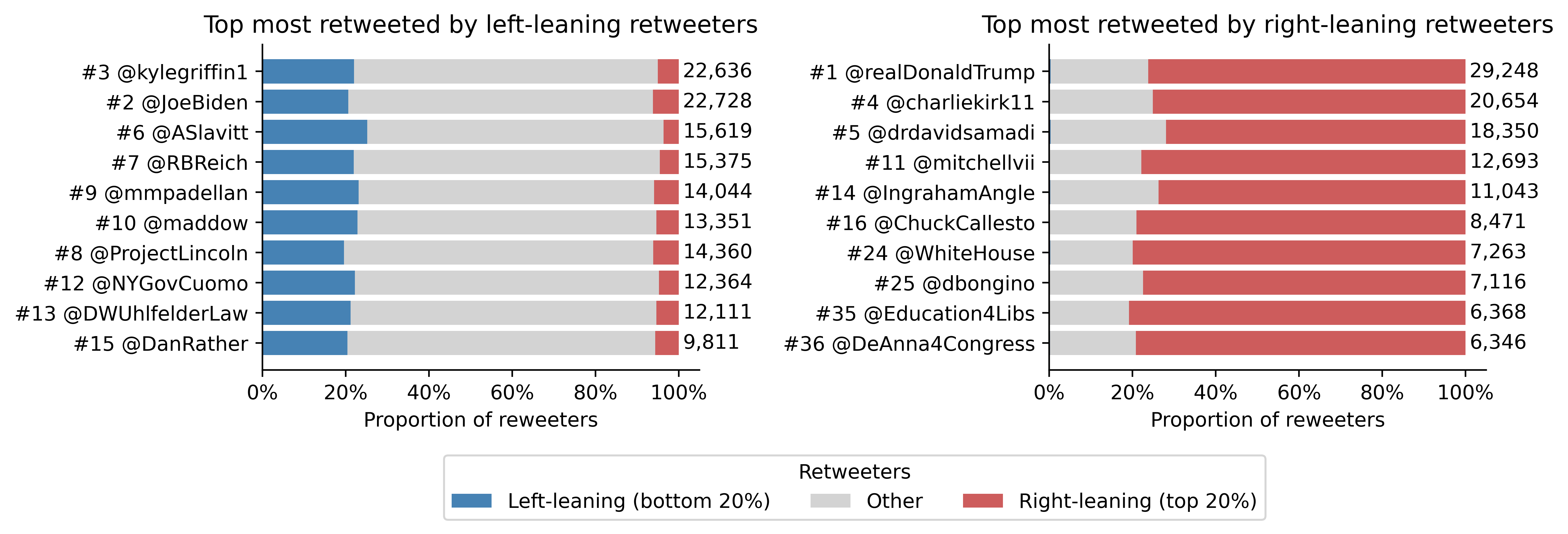}
    \caption{Users in the COVID-19 dataset with the highest number of retweeters from left- and right-leaning users. The bar plots show the distribution of their unique retweeters by political leaning. Users are also ranked by their total number of retweeters (\textit{i.e.}, \#1 @realDonaldTrump means that @realDonaldTrump has the most retweeters). Numbers appended to the end of the bars show their total number of retweeters.}
    \label{fig:high_in_degree}
\end{figure}

\subsection{Popular Users Among the Left and Right}\label{sec:popular_lr}
Retweeting is the best indication of active endorsement \cite{boyd2010tweet} and is commonly used as the best proxy for gauging popularity and virality on Twitter \cite{cha2010measuring}. Figure \ref{fig:high_in_degree} shows the users who are the most popular users among the left and the right according to the number of left- or right-leaning retweeters they have. 

Analyzing the identities of the top-most retweeted users by partisans gives us the first hint at the presence of political echo chambers. There is no overlap between the most retweeted users by the left- and the right-leaning audience, and they tend to be politically aligned with the polarization of their audience. Almost all users who are most retweeted by left-leaning users are Democratic politicians, liberal-leaning pundits, or journalists working for left-leaning media. Notably, @ProjectLincoln is a political action committee formed by Republicans to prevent the re-election of the Republican incumbent President Trump. Similarly, almost all users who are most retweeted by right-leaning users are Republican politicians, right-leaning pundits, or journalists working for right-leaning media. Despite its username, @Education4Libs is a far-right account promoting QAnon, a far-right conspiracy group. As of January 2021, @Education4Libs has already been banned by Twitter.

These popular users are not only popular among the partisan users but are considerably popular overall, as indicated by the high overall rankings by the number of total retweeters. With a few exceptions, users who are popular among the left are more popular among the general public than users who are popular among the right. 

The distribution of the polarity of retweeters of these most popular users reveals another striking observation: The most popular users among the far-right rarely reach an audience that is not also right, whereas those of the far-left reach a much wider audience in terms of polarity. Users who are popular among the far-left hail the majority of their audience from non-partisan users (around 75\%)  and, importantly, draw a sizable proportion of far-right audience (around 5\%). In contrast, users who are popular among the far-right have an audience made up almost exclusively of the far-right (around 80\%) and amass only a negligible amount of far-left audience.

\section{Discussion}
In this chapter, we studied the extent of echo chambers and political polarization in COVID-19 conversations on Twitter in the US. Using Social-LLM--more specifically, the earlier version named Retweet-BERT--a model that leverages user profile descriptions and retweet interactions to effectively and accurately measure the degree and direction of polarization, we provided insightful characterizations of partisan users and the echo chambers in the Twitter-sphere to address our research questions.

\paragraph{RQ1.} \textit{What are the roles of partisan users on social media in spreading COVID-19 information? How polarized are the most influential users? } From characterizing partisan users, we find that right-leaning users stand out as being more vocal, more active, and more impactful than their left-leaning counterparts.

Our finding that many influential users are partisan suggests that online prominence is linked with partisanship. This result is in line with previous literature on the ``price of bipartisanship,'' which is that bipartisan users must forgo their online influence if they expose information from both sides \cite{garimella2018political}. In another simulated study, \citet{garibay2019polarization} shows that polarization can allow influential users to maintain their influence. Consequently, an important implication is that users may be incentivized to capitalize on their partisanship to maintain or increase their online popularity, thereby further driving polarization. Information distributed by highly polarized yet influential users can reinforce political predispositions that already exist, and any polarized misinformation spread by influencers risks being amplified.

\paragraph{RQ2.} \textit{Do echo chambers exist? And if so, what are the echo chambers, and how do they compare?} Though COVID-19 is a matter of public health, we discover strong evidence of political echo chambers on this topic on both ends of the political spectrum, particularly within the right-leaning community.
% (similar to the conclusion of \citet{barbera2015tweeting}).
Right-leaning users are almost exclusively retweeted by users who are also right-leaning, whereas the left-leaning and neutral users have a more proportionate distribution of retweeter polarity. From random walk simulations, we find that information rarely travels in or out of the right-leaning echo chamber, forming a small yet intense political bubble. In contrast, far-left and non-partisan users are much more receptive to information from each other. Comparing users who are popular among the far-left and the far-right,  we reveal that users who are popular among the right are \textit{only} popular among the right, whereas users who are popular among the left are also popular among all users.

\subsection{Implications} 

Despite Twitter's laudable recent efforts in fighting misinformation and promoting fact-checking \cite{fowler2020twitter}, we shed light on the fact that communication is not just falsely manipulated but also hindered by communication bubbles segregated by partisanship. It is imperative that we not only dispute misinformation but also relay true information to all users. As we have shown, outside information is extremely difficult to get through to the right-leaning echo chamber, which could present unique challenges for public figures and health officials outside this echo chamber to effectively communicate information. Existing research suggests that right-leaning users are more susceptible to anti-science narratives, misinformation, and conspiracy theories \cite{calvillo2020political,uscinski2020people,romer2021patterns,chen2021covid}, which, given the echo chambers they are situated in, can worsen with time. Our work has implications for helping officials develop public health campaigns, encourage safe practices, and combat vaccine hesitancy effectively for different partisan audiences.

\subsection{Future Direction} 
Though the question of whether social media platforms \textit{should} moderate polarization is debated, we note that \textit{how} they can do so remains an open problem. It is unclear how much of the current polarization is attributed to users' selective exposure versus the platform's recommendation algorithm. Moreover, whether users are even aware that they are in an echo chamber and how many conscious decisions are being made by the users to combat that remains to be studied in future work.

Another future avenue of research could focus on studying how misinformation travels in different echo chambers. Since our study highlights that there is an alarmingly small number of far-right verified users, and given that verified users are typically believed to share legitimate and authentic information, further research is required to establish if the right-leaning echo chamber is at greater risk of being exposed to false information from unverified users. Detailed content analysis on the tweets can reveal if there are significant disparities in the narratives shared by left- and right-leaning users. Crucially, our work provides a basis for more in-depth analyses of how and what kind of misinformation is spread in both echo chambers. 

\subsection{Limitations} There are several limitations regarding this work. First, we cannot exclude any data bias. The list of keywords was manually constructed, and the tweets collected are only a sample of all possible Tweets containing these keywords. Since the data was collected based on keywords strictly related to COVID-19, we only gathered data that were relevant to the virus and not tainted by political commentary. Therefore, the data provides us with a natural setting to study the polarization of COVID-19 discourse on Twitter. 

Second, our study hinges on the fact that retweets imply endorsement, which may be an oversimplification. To reduce noisy, isolated retweet interactions, we consider only retweets that have occurred at least twice between any two users. 

Finally, our political detection model is built on weakly-supervised labelings of users using politically-relevant hashtags and the polarization of news media as the sources of ground truth. We took a conservative approach and only seeded users who explicitly use politicized hashtags in their profile or have repeatedly interacted with polarized new sources.

\chapter{Social Approval and Network Homophily as Motivators of Online Toxicity}
\label{chp:hate}

\section{Introduction}

The proliferation of hate messages in social media–commonly understood as expressions of hatred, discrimination, or attacks towards individuals or groups based on identity attributes such as race, gender, sex, religion, ethnicity, citizenship, or nationality \cite{tsesis2002destructive}--has garnered considerable research attention over recent years \cite{paz2020hate,ezeibe2021hate,thomas2021sok,frimer2023incivility}. Online hate has important connections to cyberbullying \cite{chen2011detecting} and online harassment \cite{thomas2021sok} and, in extreme cases, to the incitement of violence and offline hate crimes \cite{castano2021internet,ezeibe2021hate,muller2021fanning,wang2023identifying}. Research finds that young adults, LGBTQ+ minorities, and active social media users are especially vulnerable to online hate and harassment \cite{keipi2016online,thomas2021sok}. Moreover, evidence suggests the pervasiveness of online hate is growing increasingly \cite{mathew2020hate,thomas2021sok,frimer2023incivility}, which emphasizes the urgency to address problems of online hate and toxicity.
% \footnote{In this paper, we use the terms ``online hate'' and ``toxicity'' interchangeably.} 

This research provides an initial empirical test of a new theory focusing on the motivations and gratifications associated with posting hate messages online. It posits that online hate is fueled by the social approval that hate message producers receive from others \cite{walther2022social}. The theory suggests that online hate behavior is not primarily motivated by the desire to harm prospective victims but rather to accrue validation and encouragement from like-minded others. Within this framework, one’s propensity to express hateful messages should be related to a similar propensity among one’s social network; people who share similar resentments and actions should be linked, which can facilitate mutual reinforcement of one other’s hate behavior. Further, their hate messaging would be expected to become more extreme as they obtain more reinforcement through social approval signals from others, potentially in the form of likes, upvotes, or other forms of positive feedback. To test this theory, our research asks the following questions:

\begin{itemize}
    \item \textbf{RQ1}: Is a user’s hatefulness related to how hateful their social network is?
    \item \textbf{RQ2}: Does receiving more social approval increase a user’s subsequent hateful behavior?
\end{itemize}

The contribution of this work is two-fold. First, we investigate whether online hateful behavior conforms to patterns of social network homophily--the idea that people who share similar characteristics, interests, or behaviors are also frequently associated with one another--is a phenomenon repeatedly observed in many settings \cite{mcpherson2001birds,kossinets2009origins}. Earlier research suggests that the expression of hate is also a homophilous trait \cite{nagar2022homophily}, which we corroborate in this work. Second, we show that social approvals are linked to increased toxicity, whereas insufficient approvals are linked to decreased toxicity.  Our analysis suggests that hate speech is ``networked'': the expressions of hate, hostility, or extremism are not just isolated incidents by individual users but are influenced, amplified, and sustained through the dynamics and structures of social networks. The insights from this research carry important theoretical implications to advance our understanding of how social gratifications affect the propagation of online hate and suggest alternative strategies to deter it.

\section{Related Work}
\subsection{Homophily in Toxic Behavior}

Social network homophily--the idea that people who share similar characteristics, interests, or behaviors are also frequently associated with one another--is a phenomenon repeatedly frequently observed in many settings \cite{mcpherson2001birds,kossinets2009origins}. Some suggest that the expression of hate is also a homophilous trait \cite{nagar2022homophily}. In a comparison study of hateful and non-hateful Twitter users, prior work has shown that hateful users have higher network centrality \cite{ribeiro2017like}. Additionally, semantic, syntactic, stylometric, and topical similarities exist among hateful users connected in a Twitter network. Another work on COVID-19 discourse on YouTube highlights that YouTube commentators are segregated by toxicity levels \cite{obadimu2021developing}. Relatedly, a study on moral homophily found that moral convergence in a cluster of users in an extremist social network predicts how often they spread hateful messages \cite{atari2022morally}. However, these works do not elucidate the extent to which social network cues facilitate detecting toxic users. Drawing upon notions of social network homophily, we explore whether we can utilize them to predict user toxicity labels from a smaller, more limited training set, which would suggest that homophily plays an important role in easily locating toxic users.

\subsection{Social Motivators of Toxic Behavior}

There is a significant gap in research about the influence of social media messages on toxicity. While several empirical studies report that tweets that are more uncivil, toxic, or otherwise outrageous generate more signals of social approval in the forms of likes and retweets \cite{brady2017emotion,kim2021distorting,frimer2023incivility}, scant research has examined the opposite: whether the reception of social approval to one’s toxic messages encourages yet greater incivility and toxicity. This is the core proposition of the social approval theory of online hate: The audience for an author’s hate messages (that appear nominally focused on some target minority) is like-minded online peers and friends, whose signals of approval reinforce and encourage more extreme hatred in an author’s subsequent messages. One possible illumination of this dynamic is \citet{frimer2023incivility}'s examination of 11 years of tweets by US Congress members. Results demonstrated increasing incivility among US politicians on Twitter. Moreover, they found that politicians responded to likes and retweets for uncivil tweets by escalating their toxicity going forward (\citeauthor{frimer2023incivility}, \citeyear{frimer2023incivility}; see also \citeauthor{brady2021social}, \citeyear{brady2017emotion}; \citeauthor{shmargad2022social}, \citeyear{shmargad2022social}). However, whether these findings generalize to normal users is unclear since politicians may be more pressured to respond to constituents’ approval than normal users. The present work helps to bridge this gap by examining the effects of social feedback on a more representative sample of ``average'' yet hate-generating Twitter users.

\section{Data}
For this research, we use a Twitter dataset of hateful users. Twitter is a platform where users can share tweets and potentially receive engagement and feedback from others, for instance, in the form of likes and replies. We use the \textsc{Immigration-Hate} dataset that is also described in \S\ref{sec:sllm_data_hate}. Our dataset is based on a dataset collected by \cite{bianchi2022njh}, which collected tweets from 2020-2021 referencing US and UK anti-immigration terms. These tweets are annotated for four sub-types of incivility (profanities, insults, outrage, and character assassination) and two sub-types of intolerance (discrimination and hostility). In their dataset, 18,803 tweets were annotated as uncivil, intolerant, or both. Since only the tweet IDs were provided, we recreated the dataset using our own API credentials in January 2023. We fetched 8,790 tweets (47\% of the original dataset) produced by 7,566 unique users. The rest of the tweets were unavailable for many possible reasons (e.g., a tweet was deleted, the user set their visibility to private, or the account was suspended). Using these 7,566 users who are known to have posted hateful tweets, we collected our dataset of these hateful users’ most recent tweets, up to 3,200 tweets per user (the maximum allowed by Twitter’s historical API). Of the 21 million tweets we collected, 2.9 million of them are original tweets (i.e., not retweets, quotes, or replies), which is the main focus of this research. Note that due to our data collection setup, our dataset contains \textit{only} users who expressed hate at some point during their Twitter tenure, which appropriately limits the scope of this research yet limits it from generalizing to other users.

For each tweet, our data includes the tweet text and its engagement metrics: like counts, retweet counts, quote counts, and reply counts. Due to data API limitations, we cannot systematically collect the identity of users who engaged with a tweet (e.g., which user liked a post), nor can we collect the original quote or reply messages. We additionally collect each user’s profile description and metadata, including their account age, verification status, follower count, following count, statuses count (the number of total tweets they shared), and listed count (the number of public lists of which the user is a member).

% \begin{figure}[t]
%     \centering
%     \includegraphics[width=0.6\linewidth]{figures/hate/bot_score_dist.png}
%     % \captionsetup{justification=centering}
%     \caption{Distribution of user bot scores.}
%     \label{fig:bot_dist}
% \end{figure}

\subsection{Data Preprocessing}
As described in \S\ref{sec:sllm_data_hate}, this dataset may be heavily infiltrated by inauthentic bot accounts. Therefore, for the rest of this study, we replicate all analyses at two thresholds of bot elimination: users with bot scores $<= 0.8$, which is a conservative choice given the peak in the bot score distribution (see Figure \ref{fig:hate_bot_dist}), and users with bot scores $<= 0.5$. We also remove outlier users, those whose follower counts or engagement metrics exceed three standard deviations from the mean after transformation. We further remove outlier users with transformed follower counts or engagement metric counts that exceed three standard deviations from the mean for each level of bot elimination. The follower count is log-transformed to remove skewness, with 1 added to all counts to avoid 0s when taking the log. The engagement metrics are also log-transformed and further normalized by the log of the follower count to make them comparable across users with varying numbers of followers. Following \citet{frimer2023incivility}, we also replace all 0s in the engagement metrics with 0.1. See Appendix \ref{appendix:6} for the statistics of these metrics.

After eliminating users with bot scores greater than 0.8 and other statistical outliers, 6,665 users, 2.5 million original tweets, and 19 million total tweets remain. After eliminating users with bot scores greater than 0.5 and other statistical outliers, 2,985 users, 1 million original tweets, and 8.7 million total tweets remain. In both cases, between 1-2\% of users are verified.

Using all 21 million tweets in our dataset, we compile two social networks: a retweet network and a mention network. These networks are compiled using all 21 million tweets. In these networks, users are nodes, and edges represent either a retweet or mention interaction. Mentioning refers to all acts of mentioning (using `@') that are \textit{not} a retweet, which could include quoting, replying, or otherwise referencing the user in a tweet. We disambiguate retweets from mentions because retweeting is usually considered a form of endorsement \cite{boyd2010tweet,metaxas2015retweets} while mentioning could be used to criticize publicly \cite{hemsley2018tweeting}. Each edge also comes with an edge weight, which is equal to the frequency of the retweet or mention between the two users. Details of the networks can be found in Table \ref{tab:toxicity_homophily}.

\begin{figure}[t]
    \centering
    \includegraphics[width=0.7\linewidth]{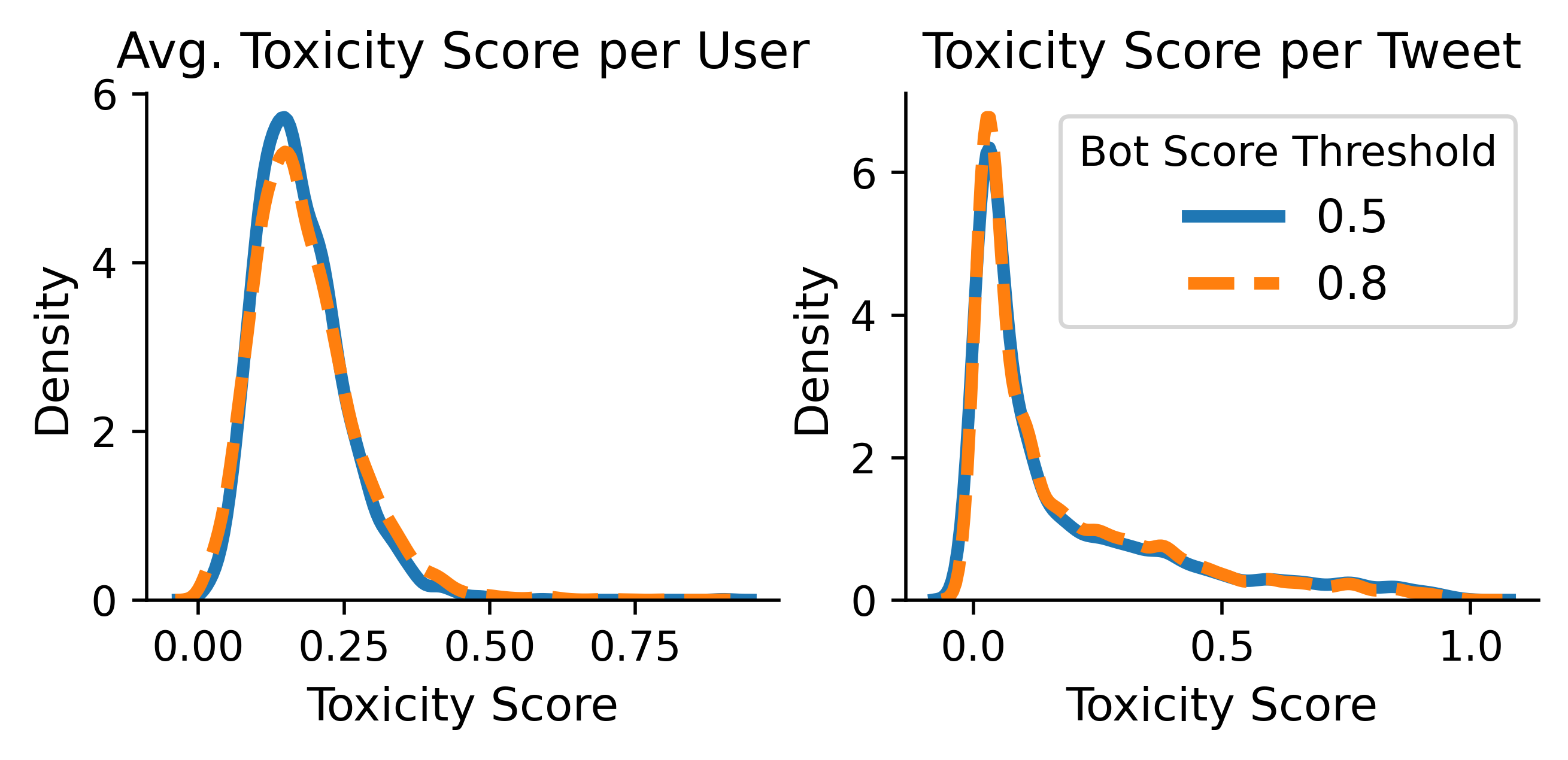}
    \caption{Distribution of the hate scores per user (average hate score of their original tweets) and per original tweet in the hate speech dataset.}
    
    \label{fig:hate_score_dist}
\end{figure}
\section{Methods}
\subsection{Measuring Toxicity} 
To detect toxicity, we apply the Perspective API\footnote{\url{https://perspectiveapi.com/}} on every original tweet. The Perspective API is a popular hate detector used in many similar studies \cite{kim2021distorting,frimer2023incivility}. Using its flagship \texttt{Toxicity} score, we compute a hate score per tweet on a scale of 0 (not toxic) to 1 (very toxic). Additionally, we compute a single hate score for each user which is the average of the score scores of all their tweets. Figure \ref{fig:hate_score_dist} shows the distributions of the tweet-level and user-level score scores.

\subsection{Measuring Network Homophily}

For the first research objective of determining whether toxic behavior is homophilous, we use the network assortativity method, commonly used to measure how related the edges of a network are in terms of some node attribute \cite{newman2003mixing}, which, in this case, is the average hate score of a user. It is computed as the Pearson correlation of the average hate scores of every pair of users connected by a retweet or mention edges, ranging from $-1$, which indicates that users are preferentially connected to users with the opposite toxicity (disassortative) to $1$, which indicates that users are preferentially connected to other users with similar hate scores (assortative). However, network assortativity does not take into account edge weights, and it neglects to consider that a user’s social network usually consists of multiple connections. Therefore, we also compute the correlation between each user’s hate score and the weighted average toxicity of all of their neighbors using the edge weights in the calculation of that weighted average.

\subsection{Measuring the Effects of Social Engagement on Toxicity }

The second research objective aims to understand how others’ social engagement with one’s hate message, in the form of likes, retweets, replies, and quotes, affects the hatefulness of one’s subsequent posts. An overly simple solution to this problem could be to binarize the likes, perhaps exploring whether \textit{any} likes or receiving more likes than some $x$ amount would impact toxicity. However, this approach is unsuitable considering that some users in our dataset receive very few likes on their tweets while others are accustomed to receiving hundreds of likes. One potential solution is to scale the social engagement metrics based on users’ follower counts, which can be a reasonable proxy for their popularity. However, we also need to adjust carefully for the varying impact of that social engagement. For instance, an additional ``like'' may carry significantly greater weight for a less popular user than a more popular one.

To address the problems, we categorize tweets into ones receiving ``high'' or ``low'' (or neither) amounts of social engagement by capturing the deviation of actual social engagement from the level of social engagement a user may have expected. We approximate the expected social engagement using the predictive model described below. We then analyze the change in a user's toxicity if the user expected \textit{much less} or \textit{much more} than expected social engagement. 

To facilitate the computational modeling of users, we use several forms of user representations that are based mainly on language features. The first is the \textit{Social-LLM} user representation model introduced in Chapter \ref{chp:socialllm}, which is a social network representation learning method. The model we use is trained with both retweet and mention edges, as well as users' profile descriptions and metadata features. Social-LLM begins with an LLM representation of the users' profile descriptions and continues to minimize the distance between two users' representations if the users are connected by a retweet or mention interaction. This model, as demonstrated in Chapter \ref{chp:socialllm}, can be very useful in detecting the hatefulness of users by combining social network cues with social media content cues. The low-dimensional Social-LLM user embeddings are useful encapsulations of the user's social network features. In addition, we also use the LLM embedding of the user's profile descriptions (Profile LLM) as well as the LLM of the tweet text (Tweet LLM). Following Chapter \ref{chp:socialllm}, we use SBERT-MPNet (\texttt{sentence-transformers/allmpnet-base-v2}) as the LLM for all embeddings.

To calculate the expected social engagement, we train a deep neural network to predict likes, retweets, etc., for every tweet in our dataset based on the following features:

\begin{itemize}
    % best 0.5 model: rt + mn, undirected unweighted, + profile and metadata
    % best 0.8 model: rt + mn weighted, undirected, + profile and metadata
    \item Social-LLM user embeddings (Chapter \ref{chp:socialllm}) containing social network and social media content cues.
    \item Profile LLM embedding of the user's profile description 
    \item Tweet LLM embedding
    \item User metadata (follower count, verified, etc.)
    \item Hate score of the tweet
    \item Average hate scores of the past 50 tweets
    \item Average social engagement metrics of the past 50 tweets
    \item Other social engagement metrics this tweet had that are \textit{not} the one being predicted (e.g., using retweets, quotes, and replies to predict likes)
\end{itemize}

% The first set of features, the user embeddings, is particularly important as it is a low-dimensional representation of each user in relation to their social network. The social network is important since, as we will show in \S\ref{sec:rq1}, users' toxic behaviors are heavily linked with their social network. We refer readers to the Appendix for model architecture and evaluation details.

With a model to estimate expected social engagement, we can then calculate the difference between the actual versus expected amount of social engagement. Since we are interested in when a user receives substantially less or more engagement than they may expect, we standardize this difference and look at instances where the z-score of the difference is smaller than $-2$ (less than expected) or more than $2$ (more than expected). We choose $|2|$ because z-scores less than 2 may represent a fluctuation in social engagement that would not be perceived by a user as other than a normal deviation within expectations. In contrast, a higher z-score threshold, such as $|3|$, would result in statistical outliers and also yield too few instances for a meaningful comparison. 

These tweets that experienced a dramatically unexpected amount of social engagement would be referred to as \textit{anchor} tweets. We then compare how the toxicity levels of a user change after the anchor tweet, using varying temporal windows of 30, 50, or 80 tweets before and after the anchor tweet. Here, we operate under the assumption that an unexpected amount of social engagement may alter a user's behavior. As our social engagement data is not timestamped, it is possible that a tweet may have received likes, retweets, etc., after the user has posted the next 30-80 tweets, but we believe this is unlikely. Our assumption is that the majority of social engagement occurred well before users posted an additional 30-80 tweets.
\section{Results}
\begin{table*}[t]
    \centering
    % \footnotesize
    
    \caption{Hate scores among users exhibit homophily in the hate speech dataset social network, as indicated by both the network assortativity and the Pearson correlation between a user's hate score and the weighted average of their neighbors' (***$p<0.001$) .}
    \label{tab:toxicity_homophily}
    \begin{tabular}{ccrrrr}
    \toprule
    \textbf{Network} &  \textbf{Bot Score}  & \textbf{\# Nodes} & \textbf{\# Edges} & \textbf{Assort.} & \textbf{Wgt. Avg. Corr.}\\
    \midrule
      Retweet & $<=0.8$ &  6,665 & 74,943 &  0.071*** &  0.310*** \\
      Retweet & $<=0.5$ &  2,985 &   9,016 &  0.195*** &  0.445*** \\
      Mention & $<=0.8$ &  6,665 & 104,802 &  0.057*** &  0.244*** \\
      Mention & $<=0.5$ &  2,985 &  14,182 &  0.185*** &  0.421*** \\
    \bottomrule
    \end{tabular}
\end{table*}
\subsection{Homophily in Toxic Behavior}\label{sec:rq1}
To investigate whether toxicity is a homophilous behavior, we compute network assortativity and weighted average correlation metrics based on the social network and the user hate scores. Both metrics use the Pearson correlation method. The results shown in Table \ref{tab:toxicity_homophily} demonstrate that the hate scores exhibit homophily in both the retweet and mention networks. We isolate users by two bot score thresholds--where the higher the bot score, the more likely the user is a bot--to eliminate the impact of bots. Both the network assortativity and the weighted average correlation of a user’s hate score and their neighbors are significantly positive.
% In particular, the weighted average correlation, which includes a weighted average calculation of a user's neighbors, is much greater, indicating the strength of connectivity among all neighbors is more correlated with a user's own toxicity. 
To ensure the validity of these findings, we perform a robustness check by comparing our results with those obtained from a null model. In the null model, we randomly shuffle the nodes on one end of the edges so that each actual edge $A\leftarrow B$ is now $A\leftarrow C$, where $C$ is a random node. This procedure yields nonsignificant Pearson correlations, all with absolute values less than 0.02 in every case, suggesting that the social network reflects homophily in toxicity. Interestingly, homophily is more evident when we consider only users with lower bot scores. One possible reason is that users who are more likely to be genuine users rather than social bots tend to create more homophilous connections (see Appendix \ref{appendix:6} for further analysis).

% The presence of homophily can be further corroborated by the work of \cite{jiang2023social}, which proposed a general-purpose user detection modeling framework \textit{Social-LLM} based on social network data. They validated their model on the same dataset used in this paper for user toxicity detection. They found that using the social network (e.g., who retweets who) combined with other simple user metadata features such as profile descriptions can predict users' hate scores much more accurately than other methods that do not utilize the network features. It additionally provides a set of user representation embeddings that can be used for any downstream tasks, which we will use in the following section to study the effect of social engagement on toxicity.

\begin{figure}[t]
    \centering
    \includegraphics[width=0.5\linewidth]{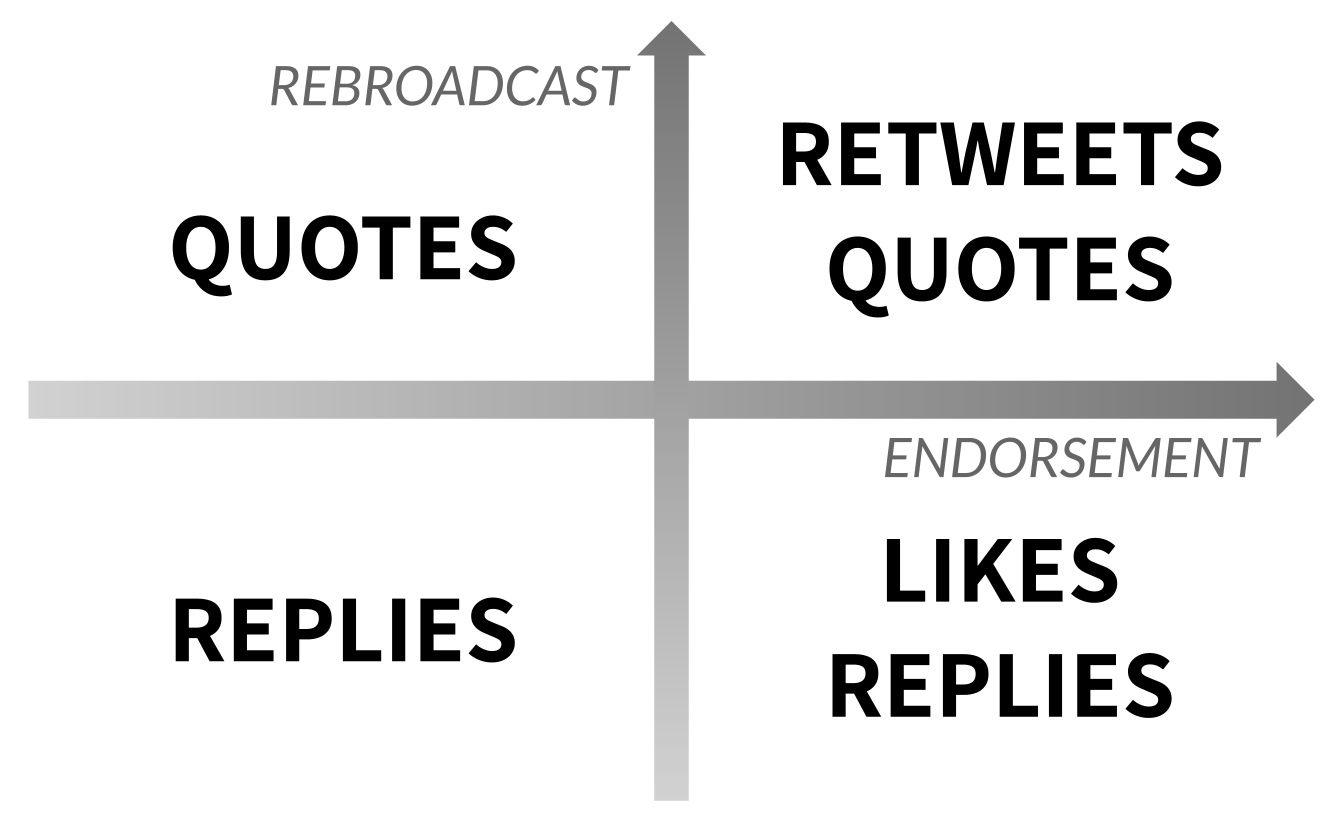}
    \caption{The four types of social engagement on dimensions of rebroadcast and endorsement. \textit{Retweets} represent rebroadcast and endorsement,  \textit{likes} represent endorsements, \textit{quotes} are rebroadcasts that can be either positive or negative, and \textit{replies} do not rebroadcast and can be either positive or negative.}
    \label{fig:social_engagement_diagram}
\end{figure}
 
\subsection{The Effect of Social Engagement on Toxicity} \label{sec:rq2}
Next, we analyze how social engagement may affect a user's propensities toward toxicity. Before presenting the results, let us review the four types of social engagement signals and what they potentially implicate. On Twitter, one user may engage with another user's post by liking, retweeting, quoting (retweeting with additional comments), or replying (commenting). As illustrated in Figure \ref{fig:social_engagement_diagram}, we can position each form of social engagement jointly on the dimensions of rebroadcast and endorsement. By retweeting, one rebroadcasts a tweet to one’s own followers on Twitter. Retweeting may connote exceptional social approval, as a retweet not only signifies endorsement but the flattering reflection of another user’s desire for the original message to be seen by their own friends and followers \cite{boyd2010tweet,metaxas2015retweets}. Liking a tweet also signifies endorsement but not the flattery implied by re-transmission. Quotes are similar to retweets, except that a quoter adds additional comments of their own that can either support or disparage the original tweet \cite{hemsley2018tweeting}. Similarly, replies can support or undermine the original tweet, but replies do not rebroadcast the original tweet. Considering these possibilities, we contend that retweeting conveys the strongest degree of social approval, followed by likes. Quotes and replies can express social approval or disapproval, but without the textual content, it is impossible to determine which sentiment it conveys. Of course, an individual can interact with a tweet in multiple ways, for instance, by retweeting \textit{and} liking, but we cannot determine this from our data. In the absence of engagement metrics that explicitly signal disapproval (such as downvoting), we attempt to distinguish social approval from disapproval by how these metrics relate to one another. For example, a tweet with relatively more replies than likes could indicate that the tweet is perceived negatively by others. Conversely, if a tweet has both a high number of quotes and retweets, the tweet may be viewed favorably.

\begin{table*}[t]
    \centering
        
    \caption{The number of anchor tweets and their corresponding number of unique users (bot score $<=0.5$) when the engagement metric that the anchor tweet received is substantially lower or higher than predicted ($k=50$) in the hate speech dataset. }
    \label{tab:n_instances_rq2}
        \begin{tabular}{@{}lrrrr@{}}
        \toprule
        & \multicolumn{2}{c}{\textbf{Lower Than Predicted}} & \multicolumn{2}{c}{\textbf{Higher Than Predicted}} \\
        \cmidrule(lr){2-3} \cmidrule(lr){4-5}
        \textbf{Metric} & \textbf{\# Ex} &  \textbf{\# Users} & \textbf{\# Ex} &  \textbf{\#  Users} \\
        \midrule
        Likes & 8,962 & 741 & 15,835 & 1,447 \\
        Retweets & 16,969 & 1,062 & 63,779 & 2,022 \\
        Replies & 16,496 & 642 & 30,743 & 1,800 \\
        Quotes & 2,462 & 347 & 52,279 & 1,883 \\
        \bottomrule
        \end{tabular}
\end{table*}
To examine the potential impact of social engagement on toxicity, we focus on instances where users experienced social approvals or disapprovals that significantly deviated from their expectations. To achieve this, we construct four machine-learning models to predict four distinct social engagement signals per tweet (see Materials and Methods). For example, in estimating the number of likes a tweet garners, we incorporate metrics such as retweets, quotes, and replies, along with an extensive set of other features derived from both user information and text content, aiming to predict the tweet's social engagement. This model can help identify tweets where the predicted engagement value substantially differs from the actual value, suggesting situations where users received unexpected levels of social approval or disapproval. We refer to these tweets as "anchor" tweets and analyze the average change in toxicity between the $k$ tweets preceding and following the anchor tweet. The following results presented focus on users with bot scores less than 0.5. For robustness, we replicate our findings for users with bot scores less than 0.8 in Appendix \ref{appendix:6} to demonstrate the consistency of our results. We display the number of anchor tweets and their corresponding unique users that received significantly higher or lower engagement counts for $k=50$ in Table \ref{tab:n_instances_rq2}.

\begin{figure}
    \centering
    \includegraphics[width=0.7\linewidth]{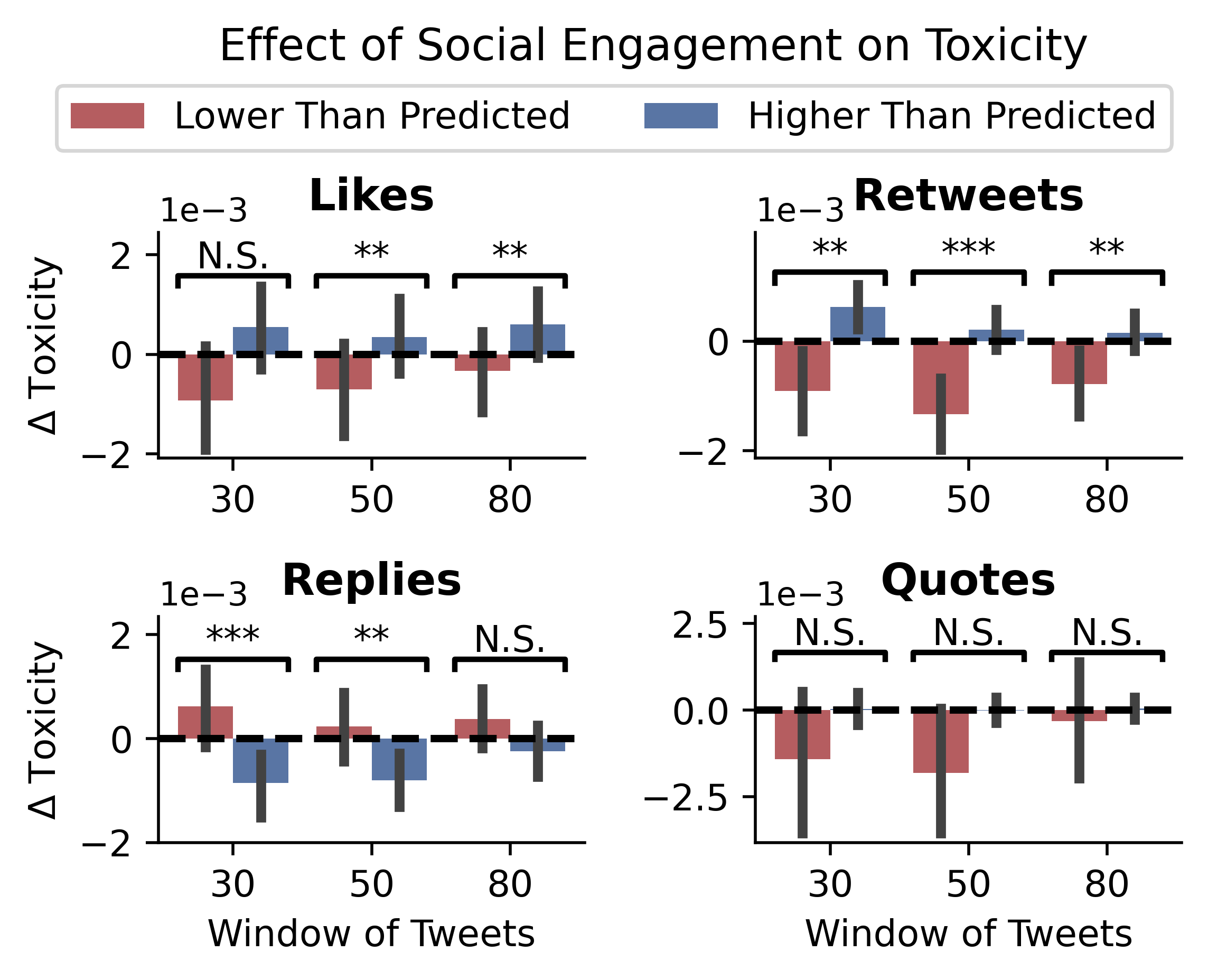}
    \caption{Changes in toxicity (y-axis) when an anchor tweet received lower (red bars) or higher (blue bars) than the predicted amount of social engagement at different windows $k$ (x-axis) in the hate speech dataset. Changes that are significantly different between the lower- and the higher-than-predicted groups are indicated (Mann-Whitney U test, ** $p<0.01$, *** $p<0.001$).}
    \label{fig:social_engagement_toxicity_diff}
\end{figure}

\subsubsection{Likes and Retweets Increase Toxicity, but Replies Reduce It}
Figure \ref{fig:social_engagement_toxicity_diff} illustrates the changes in a user’s toxicity when an anchor tweet experienced a substantially higher or lower amount of social engagement. Let $k=50$ indicate that we compare the past 50 tweets before the anchor tweet with the 50 tweets following the anchor tweet, we see that anchor tweets that received substantially more likes, more retweets, or fewer replies than expected lead to a \textit{significantly greater increase in toxicity} than when the anchor tweet received substantially fewer likes, fewer retweets, or more replies than expected. Though the net effect is relatively small, we emphasize the statistical significance of our results. In particular, not receiving enough retweets seems to have a much more dramatic effect on users: users who received substantially fewer retweets than expected generated a larger net decrease in subsequent toxicity (red bars) than if they received more retweets than expected (blue bars).

The effects of social engagement on toxicity may also reflect different temporal durations. While the increased toxicity due to retweets replicates for all other values of $k$, it differs for likes and replies. For likes, we observe a sustained increase in toxicity when the window is larger $k=50, 80$ but not smaller; for replies, the reduction of toxicity persists for smaller $k= 30, 50$ but not larger windows. One potential explanation is that giving likes requires little effort, and therefore ``likes'' connote less potent social approval compared to more effortful verbalized approval messages in the form of quotes or replies, so the effect of likes is more gradual and long-term. If a reply is negative, the explicit criticism can immediately affect a user’s behavior, but the effect may not be long-lasting.

\begin{table*}[t]
    \centering
    
    \caption{The average change in toxicity at $k=50$ when an anchor tweet received substantially higher or lower likes-per-quotes or retweets-per-quotes than expected in the hate speech dataset. Statistical significance from a Mann-Whitney U test is indicated (** $p<0.01$, *** $p<0.001$).}
    \label{tab:relative_social_engagement_toxicity}
    \begin{tabular}{lrrc}
        \toprule 
         & \multicolumn{2}{c}{\textbf{Actual vs. Expected}} \\
         \cmidrule(lr){2-3}
        \textbf{Relative Metric} & \textbf{Lower} & \textbf{Higher} &  \textbf{Sig.}\\
        \midrule
         % Likes-per-Retweets & 0.0014 & 0.0001 & N.S.\\ 
         % Likes-per-Replies & $-0.0001$ &0.0009 & N.S.\\
         % \rowcolor{ACMYellow} 
         Likes-per-Quotes & $-0.0007$& 0.0005	& \textasteriskcentered\textasteriskcentered\\
         % Retweets-per-Replies &$-0.0058$& 0.0003& N.S. \\
         % \rowcolor{ACMYellow} 
         Retweets-per-Quotes  & $-0.0019$ &	0.0005	& \textasteriskcentered\textasteriskcentered\textasteriskcentered\\
         % Replies-per-Quotes & $-0.0005$& $-0.0009$& N.S. \\ 
         \bottomrule
    \end{tabular}

\end{table*}

\subsubsection{Relatively Fewer Quotes Increase Toxicity}
It appears that quotes lead to no discernable changes in toxicity (Figure \ref{fig:social_engagement_toxicity_diff}). However, it could be the case that they \textit{do} impact toxic behavior, yet because their impact could be both extremely negative or extremely positive, the net effect may be canceled out. To test this possibility, we look at the relative number of quotes compared to the number of likes or retweets, two engagement methods that we are quite certain to be approvals and not disapprovals. We train a new social engagement model to predict $x$-per-quote, where $x$ could be likes, retweets, or replies. We observe that when there are more likes-per-quotes or retweets-per-quotes, a user displays a significantly greater increase in toxicity (Table \ref{tab:relative_social_engagement_toxicity}). The effect is not significant when we compare replies-per-quotes. One possibility is that since likes and retweets are relatively unambiguous positive social approval, a high number of likes/retweets in conjunction with a low number of quotes indicates more social approval than disapproval, and vice versa.  

\begin{figure}
    \centering
    \includegraphics[width=0.5\linewidth]{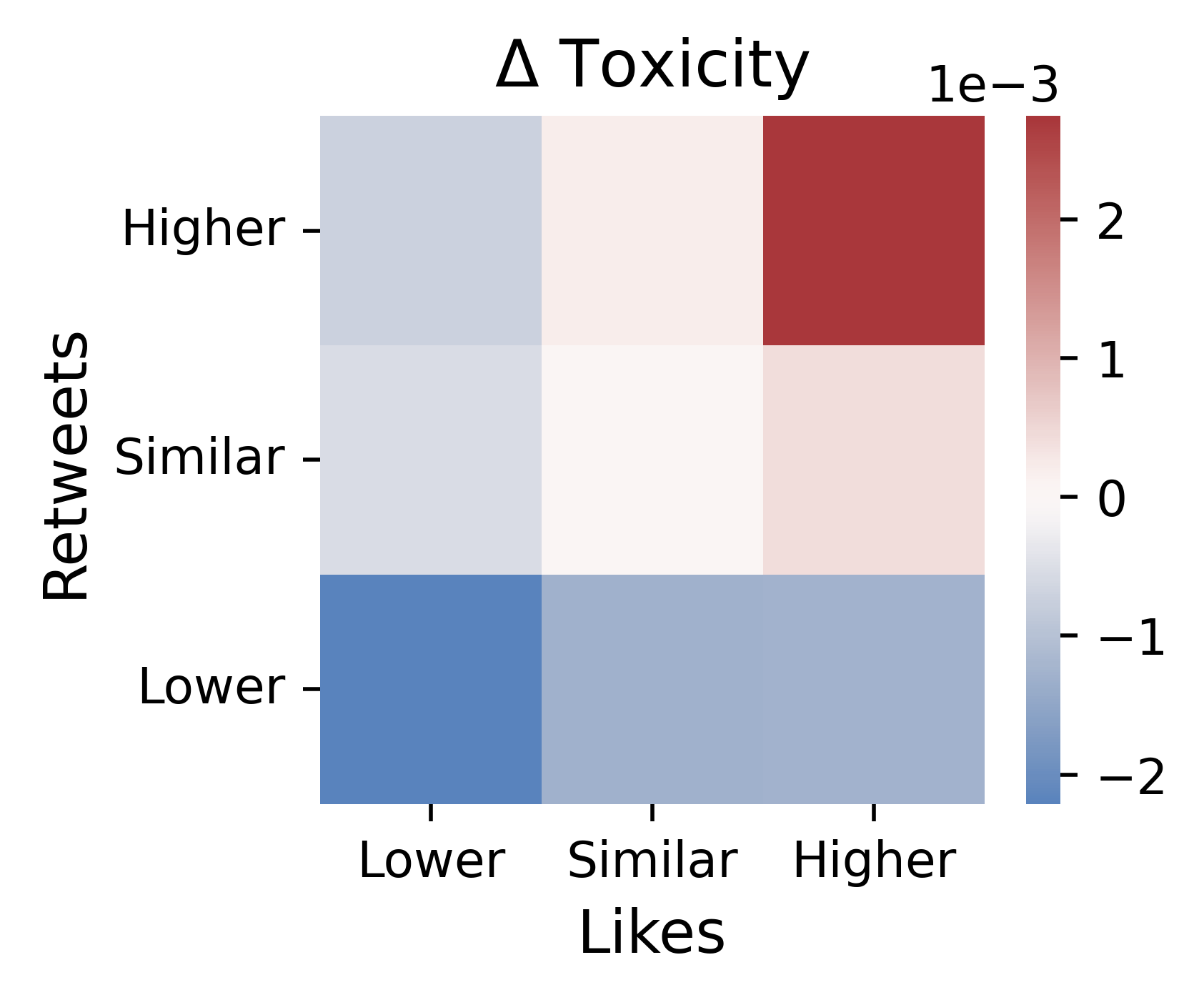}
    \caption{Having higher amounts of likes \textit{and} retweets than predicted would result in the biggest increase in future toxicity in the hate speech dataset, and vice versa ($k=50$).}
    \label{fig:like_and_retweet_toxicity}
\end{figure}

\subsubsection{The Compounded Impact of Both Likes and Retweets}
Thus far, we have only considered engagement metrics in isolation. However, a user would presumably be impacted by \textit{all} forms of social engagement on their post at once. Therefore, we analyze the combined effects of likes and retweets (Figure \ref{fig:like_and_retweet_toxicity}). When both likes and retweets are greater than expected, the increase in subsequent toxicity doubles compared to when only one form is greater. The opposite is also true: fewer likes in conjunction with fewer retweets reduce toxicity by a greater amount. 

\begin{figure}
    \centering
    \includegraphics[width=0.5\linewidth]{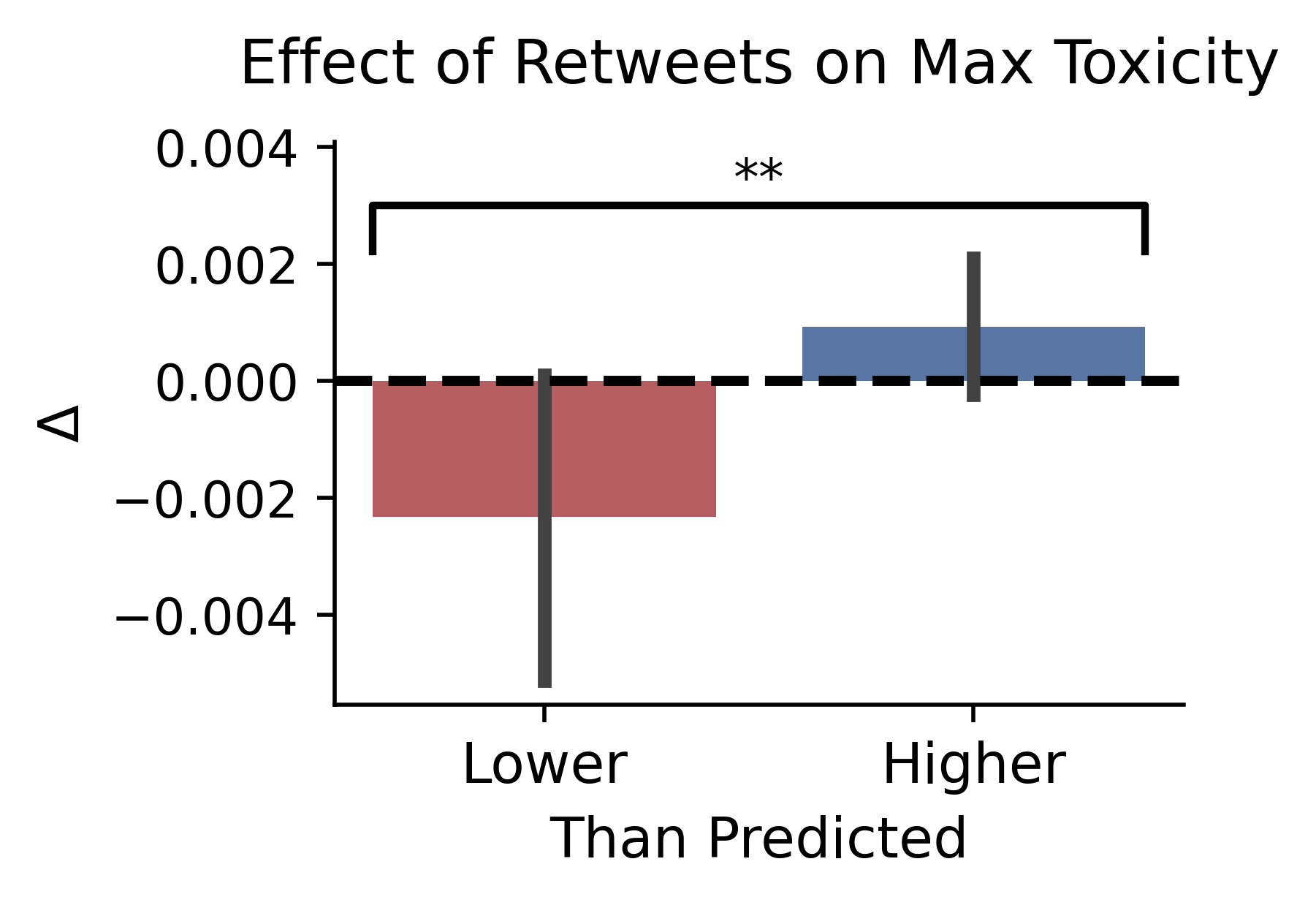}
    \caption{When an anchor tweet in the hate speech dataset receives substantially lower (red) or higher (blue) amount of retweets than expected, the difference in maximum toxicity ($k=50$) is statistically significant (Mann-Whitney U test, $**p<0.01$). More retweets lead to an increase in maximum toxicity, and vice versa}
    \label{fig:retweet_max_toxicity}
\end{figure}
\subsubsection{Retweets Escalate Maximum Toxicity} 

While we have shown that the toxicity levels are raised by social approval, we wonder whether social approval influences how hateful one is in one's most hateful tweet. Similar to our previous approach, we compare the maximum toxicity in the tweets before and after an anchor tweet. Figure \ref{fig:retweet_max_toxicity} shows that the change in maximum toxicity is statistically significant when the anchor tweet experiences substantially fewer or greater retweets. After an anchor tweet gets more retweets than expected, the user escalates their maximum toxicity. In contrast, after anchor tweets receive fewer retweets than expected, users decrease their maximum toxicity. This trend is only observed with retweets and not with likes, replies, or quotes (which are not significantly different, not shown), suggesting that retweets hold unparalleled power in altering online hate.

\section{Discussion}

In this chapter, we take a comprehensive look at millions of historical tweets by a set of known, hateful users on Twitter. We make two meaningful contributions. First, we show that toxicity is homophilous on social networks. Second, we find that a hateful user’s toxicity level can rise or fall when the user experiences substantially less or more social engagement in the form of \textit{retweets} (flattering rebroadcast and endorsement), \textit{likes} (endorsement), \textit{replies} (either endorsement or criticism), and \textit{quotes} (rebroadcast and either endorsement or criticism). We find that users’ tweets grow significantly more toxic after they receive more social approval from other users: more retweets, more likes, fewer replies, and comparatively fewer quotes. Conversely, users' tweets become less toxic after they receive insufficient social approval or social engagement that may indicate disapproval rather than approval. In particular, retweets, which signal both rebroadcast and endorsement, consistently is linked to a profound increase in users’ toxic behavior. These findings can be extended by analyzing the relation between social engagement signals and higher-level behavioral cues to explain the incentives and motivations of hateful actors through modeling techniques such as inverse reinforcement learning (e.g., \citeauthor{luceri2020detecting}, \citeyear{luceri2020detecting}).

\subsection{Implications}
Our results support the social approval theory of online hate \cite{walther2022social}: hateful users could be motivated to appeal to their supporters and respond suitably to positive social reinforcement. It may be hateful users are \textit{not} primarily incentivized to harm the nominal target of their disparaging messages but rather to gain favor from their hateful peers. In addition to advancing our understanding of toxic online behavior, we believe this work has tremendous potential to inform strategies to combat hate speech on not only Twitter but also other online platforms. For one, this empirical evidence supports efforts to moderate hateful content (e.g., Meta’s policy; \citeauthor{fbpolicy}, \citeyear{fbpolicy}). Other directions include ``shadowbanning''--hiding users’ posts from all but the user; see \citet{jaidka2023silenced}--or disabling likes, etc., on hateful posts to reduce the effects of social reinforcement on toxicity.

\subsection{Limitations}
Our research has limitations. Most importantly, while our findings are wholly consistent with the social approval theory, we recognize that we did not conduct any randomized, controlled experiment to demonstrate causality. Conducting such an experiment on this topic would not be ethical, as it would require us to reward hateful users for discriminatory messages and thereby encourage online hatred. Associational research with temporal order provides the best estimation of potentially causal theoretical relationships under the circumstances \cite{davis1985logic}. We also note other limitations due to data availability. Our data is based on one hate-infused dataset of Twitter posts concerning specifically US/UK immigration \cite{bianchi2022njh}, and our sample may be skewed since many users from the seed dataset are likely to be suspended by Twitter or have deleted their own accounts. Additionally, in the absence of the quote and reply texts that may either endorse or criticize a tweet, we cannot examine the full impact of social approval or disapproval signals on hateful behavior. Lastly, our findings regarding the impact of social engagement on future toxicity reveal small effect sizes. However, despite these shortcomings, the statistical significance and the consistency of our findings across various robustness tests enhance the validity of our conclusions. Many other influences impinge on the tenor of hate messages, from real-world geopolitical conflicts to mainstream media stories, not to mention changes in online content moderation policies. Even if the exchange of social approval is a primary influence, there are undoubtedly many others as well.

\subsection{Ethical Statement} 
We recognize that our research could be used by malicious actors to incentivize online hate through positive social reinforcement. We believe the benefits of understanding the social motivators of online hate production outweigh the risks the research poses. This research was approved by the Institutional Review Board (IRB).

\chapter{Moral Values Underpinning COVID-19 Online Communication Patterns}
\label{chp:moral_covid}

\section{Introduction}

The COVID-19 pandemic not only brought illness and death but also catalyzed a series of important changes disrupting the lives of nearly everyone across the world. 
% With it came a of  moralization of behaviors \cite{ekici2021deciding, francis2022moral, schmidtke2022evaluating}. 
Actions that once constituted ordinary conduct, such as breathing freely without face coverings, became morally laden with accusations of selfishness and harm \cite{latimes2020}. Work swiftly transitioned from in-person to remote, changing normative expectations of work \cite{pew2022normal}. Research consistently underscores the intertwining of moral judgments with decisions pertaining to health-related behaviors, such as the intricate linkages between morality and vaccine hesitancy \cite{reimer2022moral,schmidtke2022evaluating}.

In this chapter, we use moral psychology to explain online conversations surrounding COVID-19. The Moral Foundation Theory (MFT) was created to explain the elements of human moral reasoning. It consists of five main foundations: \textit{care/harm}, \textit{fairness/cheating}, \textit{loyalty/betrayal}, \textit{authority/subversion}, and \textit{purity/degradation} \cite{haidt2004intuitive,graham2013moral}. These foundations serve as a moral compass, shaping how we perceive and engage with the world. As individuals may prioritize these foundations differently, the misalignment of moral orientations between people often leads to disagreements and moral conflicts \cite{haidt2012righteous}. Computational social scientists have also uncovered patterns of moral foundation homophily in social networks \cite{dehghani2016purity}, which could perpetuate a user's existing beliefs. 
% For instance, \citeauthor{dehghani2016purity} highlighted the predilection for social networks to exhibit \textit{purity} homophily—a preference for network connections who share similar purity moral values. 
Moral values have also been found to be linked with political ideology, with liberals and conservatives preferring different sets of moral values \cite{haidt2007morality,graham2009liberals}. 
% more related work on moral and poli: \cite{graham2012moral,haidt2012righteous,feinberg2019moral,hatemi2019ideology}.
% Liberals tend to base their morals on the two \textit{individualizing} foundations--\textit{care} and \textit{fairness}--while conservatives tend to base their morals on all five original foundations, including the \textit{binding} foundations of \textit{loyalty}, \textit{authority}, and \textit{purity}. As a result, the political divide that has long persisted in American politics can be viewed as a clash between fundamentally different sets of moralities \cite{haidt2012righteous,feinberg2019moral}.
Thus, the polarization of opinions relating to COVID-19 could be related to moral differences \cite{jiang2020political}.

There remains a notable research gap concerning the role of moral foundations in shaping communication dynamics, particularly regarding COVID-19. We know from Chapter \ref{chp:poli_covid} that COVID-19 was a politicized issue on social media leading to communication echo chambers that could undermine public health strategy efforts (see Chapter \ref{chp:poli_covid}); however, we don't know to what extent this can be explained by moral differences. Through the lens of moral psychology, this research aims to explore how moral orientations influence user groups on Twitter discussing COVID-19, moral homophily in communication patterns, and the resonance of moral themes across groups. This facilitates a deeper understanding of how moral considerations influence online discussions during times of crisis. We pose the following research questions:
\begin{itemize}
\item \textbf{RQ1}: What distinct user groups emerge based on MFT and network interactions when discussing COVID-19 on Twitter?
\item \textbf{RQ2}: Do communication patterns reflect moral homophily, and how do they compare across groups?
\item \textbf{RQ3}: Which moral themes resonate more effectively with out-group members?
\end{itemize}

Our summary of contributions is as follows. First, we discover four main groups of users with vastly different moral priorities and political partisanship. We paint a nuanced picture of the relationship between morality and political ideology, demonstrating that moral orientations do not rigidly separate users across the political spectrum. We also find that most user groups exhibit group-based homophily--the tendency to communicate with in-group members. One group of users (group IV) who are primarily right-leaning users with \textit{fairness} and \textit{authority} moral foundations also tend to only communicate with in-group members, suggesting a moral echo chamber. Finally, we find that messages with moral foundations that are not typically favored by their authors and messages with moral pluralism tend to resonate better with out-group members. We conclude with insights into user group dynamics and communication patterns, emphasizing the importance of moral diversity for effective discourse.
\section{Background and Related Work}

\subsection{The Moral Foundation Theory}
The Moral Foundation Theory was proposed to explain variations in human moral reasoning. The original MFT consists of the following five main foundations \cite{haidt2004intuitive,graham2013moral}:

\begin{enumerate}
    \item \textit{Care/harm}: This foundation relates to our ability to feel empathy and compassion for others and our willingness to alleviate their suffering. It emphasizes the importance of caring for and protecting others, especially those who are vulnerable.
    \item \textit{Fairness/cheating}: This foundation emphasizes the importance of justice, equality, and fairness in our interactions with others. It focuses on the belief that everyone should be treated fairly and equally.
    \item \textit{Authority/subversion}: This foundation relates to our respect for authority, hierarchy, and tradition. It emphasizes the importance of following rules and obeying authority figures.
    \item \textit{Loyalty/betrayal}: This foundation underlies the sense of belonging to a group and the importance of showing loyalty and allegiance to that group. It emphasizes the importance of being patriotic and self-sacrificing for the betterment of the group.

    \item \textit{Purity/degradation}:\footnote{`Purity' is sometimes referred to as `sanctity' in some literature, but for consistency, we use `purity' throughout this paper.} This foundation relates to our sense of purity and cleanliness and the importance of self-control to avoid impure or degrading actions. It emphasizes the importance of protecting sanctity and avoiding contamination. 
\end{enumerate}

There is also a sixth foundation, \textit{liberty/oppression}, proposed in \citet{haidt2012righteous}. This foundation relates to our belief in individual freedom and autonomy, as well as our opposition to oppression and coercion. However, due to its frequent omission in prior research on MFT text detection \cite{johnson2018classification,rojecki2021moral,guo2023data}, there is a lack of a suitable detector for this foundation. Hence, we do not address this foundation in our current study.

\subsection{Morality and Politics}

Research shows that morality binds and divides users by political orientation \cite{haidt2007morality,graham2009liberals}. Users with liberal ideology typically reflect higher individualizing moral foundations of \textit{care} and \textit{fairness},  foundations that support the rights and welfare of individuals, whereas conservative ideology typically endorses all five moral foundations equally, which includes the binding moral foundations of \textit{authority}, \textit{loyalty}, and \textit{purity}.  \citet{koleva2012tracing} further exemplified this by examining 20 politically salient issues, such as abortion and immigration. They found that the five moral foundations, in particular \textit{purity}, are better predictors of issue-specific opinion above ideology and demographic features. This difference in moral attitude has been attributed to growing political polarization, where users on both ends of the political spectrum are unable to resonate with each other due to moral incongruence \cite{haidt2007morality} and overexaggerate their differences \cite{graham2012moral}. Techniques of moral-based reframing have thus been proposed as a way to bridge the political divide \cite{feinberg2019moral}.

\subsection{Moral Homophily}
Moral values may also explain the connection and formation of communities through social network homophily, the phenomenon that we tend to be drawn to people we are similar to \cite{mcpherson2001birds,kossinets2009origins}.  In a large-scale analysis of social network data on the US government shutdown, moral \textit{purity} can predict network ties, indicating \textit{purity} homophily \cite{dehghani2016purity}. Another study comparing multilingual tweets explicitly mentioning morality in English and Japanese found that the \textit{care}, \textit{authority}, and \textit{purity} foundations are homophilous in English tweets, while the \textit{loyalty}, \textit{authority} and \textit{purity} foundations are homophilous in Japanese tweets \cite{singh2021morality}. We theorize that for COVID-19, some moral foundations may exhibit network homophily that informs our moral-based understanding of online conversations.

\subsection{Morality and COVID-19}

Many divisive behaviors and controversial opinions regarding COVID-19 may be rooted in moral differences. Those with a higher \textit{care} disposition are found to be more likely to follow health recommendations \cite{diaz2022reactance}. A similar study by \citet{chan2021moral} also found that higher \textit{care} and \textit{fairness} predicts compliance with the health strategies of staying at home, wearing masks, and social distancing, while \textit{purity} predicts non-compliance with wearing masks and social distancing. In the face of disease threats, \citet{ekici2021deciding} found that \textit{fairness}, \textit{care}, and \textit{purity} were the most important moral foundations that predicted people's acceptability of moral transgressions. In a study of tweets on the COVID-19 mask mandate, \citet{mejova2023authority} found that \textit{authority} and \textit{purity} were associated with anti-masking sentiment while \textit{fairness} and \textit{loyalty} were associated with pro-masking sentiment; there is also a de-emphasis on the \textit{care} foundation following the mask mandate.

By 2021, arguments loaded with moral judgments both for and against the COVID-19 vaccination take center stage. In a study based in Great Britain, \citet{schmidtke2022evaluating} found that vaccine hesitancy is associated with higher moral needs of \textit{authority}, \textit{liberty}, and \textit{purity} and less need of \textit{care}. Moral foundations were also shown to be good predictors of country-level vaccination rates in the US: \textit{purity} predicted lower vaccination rates, whereas \textit{fairness} and \textit{loyalty} predicted higher vaccination rates \cite{reimer2022moral}. Analyzing the sentiment in vaccine-related tweets revealed that pro-vaccination tweets carried more \textit{care} morals, while anti-vaccination tweets carried more \textit{liberty} morals \cite{pachec2022holistic}. On Facebook, pro-vaccination users identify more with \textit{authority}, and anti-vaccination users identify more with \textit{liberty} \cite{beiro2023moral}. The debate on vaccination is also linked to partisanship. Liberals discuss COVID vaccination on Twitter with more emphasis on the moral virtues of \textit{care}, \textit{fairness}, \textit{liberty}, and \textit{authority}, whereas conservatives leaned into the vices of \textit{oppression} and \textit{harm} \cite{borghouts2023understanding}.

Besides being moralized, COVID-19 was also politicized, which, as we previously analyzed in Chapter \ref{chp:poli_covid}, may have led to the formation of online political echo chambers. Echo chambers are harmful to online ecosystems as segregated communication can lead to radicalism and extremism \cite{o2015echo}. One research suggests that moral differences can explain partisan differences in vaccine hesitancy \cite{bruchmann2022moral}. Conservatives and liberals also differ in how they use morality in their language. Conservatives are more likely to use moral vices rather than virtues in their tweets \cite{borghouts2023understanding,rao2023pandemic}. That said, there is promising research on how moral-based reframing of messages can be used to advocate mask-wearing among liberals and conservatives \cite{kaplan2023moral,luttrell2023advocating}.

\section{Data}

\begin{figure}[t]
    \centering
    \includegraphics[width=0.7\linewidth]{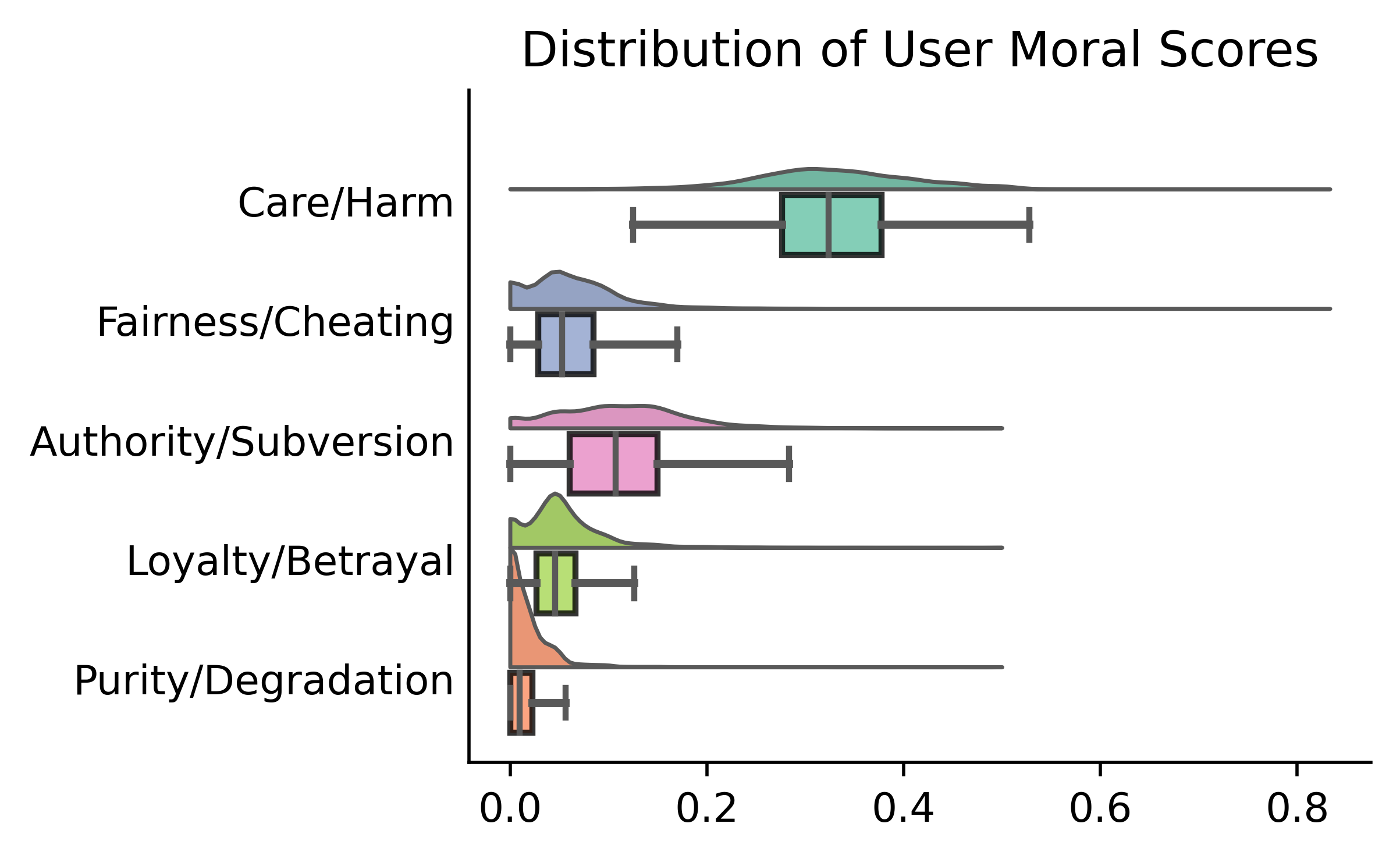}
    \caption{Distribution of raw user moral scores in the COVID-19 dataset. }
    \label{fig:user_mf_dist}
\end{figure}

% 2,169,038 of users who tweeted at least 10 tweets in a given month
% 252697816 tweets
% 318743, the 5% per month sample of the above, for sllm

For this research, we also use the large dataset of real-time COVID-19 tweets collected by \citet{chen2020covid} that we used previously in Chapter \ref{chp:poli_covid}. However, unlike the previous version of the dataset, we now use a much longer subset of the data spanning the beginning of February 2020 to the end of October 2021 for 21 full months. We then take a longitudinal set of active users who tweeted at least 10 tweets containing moral values in any given month, which means they each contain at least 10 data points of moral foundation values. We describe how we detect moral values in the Methods section below. This procedure resulted in a large dataset of 2 million users and 253 million tweets. To analyze communication patterns, we also create the retweet and mention networks. Using users as nodes, we build the retweet network by connecting users who retweet one another. Similarly, we build the mention network by connecting users who do not retweet but rather quote (retweet with additional comment) or mention (`@’) another user. We draw the distinction between the retweet and mention network as there may be different underlying motivations for each action. Retweeting usually implies endorsement \cite{boyd2010tweet,metaxas2015retweets} while mentioning could be used to endorse or criticize \cite{hemsley2018tweeting}. When identifying retweets or mentions between two target users, we use all available tweets, including those that do not have any identifiable moral foundations. The edges are weighted by the number of times one user retweets or mentions another. To allow for efficient computation, we create separate networks for each month of our dataset and aggregate the results if necessary. On average, we have 90,000 nodes, 1 million retweet edges, and 1.3 mention edges in our monthly networks. 

A sample of 150,000 users and their respective edges from this dataset is called \textsc{Covid-Morality}, used in the evaluation of Social-LLM in \S\ref{sec:data_covid_morality}. We also use this subset in the methodology section below to for user cluster detection.

% 		Care/
% Harm	Fairness/
% Cheating	Loyalty/
% Betrayal	Authority/
% Subversion	Purity/
% Degradation
% User A	{tweet}	1	0/0	0/1	1/1	1/1
% 	{tweet}	0/0				
% Aggregate		0.5				

\begin{table*}[t]
    \centering
    
    \caption{Hypothetical tweets, adapted from real ones in the COVID-19 dataset, that contain detected moral foundations values. Some tweets only contain the virtue or the vice of a foundation, and some tweets contain both.}
    \label{tab:eg_tweets}

    \footnotesize
    \begin{tabular}{@{}ll@{}}
        
    \toprule
    % & Tweet & Care/Harm & Fairness/Cheating & Loyalty/Betrayal & Authority/Subversion & purity/degradation\\
    % \midrule 
    \textbf{Tweet} & \textbf{Morals}\\
    \midrule
    \textit{The vaccine won't harm you if you are in one of the approved groups, but it will protect you.} & Care, Harm\\
    \midrule 
    \textit{The choice to not wear a mask is my right. The choice to stay at home is yours.} &  Fairness, Authority \\
    \midrule
    \textit{New Zealand just announced it will provide the new Covid vaccine to any New Zealander who wants} &  \multirow{2}{*}{Fairness, Authority}\\
    \textit{it--free of charge. They're also making the vaccine available to all their Pacific Island neighbors.}\\
    \midrule 
    \textit{It's up to us to slow the spread, save lives, \& keep businesses open. We have to work together.} & Care, Loyalty\\
    \midrule 
    \textit{I refuse to take this vaccination! It goes against my religious beliefs!} & Purity \\%
    \midrule 
    \textit{Revolting. The fact that XXX got the vaccine before healthcare workers and first responders.} & Degradation\\
     \bottomrule
    \end{tabular}
   
\end{table*}

\section{Methods}
To address our research goals, we employ several computational methods. First, we use a moral foundation model to detect the moral foundations present in each tweet and aggregate them to obtain overall moral orientation scores for each user. Next, using user content and moral features as well as the social network as input data, we apply Social-LLM (Chapter \ref{chp:socialllm}) to learn latent vector representations of the users based on their moral foundations and communication patterns, which are then used to cluster users into distinct groups. Finally, to analyze moral theme resonance across ideological lines, we apply a classifier to predict users' political leanings.
\subsection{Detecting the Morality of Tweets and Users}

We detect tweet morality using an MFT model fine-tuned specifically on this dataset \cite{guo2023data}. This model adopts a data fusion technique to account for the fundamental shifts in morality based on the topics of the dataset. It is based on a pre-trained BERT model \cite{devlin2019bert} that was fine-tuned on three different Twitter datasets annotated for morality, including one that was specifically on COVID \cite{rojecki2021moral}. See \citet{guo2023data} for a detailed evaluation of the model. This MFT detector predicts the presence of the 10 moral values---each foundation contains two opposite moral values for the virtue and the vice---in a multi-label manner for every tweet, with 1 indicating the presence of a moral value and 0 indicating the absence of it. We present some example tweets with moral foundation labels in Table \ref{tab:eg_tweets}. We then aggregate the virtues and vices of each foundation into one label. If a tweet has a score of 1 for the \textit{care} dimension but a score of 0 for the \textit{harm} dimension, then it has a combined score of 0.5 in the \textit{care/harm} foundation. We choose to do this because the morality detector looks for explicit expressions of the virtue (e.g., \textit{care}) and the vice (e.g., \textit{harm}) separately, but for our purposes, both reflect a moral disposition in that foundation. The vast majority (87\%) of tweets contain only a single moral foundation. 13\% of the tweets contain two moral foundations, of which the most popular combination is \textit{care} and \textit{authority}. Less than 1\% of tweets have 3 or 4 moral foundations, and no tweets have all five foundations.

We also compute morality scores at the user level, which is the mean moral score of each five foundations across all the tweets by one user. The distribution of the aggregated user morality scores is shown in Figure \ref{fig:user_mf_dist}. The \textit{care} foundation is the most frequently utilized foundation by a wide margin, and  \textit{purity} is the least utilized foundation. To maintain cross-comparison of each morality, we standardize (z-scores) the scores in each foundation to have a mean score of 0 and a standard deviation of 1. 

\subsection{Detecting User Groups Using MFT and Twitter Activity} 
% features  ["verified", "account_creation_date", "friends_count", "listed_count", "followers_count", "favourites_count", "statuses_count", "n_tweets", "n_orig", "n_rt", "n_qtd", "n_replies", "label_care_harm", "label_fairness_cheating", "label_loyalty_betrayal", "label_authority_subversion", "label_purity_degradation"],
% profile descripton
% base model: all-mpnet-base-v2
% edge type: both, + direct , not weighted
% batch 64, lr 2e-5, 4 epochs, 128 hidden 
In pursuit of our research goal--understanding the communication dynamics among users with various moralities--we want to find salient groups of users in our dataset based on their moral foundations and their communication preferences. For this task, we leverage Social-LLM, an unsupervised social network user representation method that combines user content features and social network features. It works by initializing user embeddings from their text and metadata features and optimizing them such that two users who share a network connection would have similar embeddings. Social-LLM is substantially easier to train than its graph neural network counterparts since it does not need (sub)graphs as inputs but rather only the pairs of edges. Despite its simplicity, we have shown that it works comparably to many state-of-the-art baselines, including various fine-tuned LLMs or complex graph neural network methods, in its ability to recover user moral foundation scores. See Chapter \ref{chp:socialllm} for a detailed description of the model and evaluations on this dataset. Given the experiment results from this dataset, we use the best hyperparameter configuration to train a Social-LLM model, which learns 128-dimensional embeddings using directed retweet and mention edges, as well as a base language model of SBERT-MPNet \cite{reimers2019sbert,song2020mpnet}.

In addition to profile descriptions and user metadata features, we also include the user moral foundation scores in this version of the model so that the user representations include moral cues. Then, we apply this trained model to our full dataset to obtain user embeddings for our 2 million users. 

These learned user embeddings contain important information about users' moral foundation values and Twitter activities (including social network interactions, profile descriptions, and other user metadata features). We then apply the $k$-means algorithm to the embeddings to uncover distinct user groups. After experimenting with cluster numbers ranging from 2 to 10 by using the elbow methods on the inertia and the silhouette scores, we select $k=4$ as the most appropriate number of clusters. These four distinct user groups, encapsulating the distinct moral orientations and Twitter behaviors within our dataset, warrant further exploration and analysis.

\subsection{Detecting User Partisanship}
As morality is often linked to differences in political partisanship \cite{graham2009liberals,koleva2012tracing}, and because numerous studies on COVID have shown that user opinions are politically divided \cite{jiang2020political,rao2023pandemic}, we also want to capture users' political partisanship in this study. To this end, we utilize another Social-LLM model trained for user political leaning detection. Compared to other political leaning detection methods on Twitter, Social-LLM has the advantage of learning crucial cues from not only the textual features of the tweets but also social network interaction features. The latter is particularly preferable since Twitter users are often politically segregated, especially on the topic of COVID-19 (Chapter \ref{chp:poli_covid}). Applying the Social-LLM model for political leaning detection on COVID-19 datasets, we obtain political-leaning labels for every user in our dataset. This particular model is also the one documented extensively in the Retweet-BERT chapter (Chapter \ref{chp:retweetbert}). Similar to prior work on political polarization analysis in Chapter \ref{chp:poli_covid}, we bin the scores into quintiles to adjust for the left bias. Users falling within the 0-20\% range are labeled as very left-leaning, those in the 20-40\% range as left-leaning, the middle 40-60\% as moderate, the 60-80\% as right-leaning, and those in the 80-100\% range as very right-leaning.

% \subsection{Detecting Moral Topics}

\section{RQ1: User Groups by Morality}
\begin{figure}[t]
    \centering
    \includegraphics[width=0.6\linewidth]{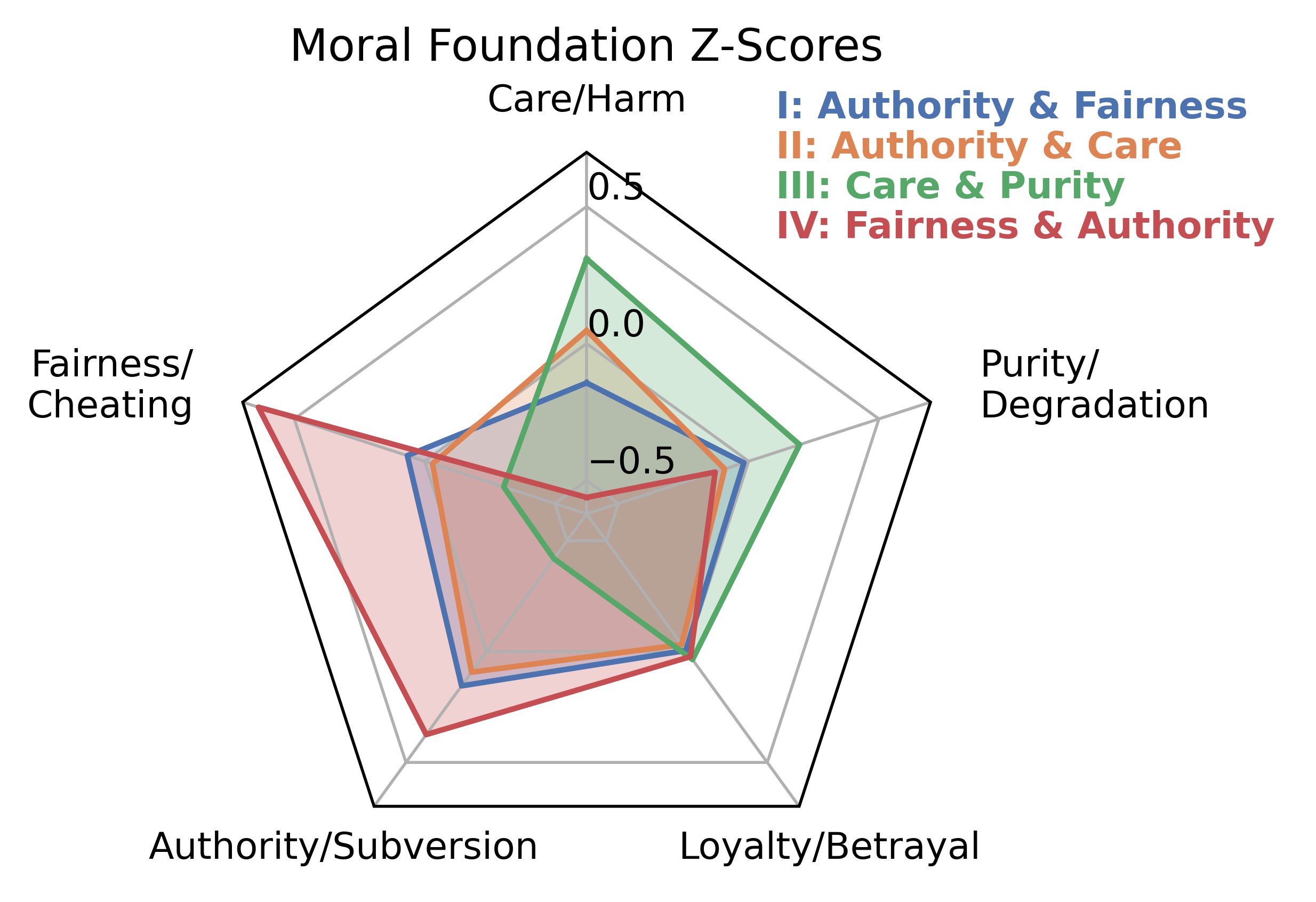}
    \caption{The average moral z-scores of each foundation for the four user groups in the COVID-19 dataset.}
    \label{fig:user_clusters}
\end{figure}
\begin{figure}[t]
    \centering
    \includegraphics[width=0.6\linewidth]{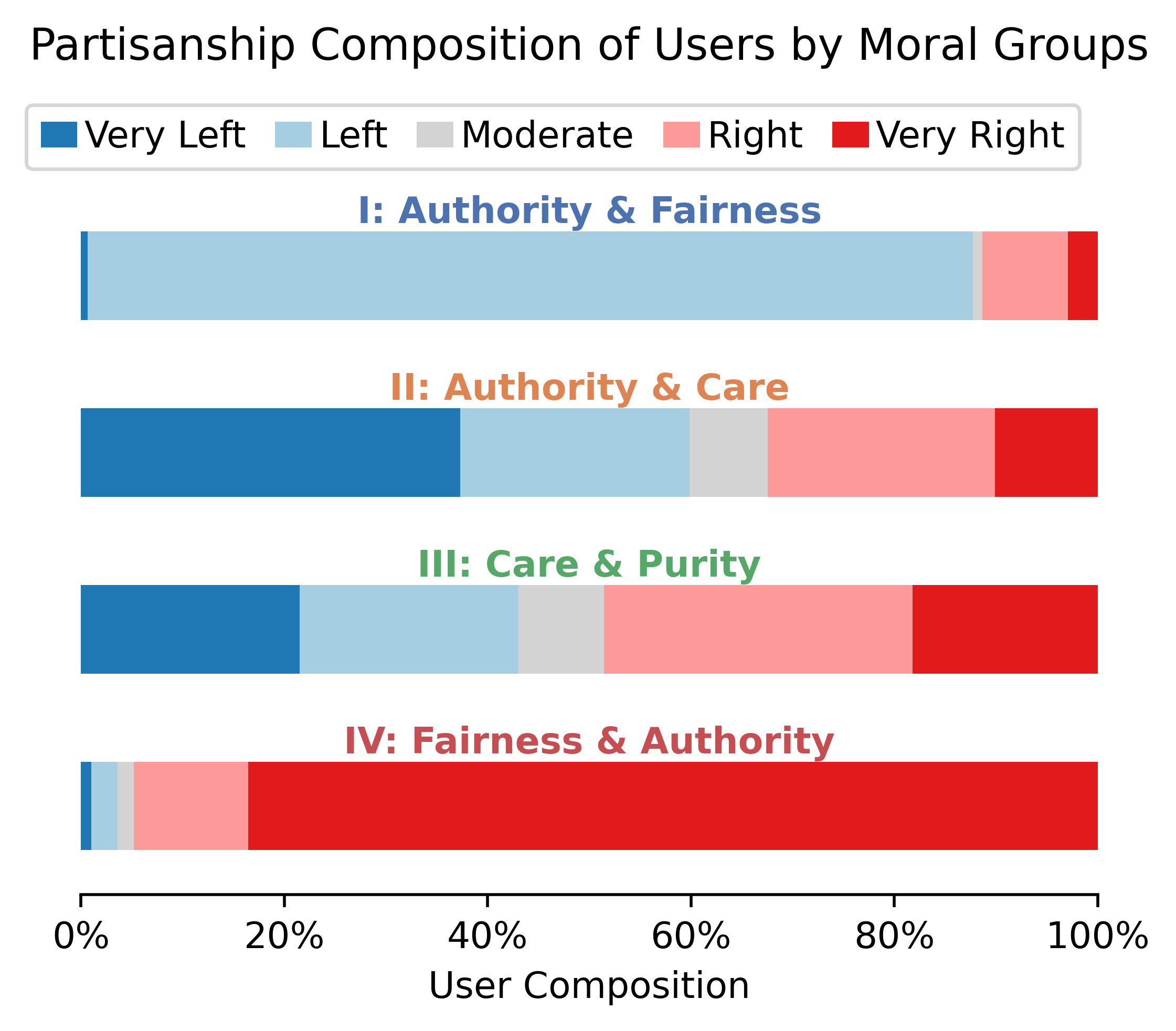}
    \caption{Partisanship breakdown of the four user groups in the COVID-19 dataset. Blue bars represent left-leaning users, and red bars represent right-leaning users.}
    \label{fig:moral_poli_bar}
\end{figure}

\begin{figure*}[t]
    \centering
    \includegraphics[width=\linewidth]{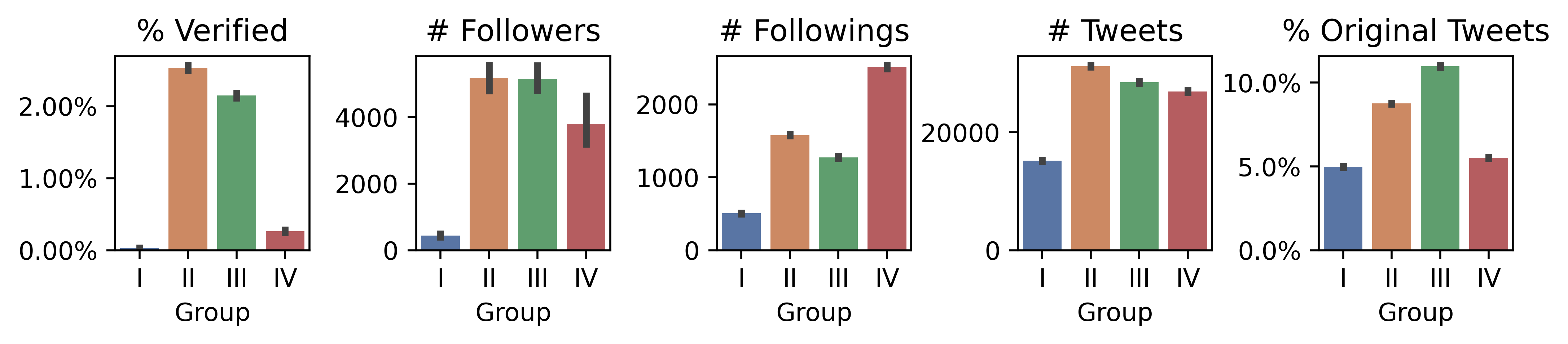}
    \caption{Distribution of the user metadata features for the four user groups in the COVID-19 dataset.}
    \label{fig:user_meta}
\end{figure*}

We first answer the research question: What are the characteristics of the groups of users tweeting about COVID-19 based on their moral values, Twitter profiles, and network interactions? We show the average moral scores of each group in Figure \ref{fig:user_clusters}, a political partisanship breakdown in Figure \ref{fig:moral_poli_bar}, and some key user metadata statistics in Figure \ref{fig:user_meta}. Below, we discuss the characteristics of each group.

\paragraph{Group I: Authority \& Fairness.} This group contains 493, 000 users (23\%) who reflect a relatively stronger morality in the \textit{authority} ($z=0.15$) and \textit{fairness} foundations ($z=0.07$). Its scores in the \textit{care} foundation are below average ($z=-0.14$). Considering user partisanship, we find that this group is predominantly occupied by left-leaning users (87\% left-leaning), but very few are extremely left-leaning. The metadata features of group I users stand out in several important ways. They have the fewest proportion of verified users. On average, they have the least number of followers, followings, posts, and proportion of tweets that are original. In sum, group I users appear to be the least active user group and most likely do not have as many influential or popular users as the other groups. As Twitter users reflect an overall left bias (Chapter \ref{chp:poli_covid}), we theorize that group I users are average users with low influence.

\paragraph{Group II: Authority \& Care.} With 800,000, or 37\% of users, this group is the largest user group in our dataset. Its users are characterized by higher moral scores in the \textit{authority} ($z=0.09$) and \textit{care} foundations $(z=0.05$). It scores comparatively lower in the \textit{purity} foundation ($z=-0.09$). In terms of partisanship, this group has a good balance of left- and right-leaning users, although it has a substantial amount of far-left users (38\%),

\paragraph{Group III: Care \& Purity.}  The third group contains 603,000 (28\%) users and exhibits stronger than average morality in the \textit{care} ($z=0.30$) and \textit{purity} ($z=0.19$) foundations, along with a weaker than average morality in the \textit{authority} ($z=-0.42$) and \textit{fairness} ($z=-0.30$) foundations. It has a balanced representation of users across all political leanings, comprising 48\% users on the right and 44\% users on the left.

\paragraph{Group IV: Fairness \& Authority.} The fourth and final group is the smallest user group, made up of only 274 (13\%) users. This group is characterized by stronger than average foundations of \textit{fairness} ($z=0.64$) and \textit{authority} ($z=0.37$), coupled with weaker than average foundations of \textit{care} ($z=-0.56$) and \textit{purity} ($z=-0.13$). This group is also made up of predominantly right-leaning users (95\%). Users who are far-right occupy 84\% of this group alone.

\bigskip
We note that the characteristics of these groups are discussed in relation to each other, not in absolute terms. That is, for example, Group I users display lower \textit{care} moral values than Group III users, but that does not mean Group I users don't utilize \textit{care} morality. Of the four user groups, we see that Groups III and IV users have very strong preferences in some moral foundations, whereas Group I and Group II have more modest preferences. Further, We note that Group III and Group IV users have almost polar opposite moral foundation inclinations. While Group III prefers \textit{care} and \textit{purity}, these are exactly the two moral foundations that are least utilized by Group IV, and vice versa for Group IV's preferred morality of \textit{fairness} and \textit{authority}. Additionally, though both Group I and Group IV prefer the foundations \textit{authority} and \textit{fairness}, they differ considerably in what foundations they don't prefer, the strength of their morality, and their political partisanship breakdown. Finally, while most of the five moral foundations are prominently favored or unfavored by at least one of the four user groups, the \textit{loyalty} foundation usage is used almost indistinguishable among the user groups.
\begin{figure}[t]
    \centering
    \includegraphics[width=0.7\linewidth]{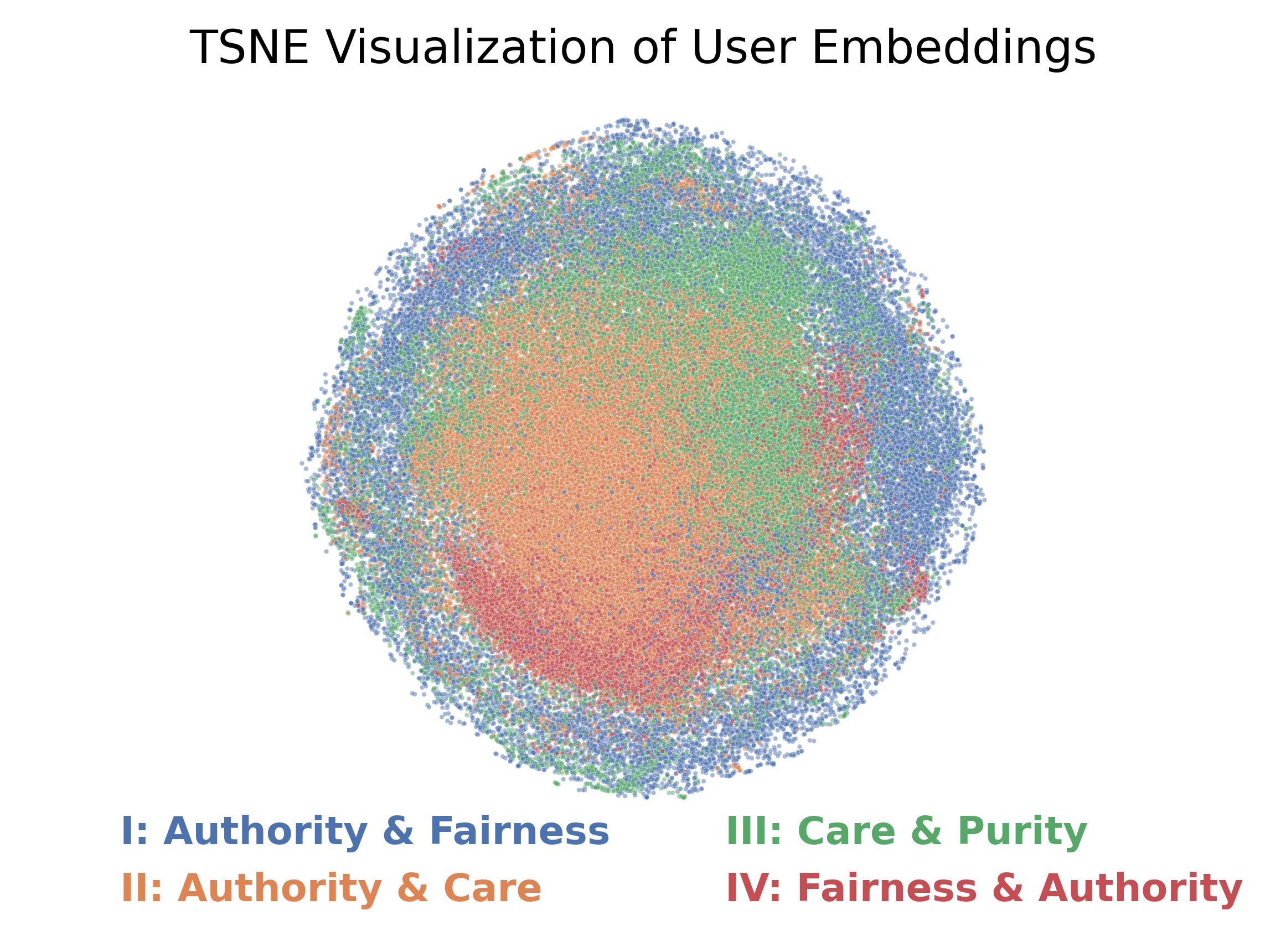}
    \caption{TSNE visualization of 100,000 sampled Social-LLM user embeddings from the COVID-19 dataset of the four user groups. These Social-LLM embeddings are learned from the users' network cues, content cues, and moral leanings.}
    \label{fig:tsne}
\end{figure}

\subsection{User Visualization} In Figure \ref{fig:tsne}, we present a TSNE \cite{tsne} visualization of 100,000 sampled user embeddings in each group. TSNE is a popular dimension-reduction technique of high-dimensional data points to reveal structural proximity among users in each group. This plot shows a good separation of every user group, with user groups II, III, and IV forming visible clusters. This may indicate that these groups form homophilous communication bubbles. However, Group I users form a circular ring enclosing other user embeddings.  As we will see in the next section, this appears to be because users in this group do not preferentially communicate with in-group members but rather interact equally with all users.

\section{RQ2: Moral Homophily}
In this section, we continue our analysis by examining whether there is a communication homophily within in-group members, which may lead to communication echo chambers among people sharing similar moral profiles.

\subsection{Homophily of Users}
As a preliminary analysis, we examine whether there is moral homophily in individual moral foundation values at the user level. That is, do users share moral foundation values similar to those of their network? We use network assortativity scores \cite{newman2003mixing} to measure how (dis)similar two sets of nodes' attributes are, given that each pair of nodes is connected by an edge. Using the standardized moral scores as the users' node attributes and the retweet or mention interaction as edges, we compute the network assortativity values of every moral foundation on monthly subgraphs. 

\begin{figure}[t]
    \centering
    \includegraphics[width=0.7\linewidth]{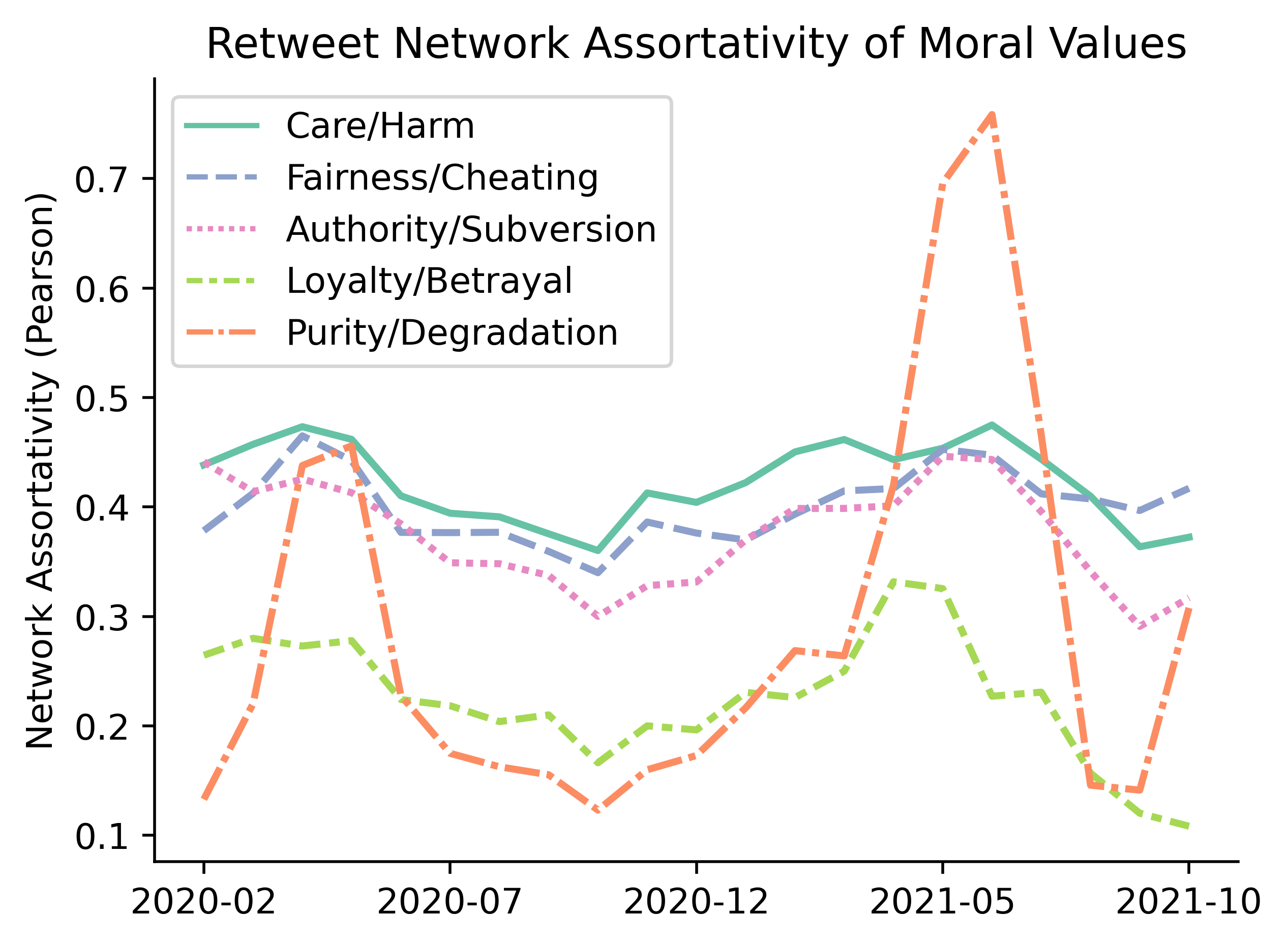}
    \caption{Retweet network assortativity of users' moral scores over time in the COVID-19 dataset. High assortativity indicates homophily.}
    \label{fig:rt_assort_time}
\end{figure}

Figure \ref{fig:rt_assort_time} shows the results from the retweet networks, which are largely similar to the mention network. All of the Pearson correlation coefficients are positive and significant (all $p<0.001$), indicating moral assortative mixing or moral homophily. In particular, the \textit{care} $(\mu=0.43$), \textit{fairness} ($\mu=0.40$), and \textit{authority} ($\mu=0.38$) foundations indicate strong, consistent moral homophily. However, \textit{loyalty} $(\mu=0.22)$ reflects much lower assortativity. Intriguingly, while most foundations reflect consistent levels of homophily over time, the \textit{purity} foundation $(\mu=0.29)$ showcases fluctuations, at times recording both the lowest and highest homophily values. This variance can be attributed to increased discussions related to notable events surrounding masking (April 2020) and vaccines (early 2021), topics related to the \textit{purity} foundation \cite{chan2021moral,reimer2022moral}. Notably, peaks in \textit{purity} assortativity align with significant events, such as the CDC's official recommendation of face coverings in April 2020 and the widespread discussion of vaccines during the first half of 2021. While these findings align partially with prior research on \textit{purity} homophily \cite{dehghani2016purity}, we emphasize the consistent and strong homophily observed in other moral foundations, particularly \textit{care}, \textit{fairness}, and \textit{authority}.

\subsection{Homophily of User Groups}
Next, we evaluate whether there is homophily at the moral group level: do users preferentially communicate with in-group members, and if so, how do they cross-compare? Given that user group membership is determined by their Social-LLM embeddings, which are learned from social network connections, it is reasonable to expect group homophily. This might initially render the research question ill-posed. However, the importance of this section lies in comparing the degree of moral homophily among user groups to illuminate differences in group-based communication strategies.

We cannot use assortativity to measure group homophily because group membership is categorical, not numerical. However, we can empirically compute how often users from group $X$ will retweet or mention users from group $Y$ compared to a null model. Let $P(X\xleftarrow{rt} Y)$ be the actual proportion of tweets published by user group $Y$ that are retweeted by user group $X$ out of all the retweets by user group $X$. We then randomly re-assign the group identity of the users to compute the null model $P_{rand}(X\xleftarrow{rt} Y)$. This procedure controls for the fact that the moral groups are not even in size, so some user groups don't appear to receive more communication only because they have more users. We then examine $P/P_{rand}$, which would be $>1$ if $X$ communicates with $Y$ more frequently than the null baseline and $<1$ if $X$ communicates less frequently with $Y$.  

\begin{figure}[t]
    \centering
    \includegraphics[width=0.6\linewidth]{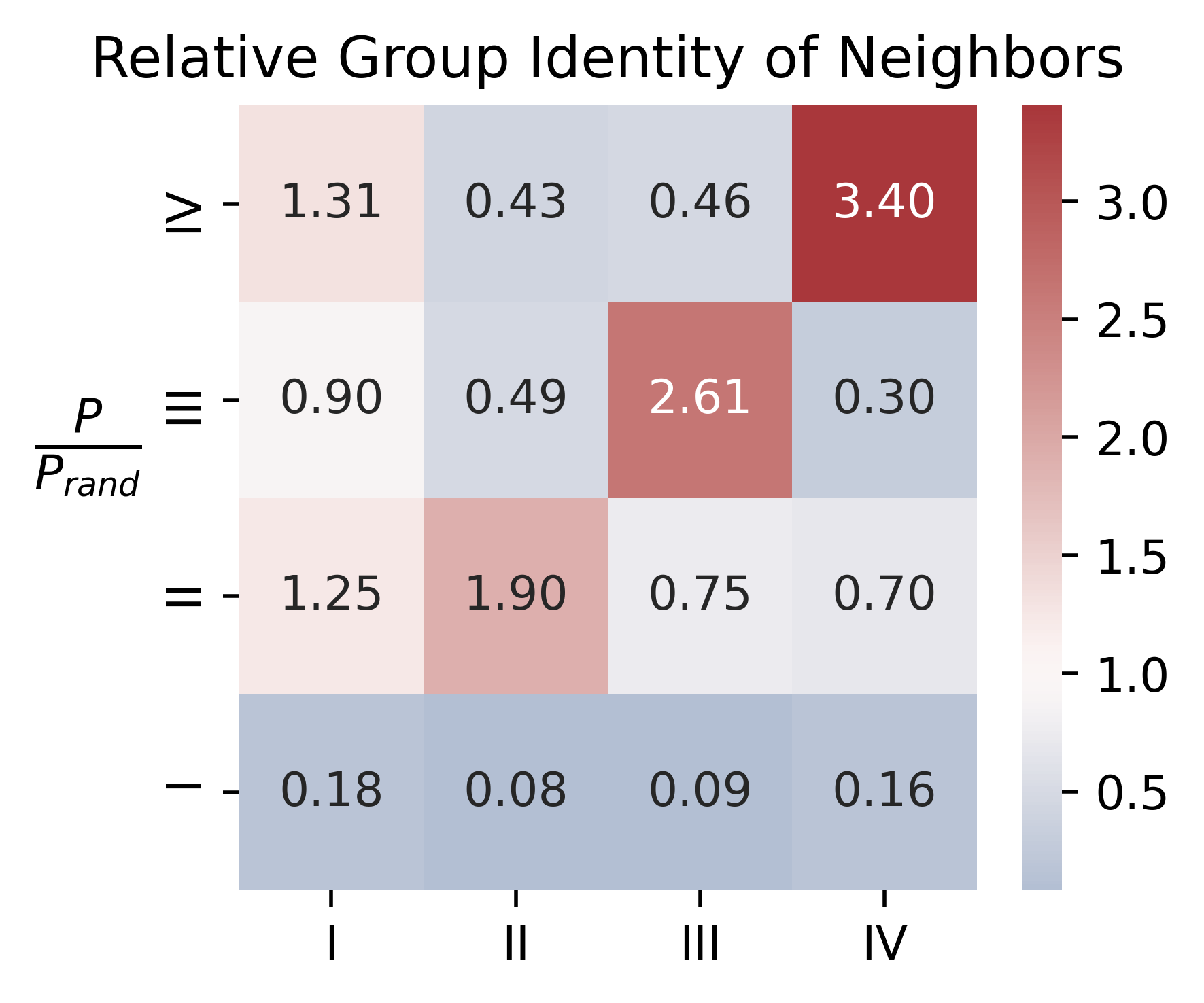}
    \caption{The ratio of how often COVID-19 dataset users in group $X$ retweet users in group $Y$ divided by the null baseline amount. $>1$ (red) cells indicate $X$ is more likely to retweet from $Y$ than the null baseline, and $<1$ (blue) cells indicate the opposite.  }
    \label{fig:rel_neighbor_group_rt}
\end{figure}

The results for the retweet network are shown in Figure \ref{fig:rel_neighbor_group_rt}. We omit the results from the mention network, which are very similar. Users in group IV (right-leaning users with \textit{fairness} and \textit{authority} morals) display the strongest homophily, preferentially retweeting from users in the same moral group more than $3x$ as much as the null baseline. This trend is followed by users in group III (politically balanced users with \textit{care} and \textit{purity} morals) and II (politically balanced users with \textit{authority} and \textit{care} morals) but to a lesser extent. Users in groups II, III, and IV also retweet from the other three groups much less frequently than the null baseline. Interestingly, group I (left-leaning users with \textit{authority} and \textit{fairness} morals) users do not display preferential communication with in-group members. In fact, they retweeted themselves very infrequently and retweeted from other groups (II and IV) more than expected. As we have seen in Figure \ref{fig:tsne}, this could indicate that group I users are not characterized by strong homophilous social network ties but rather serve as peripheral members of the Twittersphere interacting with all users.
\begin{figure}[t]
    \centering
    \includegraphics[width=\linewidth]{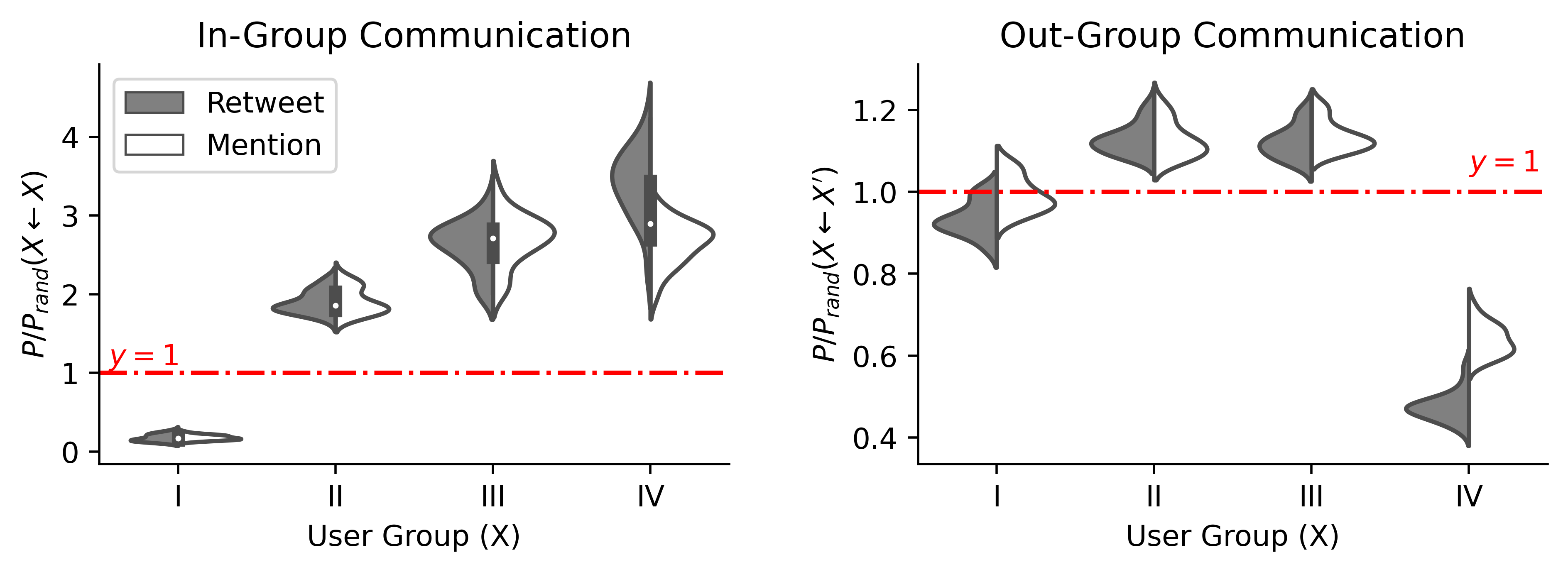}
    \caption{The ratio of how often COVID-19 dataset users in group $X$ communicate with in-group users (left) or out-group users (right) divided by the null baseline amount. $>1$ indicates $X$ is more likely to retweet from $Y$ than the null baseline.}
    \label{fig:in_out_group}
\end{figure}
\begin{sloppypar}
Figure \ref{fig:in_out_group} compares in-group communication $P/P_{rand}(X\leftarrow X)$ with out-group communication $P/P_{rand}(X\leftarrow X')$, where $X'$ include all the users not in $X$. We observe that group IV users have much higher in-group communication compared to the null baseline, followed by groups III and II, whereas group I does not favor in-group communication. However, we also see that groups II and III, the politically balanced user groups, have higher out-group communication compared to the null baseline. This is not true for group IV, which clearly prefers in-group communication and not out-group communication. This finding may signal that group IV, who are predominantly right-leaning extremists, is falling into a communication echo chamber, a potentially harmful manifestation of online communication. Our findings align with our prior research considering only user political orientation in Chapter \ref{chp:poli_covid}), which showed that right-leaning users discussing COVID on Twitter are situated in a tight-knit political echo chamber.
\end{sloppypar}

\section{RQ3: Bridging Moral Divides}

\begin{table}
    \centering
    
    \caption{For every user group, we show the top five moral combinations used in messages that are retweeted more often by out-group members than in-group members. The list is sorted by the ratio $ C(X, X',m)/C(X, X,m)$. Highlighted moral foundations are the ones that the user group does not favor (cf. Figure \ref{fig:user_clusters}). }
    \label{tab:out_mf_combo}
    \begin{tabular}{lwl{4cm}c}
         \toprule 
         \multicolumn{2}{l}{\textbf{Moral Foundations}} & \textbf{Ratio} \\
         \midrule
         \multicolumn{3}{l}{\textit{I: Authority \& Fairness}}\\
         & \textcolor{purple}{\textbf{Care}}, Purity & 2.316 \\
         & Fairness & 2.010 \\
         & \textcolor{purple}{\textbf{Care}}, Authority, Purity & 1.895 \\
         & \textcolor{purple}{\textbf{Care}}, Loyalty & 1.558 \\
         & \textcolor{purple}{\textbf{Care}}, Fairness, Loyalty & 1.293 \\
         \midrule 
         \multicolumn{3}{l}{\textit{II: Authority \& Care}}\\
        & Care, Loyalty, \textcolor{purple}{\textbf{Purity}} & 1.299 \\
        & \textcolor{purple}{\textbf{Fairness}}, Loyalty & 1.151 \\
        & Care, \textcolor{purple}{\textbf{Fairness}}, Loyalty & 1.137 \\
        & Authority, Loyalty, \textcolor{purple}{\textbf{Purity}} & 1.114 \\
        & Authority, Loyalty & 1.099 \\
        \bottomrule 
        \end{tabular}
        \quad
        \begin{tabular}{lwl{4cm}c}
        \toprule
         \multicolumn{2}{l}{\textbf{Moral Foundations}} & \textbf{Ratio} \\
         \midrule 
         \multicolumn{3}{l}{\textit{III: Care \& Purity}}\\
        & \textcolor{purple}{\textbf{Fairness}}, \textcolor{purple}{\textbf{Authority}}, Purity & 5.049 \\
        & Care, \textcolor{purple}{\textbf{Authority}}, Purity & 3.098 \\
        & Care, Purity & 2.059 \\
        & Care & 1.566 \\
        & \textcolor{purple}{\textbf{Fairness}}, \textcolor{purple}{\textbf{Authority}} & 1.177 \\
         
         \midrule 
         \multicolumn{3}{l}{\textit{IV: Fairness \& Authority}}\\
         & \textcolor{purple}{\textbf{Care}}, Loyalty & 4.053 \\
        & \textcolor{purple}{\textbf{Care}}, Authority, Loyalty & 2.303 \\
        & \textcolor{purple}{\textbf{Care}}, Authority, \textcolor{purple}{\textbf{Purity}}& 1.822 \\
        & \textcolor{purple}{\textbf{Care}}, Loyalty & 1.614 \\
        & Authority & 1.158 \\
        \bottomrule 
    \end{tabular}
\end{table}
Previously, we found that users of certain moral typologies communicate frequently with in-group members (groups II, III, and IV; group IV also displays a lack of communication with out-group members. In this section, we consider whether there is any pattern in the moralities or the combination of moralities in messages that tend to travel across groups. 
% Here, we also want to contend with the fact that messages can contain more than one moral foundation. 
Let $m$ be the 5-dimensional multilabel indicator vector of the moral foundations present in a tweet. We then count the frequency of the moral foundation combinations of every tweet published by user group $X$ that was retweeted or mentioned by a user that is not from group $X$, denoted by $C(X, X', m)$. To draw a suitable comparison, we also count the moral foundation combinations' occurrences of retweets or mentions within a group, denoted by $C(X, X, m)$. The ratio $C(X, X',m)/C(X, X,m)$ thus reflects the moral foundation combinations that were retweeted more frequently by in-group members than out-group members. Sorting by this ratio, we identify the top five moral foundation combinations that were retweeted by out-group members for every group in Table \ref{tab:out_mf_combo}, highlighting the moral foundations that are not the preferred moral foundation by that specific user group.

Key insights emerged from this analysis. First, messages with moral foundations that were less favored by the user group were retweeted much more frequently by out-group members. Second, combinations featuring multiple moral foundations garnered heightened traction among out-group members, especially since tweets with multiple moral foundations are extremely rare: tweets with two moral foundations only make up 13\% of all tweets, and tweets with three or more moral foundations make up less than 1\% of all tweets. The implications are profound, suggesting avenues of facilitators of social network connections by enhancing one's moral diversity and moral plurality. We also offer support for research in moral reframing as a tool to bridge ideological gulfs \cite{feinberg2019moral}.

However, it's important to acknowledge the limitations of this analysis. While we shed light on existing communication patterns, we cannot definitively causal relationships. There might be untapped moral combinations that could potentially resonate well with out-group members but remain unexplored within our dataset. Nonetheless, these findings offer valuable insights into understanding the dynamics of moral communication using observable data.

\section{Discussion}

In this paper, we present a large-scale empirical investigation of COVID-19 online conversation through the lens of moral psychology. Using a large dataset of COVID-19 tweets spanning nearly two years, we offer insights into the characteristics of the main types of users from a moral standpoint (RQ1), communication patterns among users of different moral typologies (RQ2), and how the morality of messages leads to more effective diverse communication (RQ3).

Based on users' moral foundation profiles, user metadata features, and social network data, we use social network representational learning to help uncover four distinct groups of users that differ in moral preferences, strength, and political affiliations. Group I users represent low-influence, left-leaning users who care mainly about \textit{authority} and \textit{fairness}, positioning themselves mostly at the peripherals of social network communication. Group II users is a politically balanced group who care about \textit{authority} and \textit{care}. The biggest contrast occurs between group III, who highly value \textit{care} and \textit{purity}, and group IV, who highly value \textit{fairness} and \textit{authority}; they are diametrically opposed in that each group values exactly the moral foundations that the other group does not. The two groups also differ considerably in their political partisanship composition: group III users are politically balanced, while group IV users are predominantly far-right. 

Analyzing communication patterns based on morality, our results illustrate patterns of moral homophily. We consistently find high homophily in the \textit{care}, \textit{fairness}, and \textit{authority} foundations. We also partially align our results with findings on \textit{purity} homophily \cite{dehghani2016purity}, although we observe \textit{purity} homophily only at specific time points that may relate to real-world events. We also find group homophily in groups II, III, and IV since they display preferential communication with their respective groups. Notably, group IV also demonstrates a substantial lack of communication with out-group members, hinting at the potential existence of harmful moral echo chambers. These findings provide valuable insights into the dynamics of moral homophily, emphasizing its presence across different moral foundations and its potential impact on group communication dynamics.

Finally, upon analyzing the moral foundation of messages that could bridge moral divides, we find that moral diversity and moral pluralism may be useful approaches. Messages that contain moral foundations their users don't usually prefer, and messages that contain multiple moral foundations tend to be more likely retweeted by out-group members. 

\subsection{Implications}
The findings of our study offer insights and recommendations for practitioners, health agencies, and researchers. We emphasize significant communication variations among users of various moral preferences regarding COVID-19. Importantly, we identify a group of users (IV), characterized by a preference for the \textit{fairness} and \textit{authority} foundations and a right-leaning political orientation, as potentially more susceptible to morally homogeneous messages. Adding to the wealth of literature on moral ideals separating the political left and right \cite{haidt2007morality,graham2009liberals}, we observe that the moral lines are not always so clear cut on this topic; users on both sides of the political spectrum can prioritize the same moralities that are both individualizing (\textit{care} or \textit{fairness}) and binding (\textit{authority} or \textit{purity}). A deeper dive into the complexity of our moral differences, in addition to political ideology factors, can lead to a more comprehensive understanding of online communication. Finally, our research sheds light on the potential usefulness of rhetorical tools focused on moral reframing to enhance the diversity of communication on online platforms.

\subsection{Limitations}
The research conducted in this study is strictly observational, and no causal relationships can be implied from the findings. The validity of the results is contingent upon the accuracy of various machine learning models utilized in the study, including those for morality detection and political partisanship detection. Additionally, it's essential to note that the scope of the results is limited to the topic of COVID, and generalizability to other topics may not be assured. These methodological considerations and limitations should be taken into account when interpreting and applying the study's results.

\subsection{Ethical Statement}
The data used in our study is publicly accessible \cite{chen2020covid}. Our study was exempt from review by the Institutional Review Board (IRB) as it solely relied on publicly available data. During our analysis, we protect user privacy by utilizing user IDs instead of screen names. Further, in the interest of user confidentiality, we present only aggregated statistics in this chapter. In conclusion, we put forth ethical recommendations, proposing the integration of moral psychology in comparable research endeavors and the development of rhetorical tools specifically designed for moral reframing. This initiative aims to enrich the diversity of communication on online platforms. Acknowledging a potential risk, we recognize the possibility of malicious actors manipulating public opinion by manipulating moral foundations. However, we contend that the positive outcomes and contributions of our research far outweigh this risk. The authors declare no competing interests.

%%% conclusions %%%%%%%%%%%%%%%%%%%%%%%%%%%%%%%%%%%%%%%%%%%%%%%%%%%%%%%%
\chapter{Conclusion}
\label{chp:conclusion}

Social media provides an online space where we can connect digitally with people everywhere in the world and spread content almost instantaneously. Despite the benefits of faster communication, social media also presents significant challenges, such as the spread of misinformation, online hate speech, privacy concerns, and political polarization. Through this dissertation, I identify problems on social media and offer data-driven insights by leveraging computational social science methodologies and the concept of network homophily.

To address the technical challenge of modeling large and sparse real-world social network data, I propose a novel social network representation learning method called Social-LLM in Part \ref{part1}. Social-LLM integrates user content with social connections to create unified representations, enabling accurate prediction and insightful visualizations of user attributes, communities, and behavioral propensities. The scalability, inductiveness, and task-agnostic nature of Social-LLM render it an effective tool for modeling extensive and complex social network data, thereby addressing a broad spectrum of open-ended research questions. I conduct a comprehensive evaluation of this method across seven real-world social network datasets, covering a wide variety of topics and detection tasks. This demonstrates Social-LLM's versatility and its potential to propel research in computational social science forward.

Part \ref{part2} of this dissertation explores the complex dynamics of online communication and behavior, particularly in the context of pressing social and political issues. I shed light on the ways in which social media platforms can amplify and reinforce certain behaviors, ideologies, and polarization.

In Chapter \ref{chp:poli_covid}, I investigate the politicization of COVID-19 discourse on Twitter, revealing the existence of echo chambers, particularly within the right-leaning community. This research highlights the relationship between information dissemination and political preferences, emphasizing the need for effective public health communication strategies.

Chapter \ref{chp:hate} delves into the role of social approval in driving online hate messaging, suggesting that the pursuit of social approval, rather than a direct desire to harm, may be the primary motivator for users engaging in toxic behavior. The study establishes a connection between receiving social approval signals and increases in subsequent toxicity, with retweeting playing a particularly prominent role in escalating toxicity.

In Chapter \ref{chp:moral_covid}, I examine the moral underpinnings of online discussions surrounding COVID-19, identifying distinct user groups characterized by differences in morality, political ideology, and communication styles. This study uncovers patterns of moral homophily and the existence of a potential moral echo chamber while also highlighting the effectiveness of messages that incorporate diverse and multitude of moral foundations in resonating with out-group members.

The unified theme connecting these studies is the exploration of how social media platforms shape and amplify user behavior, ideologies, and polarization. Each study focuses on a specific aspect of online communication--political discourse, moral foundations, or hate messaging--but they all contribute to a broader understanding of the complex social dynamics at play within online environments. Collectively, these studies serve the purpose of providing valuable insights into the mechanisms that contribute to unhealthy and unproductive communication online.

\section{Limitations}

Certainly, this research is not without its constraints. A notable limitation is the data. The exclusive use of a Twitter dataset might not encapsulate the entirety of social media users, presenting a significant limitation in the generalizability of my research. For instance, US Twitter users are more likely to be younger and more liberal-leaning than the general US population \cite{pew2019sizing}. However, given the inaccessibility of other datasets for academic researchers, and the popularity of Twitter as one of the most utilized social media platforms, this research still holds significant merit. Additionally, my focus is narrowly placed on retweets and mentions, sidelining potentially insightful social network dynamics like follower relationships or other forms of social connections due to API rate limit challenges. Despite these challenges, Twitter remains a valuable data source that offers a glimpse into large-scale human interactions and user relationship dynamics.

Another considerable limitation lies in the scope of interpretation of the findings. The observational nature of the data, coupled with the limited information on each user, makes it difficult to establish any causal relationships. Nevertheless, the insights garnered from this dissertation are of substantial importance and lay the groundwork for future studies to further examine and validate these observations. Moreover, given the ethical concerns surrounding certain behaviors, such as hate speech, observational studies often represent the most ethical approach. Deliberately exposing individuals to harmful content for research purposes would be highly unethical, thus reinforcing the value and necessity of observational research despite its limitations.

\section{Broader Impact}

More broadly, my research offers insights into the intricate relationship among social media, user engagement, and societal challenges. These findings have significant implications for a wide array of stakeholders, including academic researchers, social media platforms, lawmakers, and the general public. 

Academics have the opportunity to use this knowledge to address the adverse impacts of online divisiveness, echo chambers, and harmful behaviors. By uncovering the root causes of these issues, scholars can pave the way for more effective approaches to foster healthier digital interactions and more enlightened societal conversations.

Social media platforms can use these insights to refine their operational policies, content moderation practices, and overall platform architecture. Gaining a deeper understanding of how their platforms influence user actions and propagate certain viewpoints enables these firms to proactively create a more welcoming and positive digital space. This might involve the development of machine learning algorithms, guardrails, or moderation techniques that promote a diversity of viewpoints, diminish the dissemination of false information, and prevent harmful conduct.

Lawmakers can leverage this research to craft regulations and laws that address the complexities of today's digital environment. As social media continues to mold public sentiment and dialogue, it's vital for legislative bodies to grasp the social mechanics within these platforms. The knowledge garnered from this research offers a solid base for formulating policy decisions that uphold transparency, responsibility, and the safeguarding of online user welfare.

The general public also stands to gain from the knowledge disseminated through this research. Highlighting the hazards associated with digital communication, such as informational echo chambers, societal polarization, and the proliferation of hostile behavior, I highlight the need to educate the general public on digital literacy. We should be considerate of our digital footprints and our digital consumption.

Through this dissertation, I attempt to decipher human behavior and social interactions in our rapidly digitizing world. I emphasize the pivotal influence of social networks in shaping both personal and collective actions in online communication. These insights are instrumental in guiding research and initiatives across diverse disciplines, such as public health, political science, and social psychology, illuminating future research into the complexity of human behavior in the digital age. While my focus here is on social media applications, it's essential to recognize that the significance of this work goes beyond social media alone. It extends into the broader realm of computational social science, aiming to harness computational methods to address the complex challenges of the digital era. Although social media is a central focus, the methodologies and insights developed in this research hold applicability across any digital domain where human interaction intersects with digital platforms.

\graphicspath{}

\clearpage
\begingroup
    \setstretch{1}
    \bibliography{main}
\endgroup

\addcontentsline{toc}{chapter}{References}

% \begin{singlespace}
%   % increase penalty such that we don't break entries over pages
%   % source: https://tex.stackexchange.com/a/43275
%   \patchcmd{\bibsetup}{\interlinepenalty=5000}{\interlinepenalty=10000}{}{}

%   % reduce spacing between each bibentry
%   \setlength\bibitemsep{0.9\baselineskip}

%   % don't justify-align entries: this prevents stretching out each line
%   \raggedright
%   \printbibliography[heading = none]
% \end{singlespace}
\appendix
% From mitthesis package
% Version: 1.01, 2023/07/04
% Documentation: https://ctan.org/pkg/mitthesis

% \chapter{Appendix for Chapter 2}
\chapter{}
\label{appendix:2}
This appendix provides supplementary material for Chapter \ref{chp:retweetbert}.

% \chapter{}

\section{Heuristics-based Pseudo-Labeling Details}
We show the exact hashtags in Table \ref{tab:top_50_hashtags} and media bias ratings in Table \ref{tab:media_all_sides} used in the heuristics-based pseudo-labeling of user political leanings. In the labeling process, all hashtags are treated as case insensitive.

\begin{table}
    \centering
    \caption{Hashtags that are categorized as either left-leaning or right-leaning from the top 50 most popular hashtags used in user profile descriptions in the COVID-19 dataset. }
    \label{tab:top_50_hashtags}
    \begin{tabular}{ll}
        \toprule
         \textbf{Left}& \textbf{Right} \\
         \midrule
         Resist & MAGA \\
         FBR & KAG \\
         TheResistance & Trump2020\\
         Resistance & WWG1WGA\\
         Biden2020 & QAnon \\
         VoteBlue & Trump \\
         VoteBlueNoMatterWho & KAG2020\\
         Bernie2020 & Conservative \\
         BlueWave & BuildTheWall \\
         BackTheBlue & AmericaFirst \\
         NotMyPresident & TheGreatAwakening \\
         NeverTrump & TrumpTrain \\
         Resister &\\
         VoteBlue2020\\
         ImpeachTrump \\
         BlueWave2020\\
         YangGang\\
         \bottomrule         
    \end{tabular}
    
\end{table}
\begin{table}
    \centering
    
    \caption{The Twitter handles, media URLs, and bias ratings from AllSides.com for the popular media on Twitter.}
    \label{tab:media_all_sides}
    % \footnotesize
    \begin{tabular}{@{}llll@{}}
        \toprule
        \textbf{Media (Twitter)} & \textbf{URL} & \textbf{Rating} \\
        \midrule
        @ABC & abcnews.go.com & 2 \\
        @BBCWorld & bbc.com & 3\\
        @BreitbartNews & breitbart.com & 5 \\
        @BostonGlobe & bostonglobe.com & 2 \\
        @businessinsider & businessinsider.com & 3\\
        @BuzzFeedNews & buzzfeednews.com & 1\\
        @CBSNews & cbsnews.com & 2 \\
        @chicagotribune & chicagotribune.com & 3\\
        @CNBC & cnbc.com & 3 \\
        @ CNN & cnn.com & 2 \\
        @DailyCaller & dailycaller.com & 5\\
        @DailyMail & dailymail.co.uk & 5 \\
        @FoxNews & foxnews.com & 4 \\
        @HuffPost & huffpost.com & 1 \\
        InfoWars* & infowars.com & 5\\
        @latimes & latimes.com & 2 \\
        @MSNBC & msnbc.om & 1 \\
        @NBCNews & nbcnews.com & 2 \\
        @nytimes & nytimes.com & 2 \\
        @NPR & npr.org & 3 \\
        @OANN & oann.com & 4\\
        @PBS & pbs.org & 3\\
        @Reuters & reuters.com & 3 \\
        @guardian & theguardian.com & 2 \\
        @USATODAY & usatoday.com & 3 \\
        @YahooNews & yahoo.com & 2 \\
        @VICE & vice.com & 1 \\
        @washingtonpost & washingtonpost.com & 2 \\
        @WSJ & wsj.com & 3\\
        \bottomrule
        \multicolumn{3}{l}{*The official Twitter account of InfoWars was}\\
        \multicolumn{3}{l}{banned in 2018.}\\
    \end{tabular}
\end{table}
\section{Data Preprocessing}
We restrict our attention to users who are likely in the United States, as determined by their self-provided location \cite{jiang2020political}. Following \citet{garimella2018quantifying}, we only retain edges in the retweet network with weights of at least 2. Since retweets often imply endorsement \cite{boyd2010tweet}, a user retweeting another user more than once would imply a stronger endorsement and produce more reliable results. As our analyses depend on user profiles, we remove users with no profile data. We also remove users with degrees less than 10 (in- or out-degrees) in the retweet network, as these are mostly inactive Twitter users.

\section{Hyperparameter Tuning}
All models producing user (profile and/or network) embeddings are fit with a logistic regression model for classification. We search over parameter \{\texttt{C}: [1, 10, 100, 1000]\} to find the best 5-fold CV value. We also use randomized grid search to tune the base models. For node2vec, the search grid is \{\texttt{d}: [128, 256, 512, 768], \texttt{l}: [5, 10, 20, 80], \texttt{r}: [2, 5, 10], \texttt{k}: [10, 5], \texttt{p}: [0.25, 0.5, 1, 2, 4], \texttt{q}: [0.25, 0.5, 1, 2, 4]\}. For GraphSAGE, the search grid is \{activation: [\texttt{relu}, \texttt{sigmoid}], $S_1$: [10, 25, 50], $S_2$: [5, 10, 20], negative samples: [5, 10, 20]\}. Both node2vec and GraphSAGE are trained for 10 epochs with hidden dimensions fixed to 128. Retweet-BERT is trained for 1 epoch.

% \chapter{Appendix for Chapter 5}
\chapter{}
\label{appendix:6}
This appendix provides supplementary material for Chapter \ref{chp:hate}.

\section{Data Details}
\begin{table*}[t]
    \centering
    \footnotesize
    
    \caption{The statistics of the number of followers and average numbers of likes, retweets, quotes, and replies per user in the hate speech dataset.}
    \label{tab:metric_stats}
    % \footnotesize
    \begin{tabular}{lrrrrrrrrrrr}
        \toprule
        & & \multicolumn{5}{c}{\textbf{Raw Metrics}} & \multicolumn{5}{c}{\textbf{Transformed}}\\
        \cmidrule(lr){3-7} \cmidrule(lr){8-12}
        & &  Min & Max & Mean & SD & Median   & Min & Max & Mean & SD & Median  \\
        \midrule
        \multicolumn{5}{l}{\textit{Bot Score} $<=0.5$ ($N=2,985$)} \\
        % \midrule
        & Followers &  1 &  1,232,710.00 &  3,359.18 &  32,838.93 &  557.00 &  0.30 &  6.09 &  2.73 &  0.71 &  2.75 \\
        & Likes     &  0 &    13,356.06 &    19.86 &    277.85 &    1.01 & -3.00 &  1.42 & -0.03 &  0.35 &  0.00 \\
        & Retweets  &  0 &     1,327.37 &     2.91 &     30.00 &    0.17 & -3.32 &  1.13 & -0.30 &  0.38 & -0.28 \\
        & Quotes    &  0 &      116.08 &     0.30 &      3.23 &    0.03 & -4.02 &  0.48 & -0.53 &  0.36 & -0.48 \\
        & Replies   &  0 &     1,492.49 &     1.53 &     28.37 &    0.23 & -3.32 &  0.54 & -0.28 &  0.32 & -0.24 \\
        \midrule
        \multicolumn{5}{l}{\textit{Bot Score} $<=0.8$ ($N=6,664)$} \\
        % \midrule
        & Followers &  1 &  3,627,321.00 &  7,093.31 &  86,518.45 &  559.00 &  0.30 &  6.56 &  2.70 &  0.87 &  2.75 \\
        & Likes     &  0 &    37,450.54 &    30.43 &    566.38 &    0.80 & -5.39 &  1.42 & -0.11 &  0.48 & -0.04 \\
        & Retweets  &  0 &     5,320.85 &     5.71 &     88.32 &    0.17 & -7.88 &  1.13 & -0.35 &  0.53 & -0.29 \\
        & Quotes    &  0 &      489.23 &     0.50 &      7.41 &    0.03 & -8.60 &  0.48 & -0.57 &  0.52 & -0.47 \\
        & Replies   &  0 &     1,492.49 &     1.96 &     28.47 &    0.20 & -7.61 &  0.54 & -0.34 &  0.46 & -0.28 \\
        \bottomrule
    \end{tabular}

\end{table*}

\subsection{User Statistics}
Table \ref{tab:metric_stats} displays the raw and transformed social engagement metrics of the users in our dataset.

\begin{figure}
    \centering
    \includegraphics[width=\linewidth]{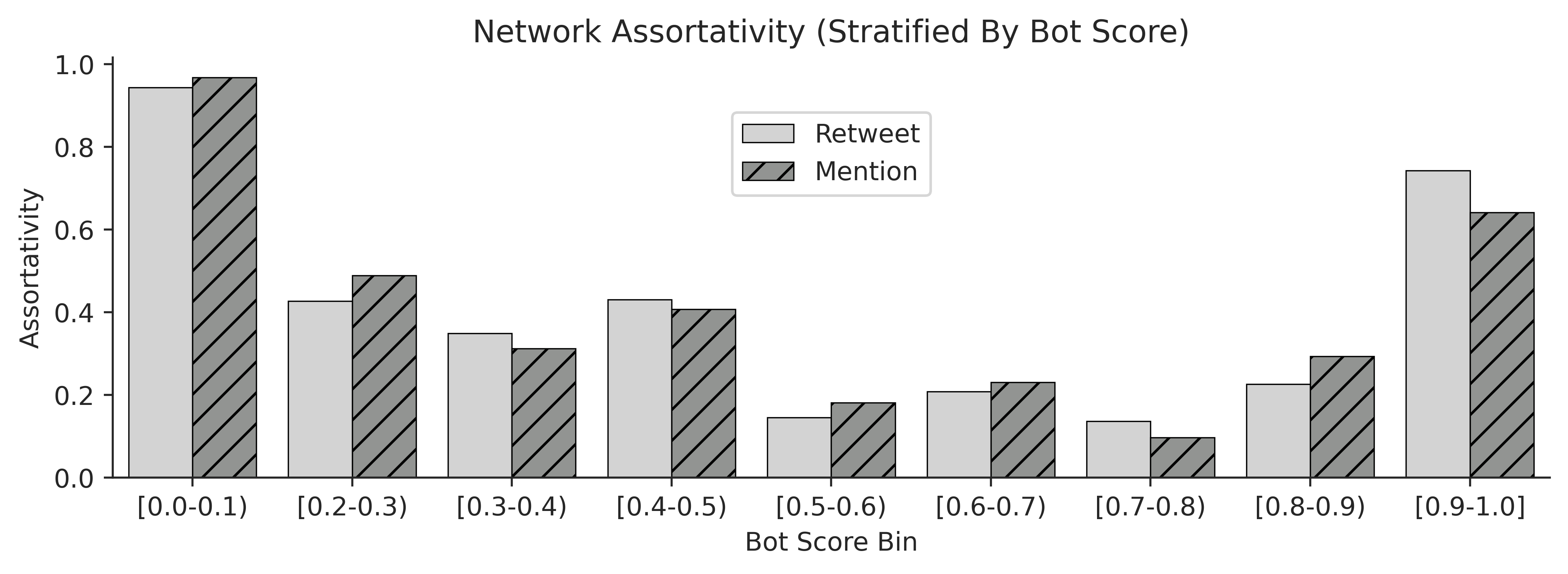}
    \caption{Hate score network assortativity of users in the same bot score bin in the hate speech dataset. All assortativity measures are significant ($p<0.001$), indicating that users are all assortative or homophilous with each other in terms of their hate scores.}
    \label{fig:bot_bin_assortativity}
\end{figure}

\subsection{Bot Score and Homophily in Toxic Behavior}
For completeness, we analyze the relationship between a user's bot tendencies and their homophily in-depth. We divide the users by their bot scores in 0.1 increments and measure the homophily in toxic behavior among users in the same bot score bin using network assortativity in Figure \ref{fig:bot_bin_assortativity}. Note that there are no users with bot scores between 0.1 and 0.2. We calculate the network assortativity of both the retweet and mention networks, which are all significant ($p<0.0001$). We find that hate score network assortativity is high for both the very human-like users (bot score bin $[0.0, 0.1)$) and the very bot-like  (bot score bin $[0.9, 1.0]$) and low otherwise. One explanation is that bots and humans do not preferentially attach themselves to each other, but rather, bots interact mostly with bots and humans interact mostly with humans (see \citeauthor{williams2020homophily}, \citeyear{williams2020homophily}). This finding motivates our choice to separate analyses in this work by bot scores.

\section{Social Engagement Experiment Details}

\subsection{Social Engagement Estimation Model}
We select the best Social-LLM models from RQ1 for use in the RQ2 modeling of social engagement estimation. The best models based on test performance all incorporated retweet edges, mention edges, profile description, and user metadata but differed regarding how the network feature was used. For users with bot threshold $=0.5$, the best model used the network edges as undirected and unweighted. For users with bot threshold $=0.8$, the best model used network edges as undirected but weighted. We use the following hyperparameters/settings when training the social engagement prediction model: learning rate $=0.001$, number of epochs $=20$, number of dense layers $=4$, hidden units $=768$, batch size $=32$, ReLU activation, and batch normalization.

% best 0.5 model: rt + mn, undirected unweighted, + profile and metadata
% best 0.8 model: rt + mn weighted, undirected, + profile and metadata

For users with bot threshold $=0.5$, our trained models produce $R^2$ values of 0.577, 0.515, 0.371, and 0.273 for predicting the number of likes, retweets, replies, and quotes, respectively, per tweet. For users with bot threshold $=0.8$, our trained models produce $R^2$ values of 0.531, 0.380, 0.167, and 0.467 for predicting the number of likes, retweets, replies, and quotes, respectively, per tweet. 

\begin{table*}
    \centering
    % \footnotesize
    
    \caption{The change in hate score when a user received lower vs. higher than expected amount of likes, retweets, replies, or quotes over a window of $k=50$ tweets in the hate speech dataset. We test the significance of the difference between the distributions using a Mann-Whitney U test (** $p<0.01$).}
    \label{tab:res_bot08}
    \begin{tabular}{lllllllllccc}
        \toprule
          & \multicolumn{3}{c}{\textbf{Lower Than Predicted}} & \multicolumn{3}{c}{\textbf{Higher Than Predicted}}& \\
         \cmidrule(lr){2-4} \cmidrule(lr){5-7}
         Metric & \# Ex & \# Users & $\Delta$ Hate & \# Ex & \# Users & $\Delta$ Hate &  Diff. in $\Delta $ Hate & Sig.  \\
         \midrule
         Likes & 2,397 & 748 & -0.001421 & 82,488 & 1,900 & 0.000204 & 0.001624  & ** &  \\
         Retweets & 67 & 52 & 0.006862 & 143,592 & 2,537 & 0.000073 & -0.006789 & N.S&  \\
        Replies & 4 & 1 & -0.004567 & 105,176 & 2,746 & -0.000169 & 0.004398 & N.S. & \\
        Quote & 14,693 & 1,034 & 0.000789 & 157,379 & 3,656 & 0.000148 & -0.000641& ** & \\
         \midrule
    \end{tabular}
    
\end{table*}
\section{Additional Results}
\subsection{Results with Higher Bot Threshold}
We report the results for users with bot scores $<=0.8$ in Table \ref{tab:res_bot08} using a time window of $k=50$. We note that due to the change in data distribution, there is no longer a sufficient number of samples where users experienced significantly fewer retweets and replies. Therefore, we cannot adequately analyze how retweets and replies affect the toxic behaviors of these users. The results for likes and quotes corroborate our main findings. When users experience more likes than expected instead of fewer likes than expected, they will become more toxic in the future. However, if they receive more quotes than expected instead of fewer quotes than expected, they will become less toxic in the future. Likes act as a positive social reinforcement, whereas quotes act as a negative social reinforcement. 

\begin{figure}[t]
    \centering
    \includegraphics[width=0.7\linewidth]{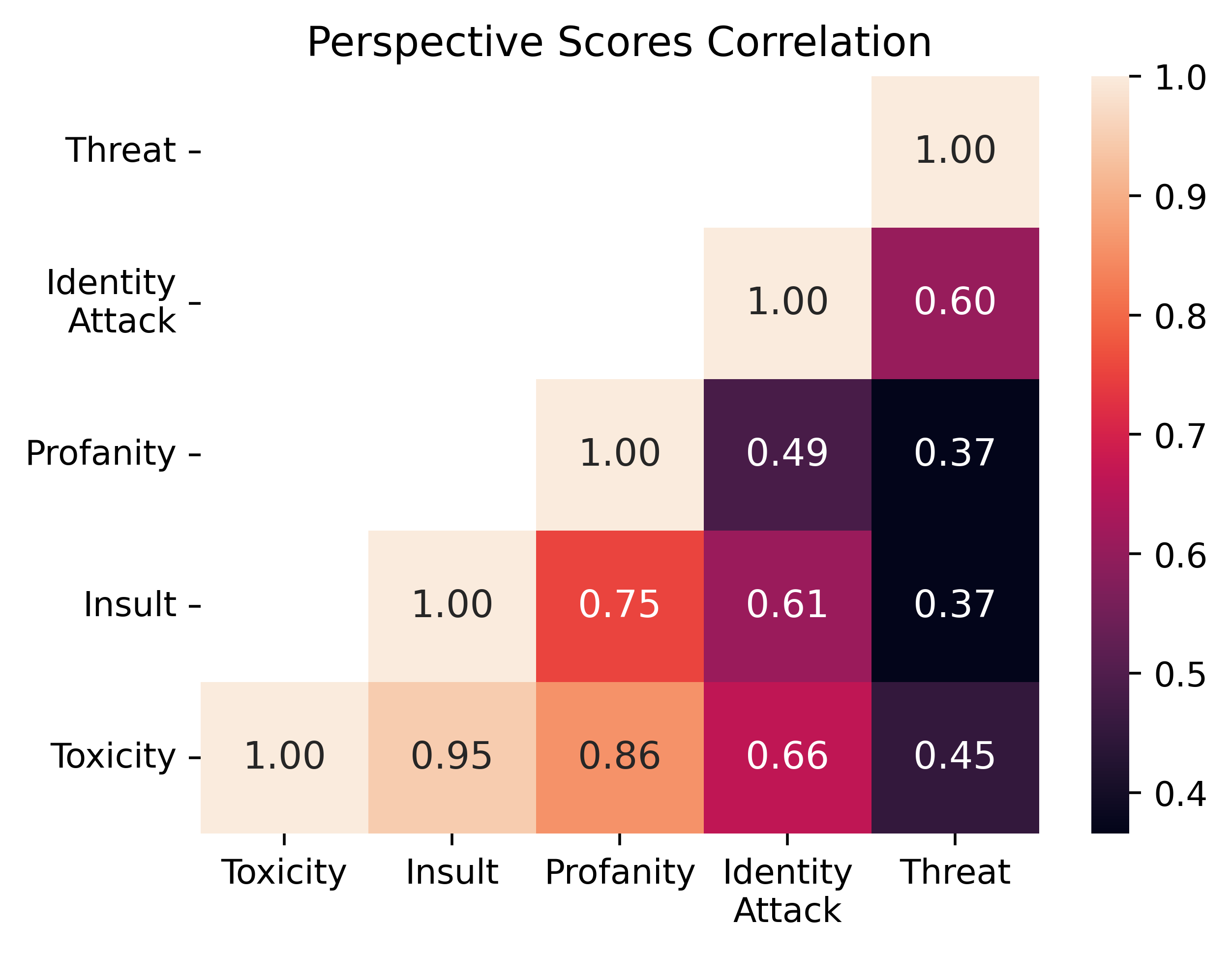}
    \caption{Pearson correlation between two Perspective (toxicity) scores of each tweet in the hate speech dataset. All values are statistically significant ($p<0.05$).}
    \label{fig:perspective_corr}
\end{figure}

\begin{figure}[t]
    \centering
    \includegraphics[width=0.7\linewidth]{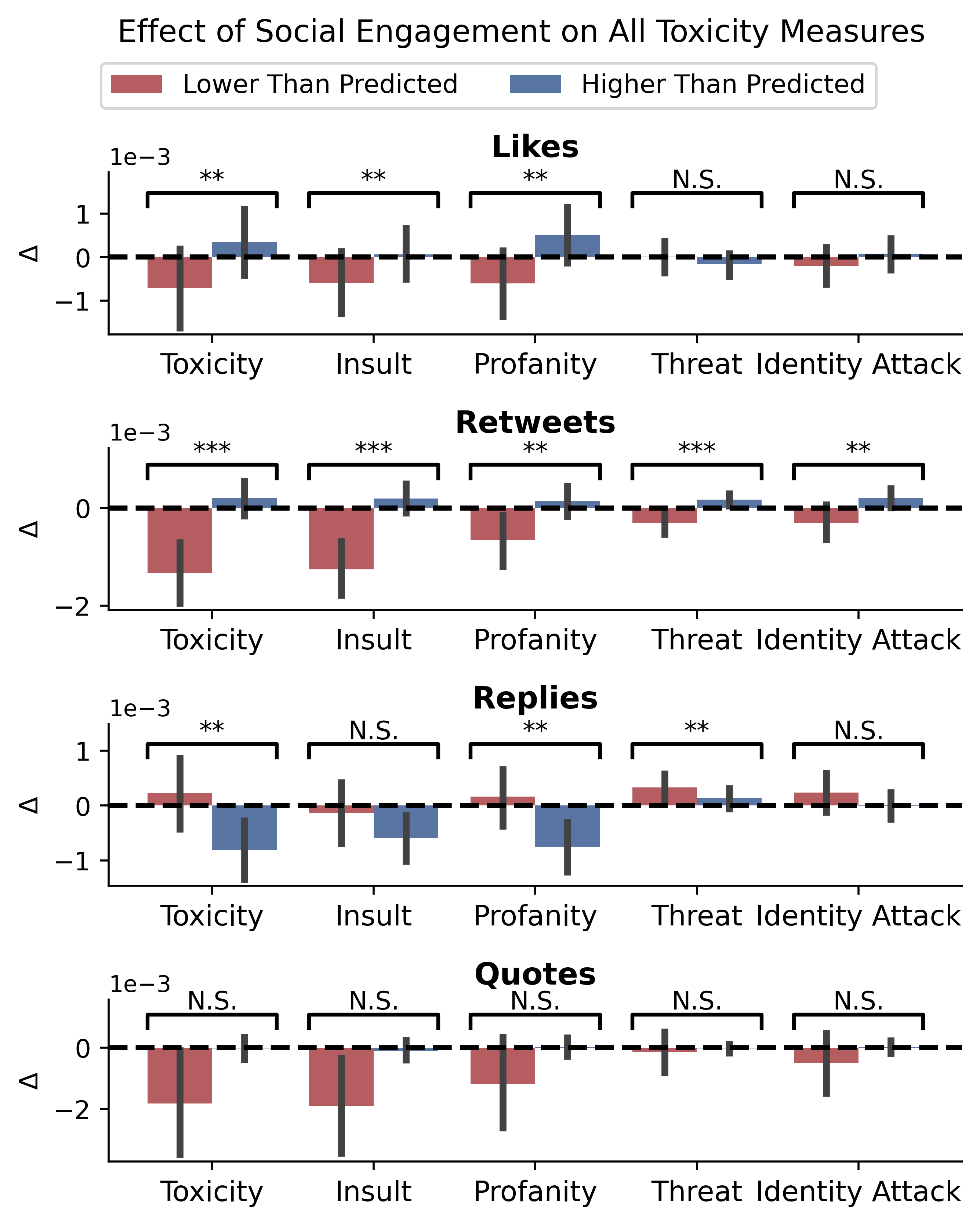}
    \caption{We display the effect of social engagement on all five toxicity attributes at $k=50$ in the hate speech dataset, similar to Figure \ref{fig:social_engagement_toxicity_diff}.}
    \label{fig:all_toxicity}
\end{figure}

\subsection{Robustness Checks with Other Toxicity Measures.} 
Besides the flagship \texttt{TOXICITY} score, the Perspective API also computes \texttt{IDENTITY\_ATTACK}, \texttt{INSULT}, \texttt{PROFANITY}, and \texttt{THREAT} scores, which are additional hateful messaging measures we use in robustness checks. We display the correlations among all five toxicity attributes of the tweets in our dataset in Fig \ref{fig:perspective_corr}. Most similar to \texttt{TOXICITY} is \texttt{INSULT} at $r=0.95$ ($p<0.001$), and the least similar is \texttt{THREAT} at a $r=0.45$ ($p<0.001$).

To ensure the validity of our results, we repeat our analysis with the four other toxicity attributes in Figure \ref{fig:all_toxicity} as a robustness check. The positive impact of likes on increased toxicity also occurs in \texttt{INSULT} and \texttt{PROFANITY}. The reduction of toxicity due to more replies is also found in \texttt{PROFANITY} and \texttt{THREAT}. Interestingly, retweets positively affect subsequent toxicity in all five toxicity attributes, further corroborating our contention about the potency of retweets.

\end{document}